\pdfoutput=1

\documentclass[
  twocolumn,
  prb,
  showpacs,
  amsmath,
  amssymb,
  superscriptaddress
]{revtex4}

\usepackage{bm}
\usepackage{bbm}
\usepackage{graphicx}
\usepackage{color}
\usepackage{bbold}

\usepackage{ulem}

\newcommand{\tr}{\text{tr}}
\newcommand{\LN}{\text{ln}}

\renewcommand{\v}[1]{\textbf{\textit #1}}
\newcommand\T{\rule{0pt}{3ex}} 
\newcommand\B{\rule[-2ex]{0pt}{0pt}} 

\definecolor{lightgray}{gray}{.9}

\begin{document}

\title{Interaction and disorder effects in 3D topological insulator thin films}

\author{E.\ J.\ K\"onig}
\affiliation{
 Inst. f\"ur Theorie der kondensierten Materie,
 Karlsruhe Institute of Technology, 76128 Karlsruhe, Germany
}
\affiliation{
 DFG Center for Functional Nanostructures,
 Karlsruhe Institute of Technology, 76128 Karlsruhe, Germany
}

\author{P.\ M.\ Ostrovsky}
\affiliation{
Max-Planck-Institute for Solid State Research, D-70569 Stuttgart, Germany
}
\affiliation{
 L.\ D.\ Landau Institute for Theoretical Physics RAS,
 119334 Moscow, Russia
}

\author{I.\ V.\ Protopopov}
\affiliation{
 Institut f\"ur Nanotechnologie, Karlsruhe Institute of Technology,
 76021 Karlsruhe, Germany
}
\affiliation{
 L.\ D.\ Landau Institute for Theoretical Physics RAS,
 119334 Moscow, Russia
}

\author{I.\ V.\ Gornyi}
\affiliation{
 Institut f\"ur Nanotechnologie, Karlsruhe Institute of Technology,
 76021 Karlsruhe, Germany
}
\affiliation{
A.\ F.\ Ioffe Physico-Technical Institute,
194021 St. Petersburg, Russia
}

\author{I.\ S.\ Burmistrov}
\affiliation{
 L.\ D.\ Landau Institute for Theoretical Physics RAS,
 119334 Moscow, Russia
}

\author{A.\ D.\ Mirlin}
\affiliation{
 Institut f\"ur Nanotechnologie, Karlsruhe Institute of Technology,
 76021 Karlsruhe, Germany
}
\affiliation{
 Inst. f\"ur Theorie der kondensierten Materie,
 Karlsruhe Institute of Technology, 76128 Karlsruhe, Germany
}
\affiliation{
 DFG Center for Functional Nanostructures,
 Karlsruhe Institute of Technology, 76128 Karlsruhe, Germany
}
\affiliation{
 Petersburg Nuclear Physics Institute,
 188300 St.~Petersburg, Russia.
}

\begin{abstract}

A theory of combined interference and interaction effects on the diffusive
transport properties of 3D topological insulator surface states is developed.
We focus on a slab geometry (characteristic for most experiments) and show
that interactions between the top and bottom surfaces are important at not too high
temperatures. We treat the general case of different surfaces (different carrier
densities, uncorrelated disorder, arbitrary dielectric environment, etc.). In order to access the low-energy
behavior of the system we renormalize the interacting diffusive sigma model in the
one loop approximation. It is shown that intersurface interaction is relevant in
the renormalization group (RG) sense and the case of decoupled surfaces is
therefore unstable. An analysis of the emerging RG flow yields a rather rich
behavior. We discuss realistic experimental scenarios and predict a
characteristic non-monotonic temperature
dependence of the conductivity. In
the infrared (low-temperature) limit, the systems flows into a metallic fixed
point. At this point, even initially different surfaces have the same transport
properties.  Investigating topological effects, we present a local
expression of the  $\mathbb Z_2$ theta term in the sigma model by first
deriving the Wess-Zumino-Witten theory for class DIII by means of non-abelian
bosonization and then breaking the symmetry down to AII. This allows us to
study a response of the system to an external electromagnetic field. Further,
we discuss the difference
between the system of Dirac fermions on the top and bottom surfaces of a topological insulator slab and its
non-topological counterpart in a double-well structure with strong spin-orbit
interaction.

\end{abstract}

\maketitle

\section{Introduction}

Topological states of matter have recently attracted immense scientific interest
which was in particular boosted by the theoretical
prediction\cite{BernevigZhang, BernevigHughesZhang, FuKaneMele, MooreBalents,
Roy} and subsequent experimental discovery\cite{KonigMolenkampetc, Hsieh} of
two-dimensional (2D) and three-dimensional (3D) time reversal invariant
topological insulators.

In their bulk, topological
insulators\cite{HasanKaneReview,QiZhang,SchnyderLudwig08,Kitaev09} (TI) are
electronic band insulators characterized by a topological invariant which
accounts for the non-trivial structure of the Bloch states. In contrast, the
interface between two topologically different phases (e.g. TI - vacuum) hosts
gapless, extended boundary states. Their appearance is topologically protected
via the bulk-boundary correspondence.\cite{Gurarie} In retrospect we understand
that the quantum Hall effect (QHE)\cite{vonKlitzing} at given quantized
transverse conductance was the first example of a topological insulator: The
Landau levels provide the bulk band gap which is accompanied by the topological
TKNN\cite{TKNN82} number and the protected chiral edge states.

In contrast to the QHE, the newly discovered 2D and
3D topological insulators require the absence of magnetic
field and rely on strong spin-orbit interaction. Further, their topological
invariant takes only values in $\mathbb Z_2$ (contrary to the TKNN integer). The
2D TI phase (also known as quantum spin Hall state) was experimentally
identified by the characteristic quantized value  $2 e^2/h$ of the
two-point conductance  in HgTe quantum wells.\cite{KonigMolenkampetc} The discriminating feature of all
3D TI is the massless Dirac states on the 2D boundary which were first
spectroscopically detected in BiSb\cite{Hsieh} alloys and subsequently in many
other materials.\cite{HasanKaneReview} To present date, various experimental
groups confirmed predominant surface state transport (for a review see Ref.
\onlinecite{Culcer}), in particular elucidating ambipolar field-effect\cite{Steinberg,CheckelskyHor2011,KimFuhrer2012,HongCui2012, KimQuiziFuhrer2012} and the typical QHE-steps of Dirac
electrons,\cite{ChengChen2010,HanaguriSasagawa2010,AnalytisFisher2010,SacepeMorpurgo2011,BruehneHgTe} Aharonov-Bohm oscillations\cite{PengCui2010,XiuWang2011,DufouleurGiraud2013} as well as weak antilocalization (WAL) corrections in the magnetoconductivity data.\cite{SteinbergPRB,ZhangLi2011,Chen} Moreover, several transport experiments reveal the importance of electron-electron interactions in 3D TI materials.\cite{Chen,Wang,Liu}

Inspired by recent experimental advances, we present here a detailed analysis of
interference and interaction corrections to conductivity in the most conventional setup for
transport experiments: the slab geometry, in which the  3D TI films are rather
thin (down to $\sim 10$ nm) although still thick enough to support well
separated surface states. As we will explain in more detail, the long-range
Coulomb interaction between the two major surfaces plays an important role. We
derive the quantum corrections to conductivity in the diffusive regime by taking
into consideration the WAL effect as well as
corrections of Altshuler-Aronov type \cite{AltshulerAronov}
induced by inter- and intrasurface interaction. We consider the
general situation of different surfaces subjected to different random potentials, mismatch in carrier densities
and unequal dielectric environment.

The present paper constitutes a natural extension of the previous
work\cite{OGM2010} by three of the authors in which a single 3D TI surface was
analyzed. It was found that the interplay of topological protection and
interaction- and interference-induced conductivity corrections drives the system
into a novel critical state with longitudinal conductance of the order of
$e^2/h$. As we show below, the intersurface interaction in a thin
TI slab makes the overall picture much more complex and crucially affects the
ultimate infrared behavior.

In another recent paper, \cite{BurmistrovGornyiTikhonov} two of us were involved in the
theoretical investigation of inter- and intrawell interaction effects in double quantum
well heterostructures studied experimentally in Ref. \onlinecite{Minkov}. Let us point out key
differences between the present paper and that work. First,
in Ref.~\onlinecite{BurmistrovGornyiTikhonov} only
equal carrier densities were considered. Second, disorder was assumed to be the
same in both quantum wells (and thus completely correlated). This affects the
soft-mode content of the low-energy theory. Third, quantum
wells host electrons with spin degeneracy which can be lifted by a
magnetic field. As a consequence, i) electrons in double quantum well fall into a symmetry class different
from that of 3D TI and ii) more interaction channels have to be included.
These {subtleties} affect in a crucial way the renormalization group (RG) flow: according to the analysis
of Ref.~\onlinecite{BurmistrovGornyiTikhonov}, the interwell interaction becomes irrelevant at
low energies, which is opposite to what we find in the two-surface TI model in
the present paper. Finally, the TI problem shows topology-related effects that
{were} absent in the double quantum well structure.

As in the two preceding works, we here use the interacting, diffusive non-linear
sigma model (NL$\sigma$M) approach to capture the diffusive low-energy physics.
Quantum corrections to the longitudinal conductivity $\sigma$ are obtained by
renormalization of this effective action in the one loop approximation (i.e.
perturbatively in $1/\sigma$ but exactly in interaction amplitudes). The
interacting NL$\sigma$M was originally developed by
Finkel'stein in the eighties \cite{Finkelstein1983,Finkelstein1984} (for review articles see Ref.s \onlinecite{Finkelstein1990,BelitzKirkpatrick,Finkelstein2010}). {In addition to perturbative RG treatment (which can also be performed diagrammatically \cite{CastellaniDiCastro}) it also allows one to incorporate topological effects} and was thus a
fundamental tool for understanding the interplay of disorder and interactions in
a variety of physical problems, including the superconducting
transition in dirty films, \cite{Finkelstein1987,FinkelsteinSITreview} the integer
QHE, \cite{PruiskenBaranov,PruiskenBurmi} and the metal-insulator transition in
Si MOSFETs. \cite{PunnooseFinkelstein}

Analyzing the RG equations for the thin 3D TI film, we find that (in contrast to the previous work on the double quantum well structure) the intersurface interaction is relevant in the
RG sense. The system flows towards a metallic fixed point at which even two
originally different surfaces are characterized by the same conductivities. As we discuss in detail below, the hallmark of the intersurface
interaction in 3D TI transport experiments is a characteristic non-monotonic
temperature dependence of the conductivity. In contrast to the case
of decoupled surfaces, due to the intersurface interaction, quantum corrections to the conductivity depend on the carrier densities. 

The paper is structured as follows. In Sec.~\ref{sec:Formalism}
we expose in detail the theoretical implications of a typical experimental slab
geometry setup, demonstrate the relevance of intersurface interaction and
introduce the microscopic fermionic Hamiltonian. Subsequently (Sec.
\ref{sec:towardsNLSM}), we use the non-Abelian bosonization technique to map the
fermionic theory on the ($\mathbf U\left (1\right )$-) gauged, interacting
NL$\sigma$M with $\mathbb Z_2$ topological term. Here we also discuss the Fermi
liquid treatment of generally strong electron-electron interactions. Next, we
renormalize the NL$\sigma$M in Sec.~\ref{sec:1loopRG}. {Sections \ref{sec:towardsNLSM} and \ref{sec:1loopRG} contain both pedagogical explanations and important details for experts. Readers purely interested in the results can jump to} Sec. \ref{sec:AnalysisofRG}, where the RG flow and
the implied phase diagram are analyzed.
Detailed predictions for typical experiments can be found in
Sec.~\ref{sec:experiment}. We close the paper by
summarizing our results and discussing prospects for
future work in Sec.~\ref{sec:conclusions}.

\section{Topological insulator slabs: Experimental setup and theoretical model}
\label{sec:Formalism}

\subsection{Setup}
\label{sec:setup}

In this work we analyze the effect of interaction on  transport properties of
strong 3D topological insulator thin films in the diffusive regime. While we
mainly focus on the theoretically most interesting case of purely surface
transport,
we also show that our theory can easily be extended to a case when only a
part of the sample is in the topological phase, i.e. one has a
conduction through a topologically protected surface spatially separated from a
thick (bulk) conducting region.

\begin{figure}
\includegraphics[scale=.44]{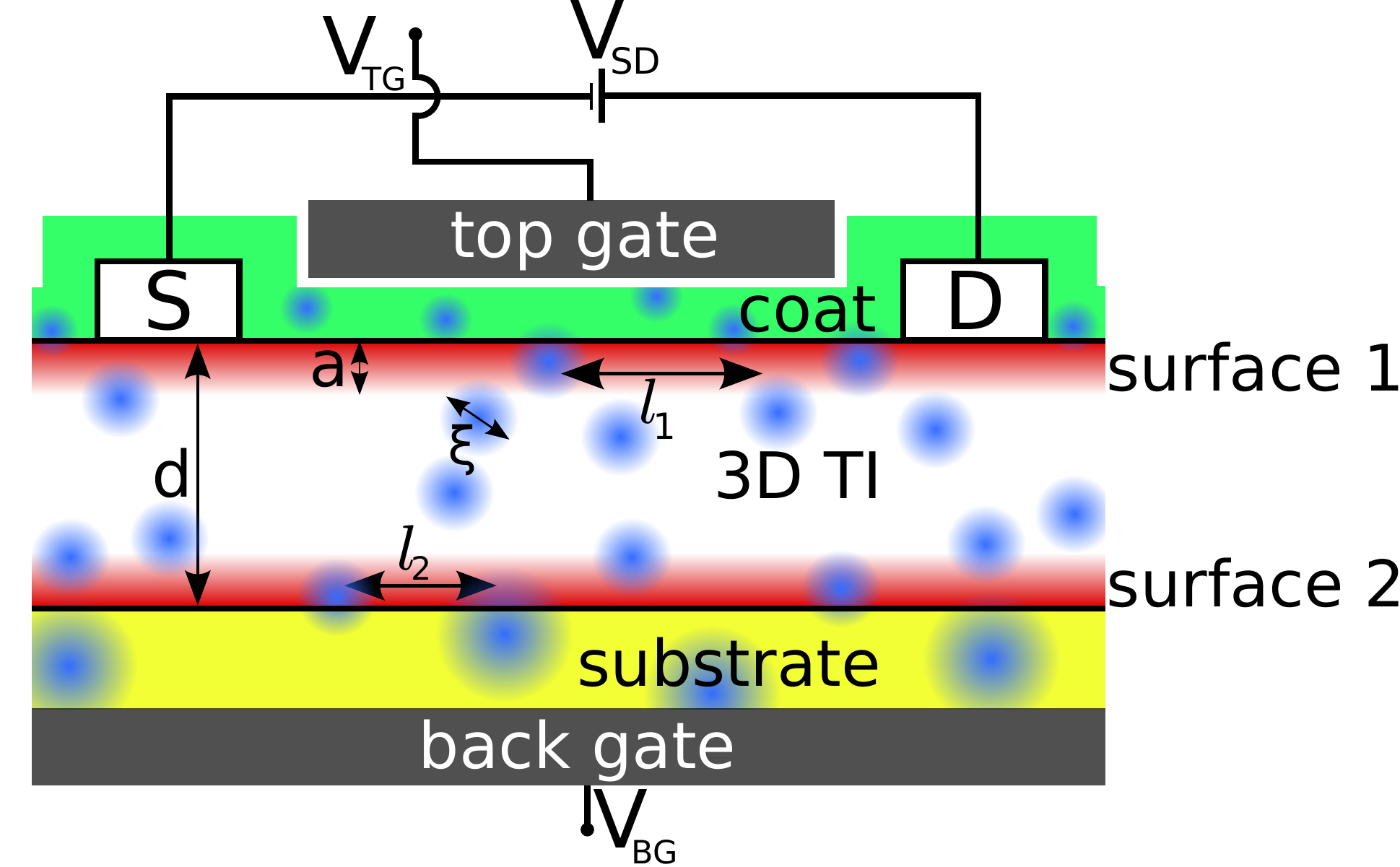}
\caption{Scheme of a typical experimental setup. The hierarchy of length scales
is explained in the main text.}
\label{fig:setup}
\end{figure}

A typical experimental setup is shown in Fig.~\ref{fig:setup}.
Our analysis is
valid in the regime where the  penetration depth of
surface states $a$ is small with respect to the film thickness $d$. We
therefore neglect intersurface tunneling (which would destroy the topological
protection).
Further, we assume the disorder correlation length (depicted by the range of
the impurity potentials) to be small $\xi \ll d$. We treat a generic case
when the vicinity to the coat or, respectively, to the substrate may induce
a different degree of disorder on the top and bottom surfaces. We thus consider the corresponding mean free
paths $l_1$ and $l_2$ as two independent parameters.
Moreover, we also allow the chemical potentials $\mu_1$ and $\mu_2$ on the two
surfaces to be different. {(By convention we set $\mu_s = 0$ at the Dirac point. Here and below $s = 1,2$ denotes the surface index.)} 
The chemical potentials may be experimentally controlled by means of
electrostatic gates. As has been stated above, we mostly focus on the situation
where both
$\mu_1$ and $\mu_2$ lie well within the bulk gap $\Delta_{\rm bulk}$. The
extension
of our results to the experimentally important regime when only one of
chemical potentials is located within the bulk gap, $\vert \mu_1 \vert \ll
\Delta_{\rm bulk} \lesssim \vert \mu_2 \vert$, can be found in section
\ref{sec:surfacebulk}.

If the electrostatic gates are present and too close\footnote{Closer than the
typical length scale $L_E$ of the system, see Eq. \eqref{eq:Diffregime2}.} to
the sample, Coulomb interaction is externally screened and the electron-electron
interaction is purely short range. However, such an experimental scenario is
a rare
exception from the rule. Therefore, in the main text we assume sufficiently
distant gates and concentrate on the limit of long-range Coulomb interaction. In
addition we derive general RG equations (Appendix \ref{sec:RGderiv})
which allow us to explore the crossover from the long-range case
to the short-range one, see Appendix \ref{sec:shortrange}. Qualitatively,
the RG flow for a sufficiently strong short-range interaction in the
case of externally screened surfaces turns out to be similar to the flow in
the absence of external screening.

Since we assume that the thickness $d$ of the sample is much smaller than its
other linear dimensions, we neglect contributions of four side faces of the
slab (whose area is proportional to $d$).

The goal of the present analysis is to study conduction properties of thin 3D TI films in the
diffusive regime, i.e., at energy scales $E$ far below the elastic
scattering rates $1/\tau_{s}$ of both surfaces,
\begin{equation}
E \ll \min_{s = 1,2} \hbar/ {\tau_{s}}\, . \label{eq:Diffregime1}
\end{equation}
In turn the elastic scattering rates are assumed to be small compared to the chemical potentials
\begin{equation}
\hbar/\tau_{s} \ll \vert \mu_s \vert. \label{eq:weakdisorder}
\end{equation}
In experiment $E$ is set
by the AC frequency ($E = \hbar \omega$) or by temperature ($E = k_B T$),
whichever of the two is larger. Equation \eqref{eq:Diffregime1} is equivalent to
the  hierarchy of length scales
\begin{equation}
l \ll L_E \label{eq:Diffregime2}\,,
\end{equation}
where we have introduced the maximal
mean free path $l = \max_{s=1,2} l_s$ and the length scale $L_E = \min_{s=1,2}
(\hbar D_s / E)^{1/2}$, with $D_s$ being
the diffusion coefficients for the two surfaces.

\subsection{Interaction}
\label{sec:IAimportant}

Can Coulomb interaction between the top and bottom surface states play an important role in
the experiment? To answer this question, we compare the sample thickness with
all natural length scales of the system: the screening length $l_{\rm scr}$, the
(maximal) mean free path $l$ and the experimentally tunable scale $L_E$.

The Coulomb interaction is (throughout the paper underlined symbols denote
$2\times 2$ matrices in the surface space)
\begin{equation}
\underline U_0 \left ( \v r \right )= \frac{e^2}{\epsilon}
\left (\begin{array}{cc}
\frac{1}{r} & \frac{1}{\sqrt{ r^2 + d^2}} \\
\frac{1}{\sqrt{r^2 + d^2}} & \frac{1}{r} \label{eq:Coulombnoelstat}
\end{array} \right ).
\end{equation}
The two dimensional vector $\v r$ connects the two dimensional positions of the
particles, $r=|{\bf r}|$, $e$ is the charge of the electrons, and $\epsilon$
denotes the effective dielectric constant.

Fourier transformation and RPA-screening leads to \cite{ZhengMacDonald,
KamenevOreg, Flensberg, BurmistrovGornyiTikhonov} ($U \equiv U\left ( \v q
\right ) \equiv 2\pi e^2/\epsilon q$)
\begin{equation}
\underline U_{\rm scr} \left ( \v q \right ) = \frac{\underline U}{1-
\left (\Pi_1 + \Pi_2 \right )U + U^2 \Pi_1 \Pi_2 \left (1 - e^{-2dq}\right )}
\label{eq:URPA}
\end{equation}
with
\begin{equation}
\underline U = U \left (\begin{array}{cc}
1-\Pi_2 U \left (1 - e^{-2dq}\right ) & e^{-dq} \\
e^{-dq} & 1-\Pi_1 U \left (1 - e^{-2dq}\right )
\end{array} \right ) . \notag
\end{equation}
Here $\Pi_s$ is the polarization operator of the surface states.

In the present section we will concentrate on the statically screened
interaction potential. In this limit the polarization operator is determined by
the thermodynamic density of states:  $ \Pi_{s}\left (\omega = 0, \v q\right ) =
-\nu_s$. 

In the diffusive regime defined by the condition \eqref{eq:Diffregime2}, the
wavevector $q$ satisfies the inequality $1/L_E \ll q \ll 1/l$.
Therefore, in a sample of thickness $d \gg L_E$ we always have $d q \gg 1$ and
the two surfaces decouple,
\begin{equation}
\underline U_{\rm scr} \stackrel{d \gg L_E}{=} 2\pi \frac{e^2}{\epsilon} \left
(\begin{array}{cc}
\frac{1}{q + \kappa_1} & 0 \\
0 & \frac{1}{q + \kappa_2}
\end{array} \right ), \label{eq:decoupledsurf}
\end{equation}
where $\kappa_s =  2\pi e^2\nu_s/\epsilon$ is the inverse Thomas-Fermi
screening length for a single surface $s$. A universal form of the
Altshuler-Aronov correction to conductivity induced by the
Coulomb interaction \cite{AltshulerAronov, OGM2010} arises in the unitary limit
when one can neglect $q$ as compared with $\kappa_s$ in
Eq.\eqref{eq:decoupledsurf}. The unitary limit is achieved if $\kappa_s^{-1} \ll
l$ (the meaning of this condition {as well as the complementary case} are
discussed in section \ref{sec:scattchannels}). 

In the opposite limit of a small interlayer distance, $d \ll l$, we can
approximate $e^{-dq} \approx 1$ in the whole diffusive regime. This implies
\begin{eqnarray}
\underline U_{\rm scr} \stackrel{d \ll l}{=} & \displaystyle \frac{2\pi
e^2}{\epsilon}\frac{1}{q +\kappa_1 + \kappa_2 + 2d\kappa_1 \kappa_2 \left
(1-qd\right )}\notag \\
& \times \left
(\begin{array}{cc}
1+2 \kappa_2 d & 1 \\
1 & 1+2 \kappa_1 d
\end{array} \right ). \label{eq:coupledsurf}
\end{eqnarray}
At the first glance, it looks as if also negative interaction potential was
possible. However, this is not the case as shall be explained in what follows.
Depending on the hierarchy of the lengthscales $\kappa_1^{-1}, \kappa_2^{-1}$
and $d$ the following scenarios are conceivable: 

First, consider $\kappa_s d \ll 1$ for both $s = 1$ and $s =2$. In this case,
the $q$ dependence of the interaction potential implies the definition of the
coupled layer screening length $l_{\rm scr}$:
\begin{equation}
\left (\underline U_{\rm scr}\right )_{ss'}\left (\v q\right ) \sim \frac{1}{q +
\kappa_1 + \kappa_2}\ \  \Rightarrow\ \  l_{\rm scr} = \frac{1}{\kappa_1 +
\kappa_2}. 
\label{eq:screeninglength}
\end{equation}
If in addition the condition $l_{\rm scr} \ll l$ is fulfilled,
the Coulomb interaction potential \eqref{eq:coupledsurf} becomes
``overscreened'' ($q$-independent) for all diffusive momenta $q \ll
l^{-1}$.

Second, assume that $\kappa_s d \gg 1$ for at least one surface. Then the
$q$-dependence of $\underline U_{\rm scr}$ is always negligible and thus the
notion of coupled layer screening length is meaningless. It is worthwhile to
remark that, as expected, the potential 
\eqref{eq:coupledsurf} reduces to the decoupled form 
\eqref{eq:decoupledsurf} in the limit when $\kappa_s^{-1} \ll d$ for both
surfaces (which also implies that $\kappa_s^{-1} \ll l$). 

In this paper we derive the conductivity
corrections in the unitarity limit of $q$-independent interaction, see Eqs. \eqref{eq:RGeqs}. As expected, in the
limit of decoupled surfaces, $\kappa_s^{-1} \ll d$, they reproduce the previous
result \cite{OGM2010}, while whenever $d \ll \kappa_1^{-1}$ or $d \ll
\kappa_2^{-1}$ novel conductivity corrections induced by
intersurface electron-electron interaction emerge.

Finally, in the intermediate regime $l \ll d \ll L_E$ the scale-dependent
conductivity can be obtained by the following two-step RG analysis. First, one
integrates the single-surface RG equations starting from the shortest scale $l$
up
to the intersurface distance $d$. After this, one uses the running
coupling constants at scale $d$ as starting values for the coupled-surface RG flow
and integrates these RG equations up to the scale $l_E$.

\begin{figure}
\includegraphics[scale=.55]{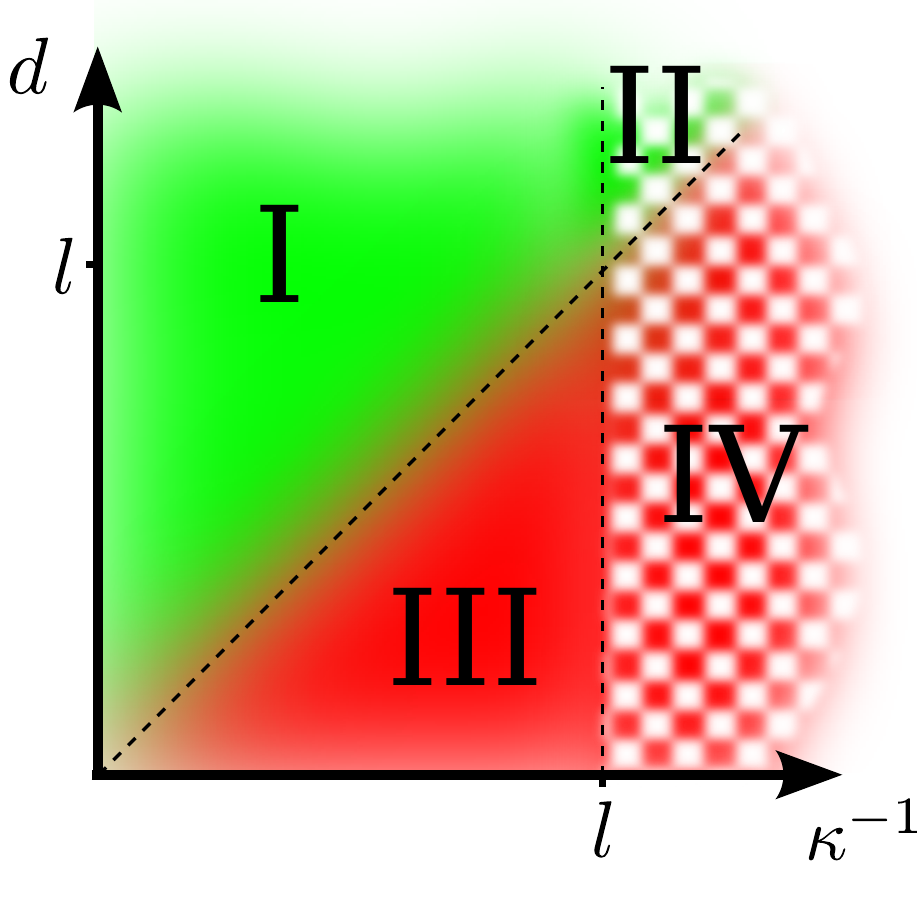}
\caption{Sketch of the regimes discussed in the main text for
the case of comparable screening lengths, $\kappa^{-1}_1 \sim \kappa^{-1}_2$
(denoted by $\kappa^{-1}$). The regimes \textbf{I} and \textbf{II} correspond to
effectively decoupled surfaces (studied in Ref. \onlinecite{OGM2010}), { while in regimes \textbf{III} and \textbf{IV} intersurface interaction is important.} The conductivity corrections in \textbf{I} {and \textbf{III}} are due to
``overscreened'' Coulomb interaction. In contrast, in \textbf{II} {and \textbf{IV}} this type of
corrections sets in only in the low-energy regime where the running length scale
(i.e. the typical scale $L_E$) exceeds the screening length.}
\label{fig:regimediag}
\end{figure}

Different regimes discussed above are shown schematically in Fig.
\ref{fig:regimediag} in the parameter plane $d$ -- $\kappa^{-1}$. For
simplicity, we assume there the two surfaces have comparable screening lengths:
$\kappa_1^{-1} \sim \kappa_2^{-1}$.

In the end of the paper, Sec.~\ref{sec:experiment}, we analyze in
detail the regions and limits of applicability of our theory with respect to
representative experimental setups. In particular, we show that the hierarchy
of scales $d \ll l \ll L_E$ is realistic.

\begin{figure}
\includegraphics[scale=.45]{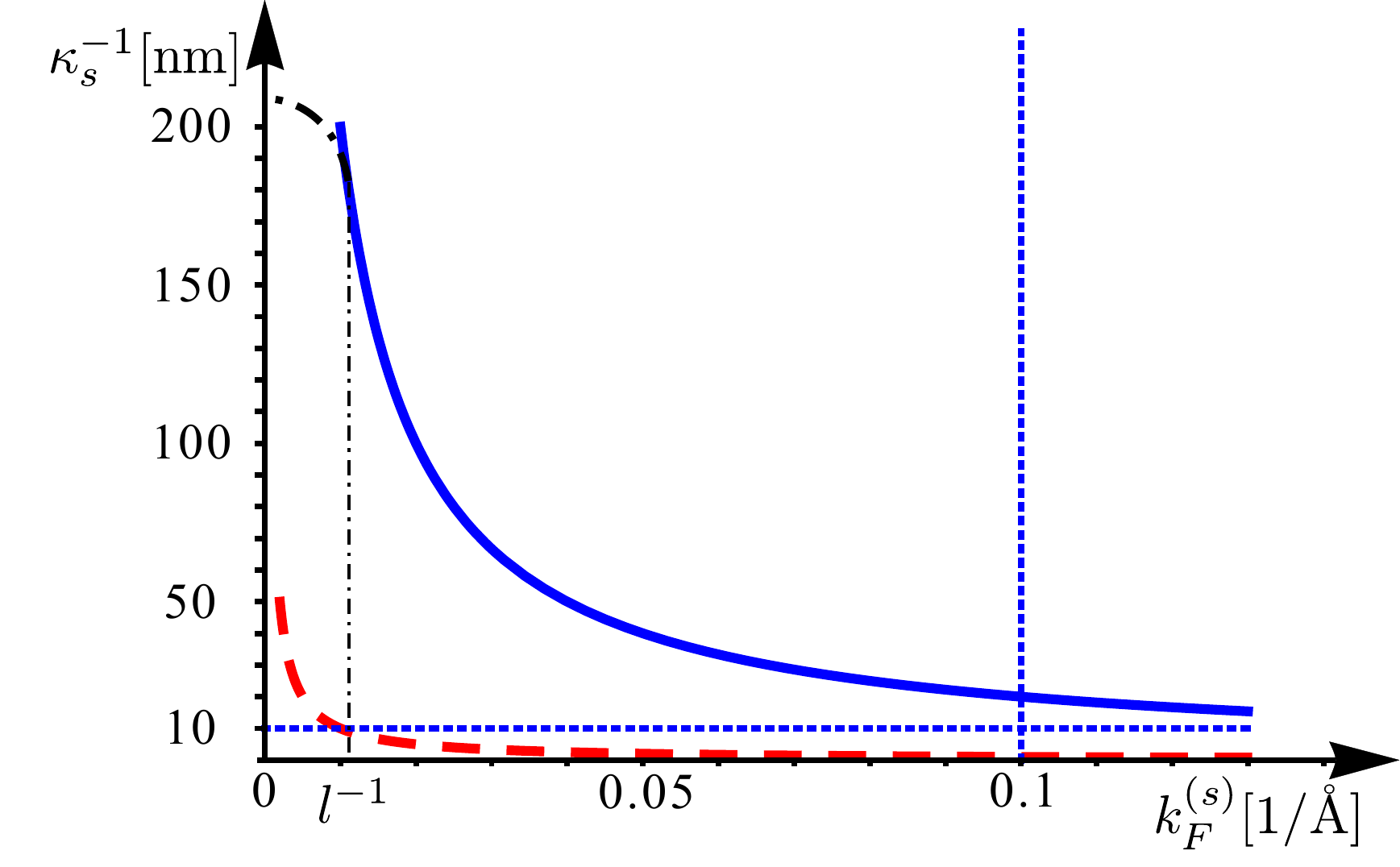}
\caption{Plot of the single surface screening length $\kappa_s^{-1}$. The red curve (large dashes) is the lower
bound (corresponding to $\alpha = 1$) of the screening length. The solid, blue curve is the screening length for Bi$_2$Se$_3$ film with experimental parameters given in Table \ref{tab:expvaluesBise} {in Sec.\ref{sec:2materials}}. For the latter, the required minimal thickness
and maximal Fermi momentum are also depicted (dotted blue lines). The
disorder-induced regularization of the divergence at small Fermi momentum is
schematically represented by the black dot-dashed curve. }
\label{fig:screeninglength}
\end{figure}

In order to illustrate the importance of intersurface
interaction (i.e. the relevance of the inequality $d \lesssim \kappa_s^{-1}$)
under realistic conditions, we show in Fig.~\ref{fig:screeninglength} a
dependence of the screening length on the Fermi momentum.

The density of states for the linear (Dirac) spectrum is $\nu (\mu_s) =
k^{(s)}_F / 2\pi \hbar v_F$, where $k^{(s)}_F$ is the Fermi wave vector of
the $s$-th surface state and $v_F$ the Fermi velocity. Therefore
\begin{equation}
\kappa_{s}^{-1} = \frac{1}{\alpha} \frac{1}{k^{(s)}_F } .
\label{eq:dcondition}
\end{equation}
We introduced the dimensionless parameter $\alpha = {e^2}/{\epsilon \hbar
v_F}$ which is the effective coupling constant of the Coulomb interaction and
is equal to ${c}/{\epsilon v_F}$ times the fine structure constant of
quantum electrodynamics.
Clearly, $\alpha$ plays the same role as the
dimensionless density parameter $r_s$ in conventional theories of electrons in
parabolic bands. We will assume that the interaction is not too
strong, $\alpha\lesssim 1$; otherwise the system may become unstable, see
a discussion at the end of Sec.~\ref{sec:Microham}.

The dashed red curve in Fig.~\ref{fig:screeninglength} represents the lower
bound (corresponding to $\alpha = 1$) of $\kappa_s^{-1}$ as a function of 
$k^{(s)}_F$. The actual value of 
$\kappa_s^{-1}$ for an exemplary case of Bi$_2$Se$_3$ (experimental parameters
can be found in Table \ref{tab:expvaluesBise} below) is depicted by the blue
solid curve. We see that the screening length can by far
exceed the thickness of the topological insulator slab. Indeed, the Bi$_2$Se$_3$
experiments \cite{Wang, Liu, Chen, Steinberg} are performed on probes of
thickness $d \simeq 1 - 100$ nm. For this material, our assumption of separate
gapless surface states (no tunneling) is both numerically\cite{Linder} and
experimentally\cite{Ando} shown to be valid down to $d \simeq 10$ nm (blue
horizontal dashed line). Thus, relevant experimental values of $d$ in the
experiments of interest range from $ d\simeq 10$ nm up to  $ d\simeq 100$ nm.
On the other hand,
surface electrons have a maximal Fermi wavevector of
$k_F \sim 0.1/\text{\AA}$ associated with $\mu = \Delta_{\text{bulk}} = 0.3$ eV,
see blue vertical dashed line. For the lowest concentration, increase of the
screening length is limited by disorder. In this way, we estimate the range
of $\kappa_s^{-1}$ as 20-200 nm, so that the condition $\kappa_s^{-1} > d$ can
be easily fulfilled. This is particularly the case for relatively thin films
($d \simeq 10$ nm) and in the vicinity of surface Dirac point.

The above analysis proves the relevance of the intersurface electron-electron interaction.
In fact, in course of this analysis we have made several simplifying
assumptions that require certain refinements; we list them for the reader's
benefit. First, in general, the
coating material ($\epsilon_1$), the topological insulator ($\epsilon_2$), and the substrate ($\epsilon_3$) are all dielectrica with different dielectric constants $\epsilon_1\neq\epsilon_2\neq\epsilon_3$. In order to
determine the exact Coulomb interaction, one has to solve the electrostatic
problem of a point charge in such a sandwich structure of dielectrica,\cite{Profumo2010,Katsnelson2011,CoulombDragCarrega} see
Appendix \ref{sec:Elstat}.
Second, the long-range Coulomb interaction is accompanied by short-range
contributions, which, in particular, induce corrections to the
polarization operator which affect the screening length. More precise
calculations taking Fermi liquid corrections into account can be found in
Section \ref{sec:cleanFLMaintext} and Appendix \ref{sec:cleanFL}.
Finally, we neglected the dependence of the Fermi velocity $v_F$ on the
chemical potential $\mu_s$, see
Sec.~\ref{sec:Microham}. However, all these refinements do not modify our
conclusion of the importance of interaction between the surface states. We now
proceed with presentation of the field-theoretical formalism that will allow us
to explore the problem.

\subsection{Microscopic Hamiltonian}
\label{sec:Microham}

\begin{figure}
\includegraphics[scale=.15]{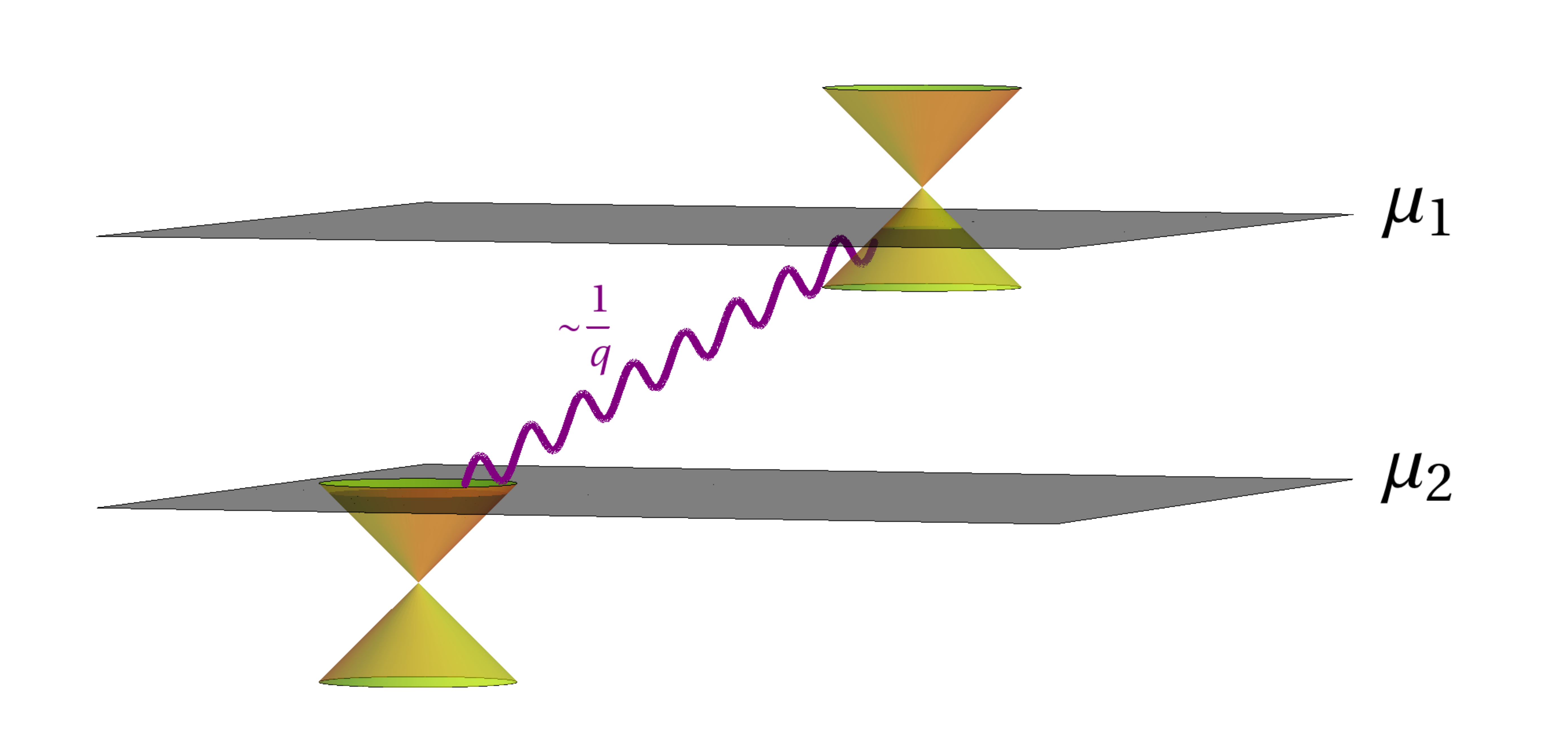}
\caption{Pictographic representation of the microscopic model: Diffusively
propagating surface states at different chemical potentials which interact with
each other by means of long-range Coulomb interaction. 
}
\label{fig:Microscopicaction}
\end{figure}

The model under consideration {is schematically depicted in Fig. \ref{fig:Microscopicaction}. It is described in path integral technique
\begin{equation}
\mathcal Z = \int \mathcal D \left [\bar \psi, \psi \right ] \; e^{-S\left [\bar \psi, \psi \right ]}
\end{equation} 
by} the following microscopic Matsubara action:
\begin{equation}
S\left [\bar \psi, \psi \right ] = \int_{\tau, \v x} \bar \psi \left
(\partial_\tau + \text H_0 + \text H_{\rm dis} \right ) \psi + S_{\rm int}.
\label{eq:Highenergyaction}
\end{equation}
The notation $\int_{\tau, \v x} = \int d^2x \int_0^\beta d \tau$ will be used throughout the article, where, as usual, $\beta = 1/T$ is the inverse temperature. If not specified otherwise, we set
Boltzmann's constant, Planck's constant, and the speed of light $k_B =
\hbar = c = 1$ in the remainder.
The fermionic fields $ \bar \psi \left (\v x, \tau\right ) = \left ( \bar
\psi^\uparrow_1, \bar \psi^\downarrow_1, \bar \psi^\uparrow_2, \bar
\psi^\downarrow_2\right )$
and $ \psi  \left (\v x, \tau\right ) = \left ( \psi^\uparrow_1,
\psi^\downarrow_1, \psi^\uparrow_2, \psi^\downarrow_2 \right )^T$
describe the spinful ($\uparrow,\downarrow$) excitations living
on surfaces $s=1$ and $s=2$. The one particle Hamiltonian which characterizes
the surface $s$ is
\begin{equation}
\left (\text H_0 + \text H_{\rm dis}\right )_s = \left( V_s(\v x ) -\mu_s \right
)
\otimes \mathbf I_{\sigma} + 
i(-)^{s}  v^{(s)}_F
\nabla \wedge \vec \sigma
, \label{eq:OneParticleHam}
\end{equation}
where $\mathbf I_\sigma$ is the unit matrix in spin space and
we define $\v a \wedge \v b = a_x b_y - a_y b_x$.
The disorder potentials $V_s\left ( \v x \right )$ for two surfaces are
assumed to be white-noise distributed and uncorrelated:
\begin{equation}
\left  \langle V_s(\v x ) V_{s'}(\v x' ) \right  \rangle = \frac{\delta \left
(\v x - \v x '\right ) \delta_{ss'}}{\pi \nu_s \tau_s}. 
\end{equation}
The disorder strengths $1/\pi \nu_s \tau_s$ may be different for two
surfaces.

It is worth emphasizing the following physical implications of this
Hamiltonian. 
\begin{itemize}

\item First, the model (and its analysis below) corresponds to the general case
in which the chemical potentials $\mu_1$, $\mu_2$ and hence the carrier
densities of the two surfaces may differ.

\item Second, since the disorder potentials are different for two
surfaces, no inter-surface diffuson and cooperon modes will arise. 
Note that the considered model of fully uncorrelated disorder correctly describes the low-energy physics of the majority of experimental setups, even in the presence of moderate inter-surface correlations of disorder. Indeed, any mismatch in chemical potentials and/or disorder configurations leads to an energy gap in the inter-surface soft modes. Two physical regimes are conceivable: 
\begin{itemize}
\item[(i)] almost identical surfaces in almost fully correlated random potentials, $\vert \mu_1 -\mu_2 \vert \ll 1/\tau_s$ and $ \left  \langle \left [ V_1(\v x ) -  V_{2}(\v x' ) \right ]^2 \right  \rangle \ll  \sum_{s=1,2} \left  \langle V_s(\v x ) V_{s}(\v x' ) \right  \rangle$; 
\item[(ii)] all other parameter regimes, when at least one of the conditions in (i) is not fulfilled. 
\end{itemize}
Our model is designed for the case (ii), where the gap is
comparable to the elastic scattering rate and inter-
surface soft modes do not enter the diffusive theory
at all. It also applies to the case (i) in the ultimate large-scale limit (i.e.  at energy scales below the gap). In this case there will be, however, an additional, intermediate regime in the temperature dependence (or AC frequency dependence) which is not considered in our work. 

\item Third, $\vec \sigma$ in Eq.\eqref{eq:OneParticleHam} in general does not describe the physical spin. For example,
in Bi$_2$Se$_3$ structures the effective spin $\sigma$ is determined by a linear
combination of real spin and the parity (band) degrees of freedom. The mixing
angle depends on how the crystal is cut. \cite{ZhangKaneMele} In this case also
the Fermi velocity becomes anisotropic. 

\item Fourth, because of interaction effects, the true dispersion relation is not
linear but contains logarithmic corrections (or more generally is subjected to ``ballistic'' RG \cite{GonzalezGuineaVozmedianoGraphene01,FosterAleiner08,SheehySchmalian07}) which leads to dependence of the Fermi
velocity on the chemical potential. This is reflected in the notation
$v_F^{(s)} \equiv v_F (\mu_s)$. 

\item Similarly, also the strength of the disorder may be substantially different for
both surfaces, so that the (quantum) mean free times  $\tau_s$  are considered
as two independent input parameters. This is primarily because the vicinity to
the substrate or, respectively, to the coating material makes the impurity
concentration on both surfaces a priori different.
In addition, $\tau_s$ acquire renormalization corrections, leading to a
logarithmic dependence on $\mu_s$. \cite{AleinerEfetovGraphene,OGM2006Graphene,SOGM09,FosterAleiner08}

\item The (pseudo-)spin
texture on the top and bottom surfaces is opposite (denoted by the factor $(-)^{s}$).

\item Finally, in some materials (in particular, in Bi$_2$Te$_3$), the Dirac cone is
strongly warped. We neglect the warping as it does not affect the main result of
this paper, namely the (universal) RG equations.
Recently,\cite{diffusionandwarping} it has been shown that warping only
influences the dephasing length (i.e., the lengthscale at which the RG flow is
stopped).
\end{itemize}
The interaction is mediated by the Coulomb potential, see
Eq.~\eqref{eq:Coulombnoelstat} and Appendix \ref{sec:Elstat}. With the
definition $\rho_s\left (\tau, \v x\right ) = \bar \psi_s\left (\tau, \v x
\right ) \psi_s\left (\tau, \v x \right )$ the corresponding contribution to the
action is given by
\begin{equation}
S_{\rm int} = \frac{1}{2} \sum_{ss'}\int_{\tau, \v x, \v x'} \rho_s\left (\tau,
\v
x\right ) U_{0,ss'}\left (\vert \v x - \v x'\right \vert)\rho_{s'}\left (\tau,
\v x'\right ). \label{eq:S_Coulsimple}
\end{equation}
For equal surfaces ($v_F^{(1)}=v_F^{(2)}$), a simple rescaling of equations
\eqref{eq:Highenergyaction} and \eqref{eq:S_Coulsimple} shows that the effective
coupling to the Coulomb interaction is $\alpha$. It can, in general, become
of the order of unity. Since the perturbation theory is insufficient in such a
case, we adopt the more general, yet phenomenological, Fermi liquid theory
to access the behavior for energies down to the elastic scattering rates
$\tau^{-1}_{1,2}$, see Sections \ref{sec:scattchannels},
\ref{sec:cleanFLMaintext} and Appendix \ref{sec:cleanFL}). This (clean)
Fermi liquid theory will then be a starting point for the interacting
diffusive problem at energies below the elastic scattering rate.

If the interaction becomes too strong, it might in principle drive the system
into a phase with spontaneously broken symmetry.\cite{SitteRoschFritz2013} Examples are the Stoner
instability \cite{Peres} as well as  more exotic phenomena such as
topological exciton condensation, \cite{TEC} which is specific to 3D TI thin
films.
Throughout our analysis, we assume that the system is not in a vicinity of
such an instability. To our knowledge, this assumption is consistent with all
transport experiments on 3D TI slabs addressed in this work.

\section{Sigma-model description}
\label{sec:towardsNLSM}

We are interested in the low-energy (low-temperature, long-length-scale) physics
of the 3D TI problem defined by Eqs.~\eqref{eq:Diffregime1} and
\eqref{eq:weakdisorder}. This physics is controlled by coupled diffuson and
cooperon modes. In this Section we derive the effective field
theory -- diffusive non-linear $\sigma$ model -- that describes
the system in this regime.

\subsection{Symmetries of the action}
\label{sec:symmetries}

The structure of the effective low-energy theory, the diffusive NL$\sigma$M,
is controlled by symmetries of the microscopic action. The information
about other microscopic details enters the theory only via the values of the
coupling constants. We thus begin by analyzing symmetries of the problem.

First, our system obeys the time reversal symmetry $H = \sigma_y H^T \sigma_y$.
Second, we assume no intersurface tunneling, i.e., the particle
number is conserved in each surface separately. This
implies invariance of the action with respect to  $\mathbf{U}(1) \times
\mathbf{U}(1)$ transformations (global in space and time).

The presence of Coulomb interaction promotes the $\mathbf{U}(1)$ symmetry in the
total-density channel, $\rho_1+\rho_2$, to transformations which are local in time but global in
space. In other words, rotations of fermionic fields, $\bar \psi_s\left (\tau, \v x
\right ) \to \bar \psi_s\left (\tau, \v x
\right ) \exp{[-i \chi_s(\tau)]}$, $\psi_s\left (\tau, \v x \right )\to\exp{[i \chi_s(\tau)]}\psi_s\left (\tau, \v x \right )$, with equal phases $\chi_1\left (\tau\right
)=\chi_2\left (\tau\right )$ leave the action
\eqref{eq:Highenergyaction} invariant. This is a special case of ``$\mathcal
F$-invariance'' \cite{MishandlingI} and has important consequences for the
present problem. The $\mathcal F$-invariance (it is intimately linked to gauge invariance) generally states that in each
channel with long-range interaction, time-dependent but spatially constant
$\mathbf{U}(1)$ rotations are symmetries of the action. In our problem, as it
follows from the $q\to 0$ limit of the Coulomb interaction:
\begin{equation}
\underline{U}\left (\v q\right ) \stackrel{q\rightarrow 0}{\propto} \frac{1}{q}
\left (\begin{array}{cc}
1 & 1 \\
1 & 1
\end{array} \right ), \label{eq:q0limitt}
\end{equation}
only the interaction between the total densities is long-ranged.
The structure of Eq. \eqref{eq:q0limitt} remains true also in the case of asymmetric dielectric
environment, see Appendix \ref{sec:appendixCoulomb}.

To make the time-reversal symmetry explicit, we define
particle-hole bispinors by
combining $\psi$ and $\bar{\psi}$ fields. \cite{EfetovLarkinKhmel, BelitzKirkpatrick} In the momentum space the
bispinors read
\begin{equation}
\Phi_{n}\left (\v k\right ) = \frac{1}{\sqrt 2} \left (\begin{array}{c}
\bar \psi_{n}\left (- \v k\right ) ^T \\
i \sigma_y \psi_{n}\left (\v k\right )
\end{array} \right ) 
\end{equation}
 and
 \begin{equation}
 \bar \Phi_{n}\left (\v k\right ) = \left [C \Phi_{n}\left (-\v k\right )\right
]^T \text{ with } C = i \sigma_y \tau_x , 
 \end{equation}
where $n$ is the index associated to the fermionic Matsubara frequency $i
\epsilon _n$, and $\tau$ matrices  act in the particle-hole space. 
This allows us to rewrite the one-particle Hamiltonian as
\begin{equation}
S^{\rm free} = -\sum_n \int_{\v k} \bar \Phi_{n} \left (\v k\right ) \left (i
\epsilon_n - H^T\left (- \v k \right ) \right )  \Phi_{n} \left (\v k\right ).
\end{equation}

It is convenient to perform a rotation of bispinors
\begin{equation}
\eta = \sqrt {\tau_x} \Phi, 
\end{equation}
where $\sqrt {\tau_x} = e^{-i\pi/4} (\mathbf I_\tau + i\tau_x)/\sqrt{2}$.
The free action then takes the form
\begin{eqnarray}
S^{\rm free} &=& -\sum_s \int_{\v x} \eta^T_s  \Big \lbrace \left [i \hat
\epsilon - V_s + \mu_s \right ] \left ( - i \sigma_y \right ) \\ 
&+&  \left. (-)^{s+1} v^{(s)}_F\left (\partial_x - i
\partial_y \sigma_z \right ) \right\} \eta_s  \label{eq:Sofeta} .
\end{eqnarray}
The Matsubara frequency summation is incorporated into the scalar product
$\eta^T \left (\dots \right ) \eta$. In these  notations,  $\hat \epsilon$ is
a diagonal matrix in the Matsubara space consisting of entries $\epsilon_n$.

In order to perform the average over disorder, we replicate the theory $N_R$
times. Furthermore, in order to implement the $\mathbf
U\left (1\right )$-gauge invariance in the framework of the NL$\sigma$M,
we apply a double cutoff truncation procedure with $N_M \ll N^\prime_M$
for the Matsubara frequencies.\cite{MishandlingI}
Here $N'_M$ and $N_M$ are the numbers of retained Matsubara harmonics for fast
(electrons of the original theory) and slow (diffusons and cooperons
of the NL$\sigma$M) degrees of freedom, respectively.
As a consequence, $\eta$ becomes a $\left (2_s \times 2_\sigma \times 2_\tau
\times 2 N'_M
\times N_R\right )$-dimensional Grassmannian vector field. Except for the
frequency term, the free action \eqref{eq:Sofeta} is manifestly invariant
under global orthogonal rotations of the kind
\begin{equation}
 \eta_s \rightarrow \left (O_s \otimes \mathbf I_\sigma \right ) \eta_s \text{
with }  O_s  \in \mathbf{O}\left (2_\tau  \times 2 N'_M \times N_R\right ).
\end{equation}
Since the surfaces are fully decoupled in the absence of interactions,
the rotations $O_1$ and $O_2$ of the fields corresponding to the top and bottom surfaces are
completely independent.

\subsection{Quasiclassical conductivity}
\label{sec:semiclassical}

To obtain the quasiclassical conductivity, we first find the fermionic
self-energy within the self-consistent Born approximation (SCBA):
\begin{equation}
\Sigma^s_n = \frac{-2i\sigma_y}{\pi \nu_s \tau_s}  \left \langle \eta_{\v x,s}
\eta_{\v x,s}^T\right\rangle_{\text{SCBA}}.
\label{eq:SCBA}
\end{equation}
Here $\left \langle \dots \right \rangle_{\text{SCBA}}$ denotes the
self-consistent treatment, i.e. a shift $\mu_s \rightarrow \mu_s + \Sigma^s_n$
in the fermionic propagator. Equation \eqref{eq:SCBA} yields for the imaginary
part of the self-energy ${\rm Im}(\Sigma^s_n) = (i / 2\tau_s) \text{sgn}(n)$.
The quasiclassical Drude DC conductance of the non-interacting problem in the
absence of a magnetic field is
\begin{equation}
\sigma^D_s = 2\pi \nu_s D_s \frac{e^2}{h} ,\label{eq:Drude}
\end{equation}
with $D_s = (v^{(s)}_F)^2 \tau_s$. Note that the transport time is twice
the quantum mean free time  $\tau_s$. In the diagrammatic language, this is
a consequence of vertex corrections.

\subsection{Fermionic currents and bosonization rules}

To derive the NL$\sigma$M, we use the method of non-Abelian bosonization.
 \cite{Witten, NersesyanTsvelikWenger94, NersesyanTsvelikWenger95, ASZ,
AltlandGraphene} An advantage of this approach is that non-trivial
topological properties of the Dirac fermions are translated
into the field theory in a particularly transparent way.

In the first step, the kinetic term (Sec. \ref{sec:NONABBOS}) is bosonized.
Subsequently, we bosonize also the terms induced by the chemical potential, disorder
and frequency (Sec. \ref{sec:freeNLSM}). Since only interaction couples the two
surfaces,  we omit the surface index $s$ in Sec. \ref{sec:NONABBOS} and Sec.
\ref{sec:freeNLSM}. This index is restored
later in Sec.~\ref{sec:interactingNLSM} where the interaction is included.

Local left ($\eta_{\uparrow} \rightarrow O_{L} \eta_{\uparrow}$) and
right ($\eta_{\downarrow} \rightarrow O_{R} \eta_{\downarrow}$) rotations define the
left and right currents. The bosonization rules for these currents as well as
for the mass term are
\begin{subequations}
\begin{align}
j_{+} =   v_F \eta_{\uparrow} \eta_{\uparrow} ^T &\leftrightarrow
\frac{1}{8\pi}\left (O \partial_+ O^T\right ),  \label{eq:leftcurrent} \\
j_{-} =   v_F \eta_{\downarrow} \eta_{\downarrow} ^T &\leftrightarrow
\frac{1}{8\pi}\left (O^T \partial_- O\right ),  \label{eq:rightcurrent} \\
\eta_{\uparrow} \eta_{\downarrow}^T &\leftrightarrow i \lambda O,
\label{eq:thirdbosonrule}
\end{align}
\label{eq:Bosonrules}
\end{subequations}
where $\partial_\pm = \partial_x \pm i \partial_y$. The energy scale $\lambda$
is of the order of the ultraviolet (UV) cutoff and is introduced here for
dimensional
reasons; see Sec.~\ref{sec:freqboson} and \ref{sec:Densityresponse} for a
discussion of its physical meaning. Note that in general, the UV cutoff is different for the top and bottom surfaces, 
$\lambda_1 \neq \lambda_2$. Further, $O$ is an orthogonal $\left (2_\tau \times 2
N'_M
\times N_R\right )\times \left (2_\tau \times 2 N'_M \times N_R\right )$
{matrix field}. Below we will need the following constant matrices in this space
\begin{eqnarray}
\Lambda^{\tau_1 \tau_2; \alpha \beta}_{nm} &=& \text{sgn}\left (n \right )
\delta^{\tau_1 \tau_2} \delta^{\alpha \beta}  \delta_{nm} ,  \notag \\
\hat \eta^{\tau_1 \tau_2; \alpha \beta}_{nm} &=& n \delta^{\tau_1 \tau_2}
\delta^{\alpha \beta} \delta_{nm}  , \label{eq:matrixdefs} \\
\left (I^{\alpha_0}_{n_0} \right )^{\tau_1 \tau_2; \alpha \beta}_{nm} &=&
\delta^{\tau_1 \tau_2} \delta^{\alpha_0 \alpha} \delta^{\alpha_0 \beta}
\delta_{n-m,n_0}. \notag
\end{eqnarray}
Here and throughout the paper we use a convention that
$\alpha, \beta \in \left \lbrace 0, N_R \right \rbrace$ denote replicas and
$m, n \in \left \lbrace - N'_M, \dots , N'_M - 1 \right \rbrace$ Matsubara
indices.
The double cutoff regularization scheme \cite{MishandlingI} prescribes that
matrices $O$ have non-trivial matrix elements $O_{nm}$ only for low-energy
excitations $n,m \in \left \lbrace - N_M, \dots , N_M - 1 \right \rbrace$ and
stay equal to the origin  $O_0$ of the $\sigma$ model manifold outside this
low-energy region. As explained below, $O_0 = \Lambda$.

\subsection{Bosonization of the kinetic part}
\label{sec:NONABBOS}

The kinetic part of \eqref{eq:Sofeta} is nothing but the Euclidean counterpart
of the model considered {in} Ref. \onlinecite{Witten}. Upon
non-Abelian bosonization it
yields the Wess-Zumino-Novikov-Witten (WZNW) action
\begin{equation}
S_{\rm WZNW} = \int_{\v x} \frac{1}{16\pi} \tr \nabla O \nabla O^{-1} +
\frac{i}{24
\pi} \Gamma_{WZ} ,
\label{eq:S-WZNW}
\end{equation}
where $\Gamma_{WZ}$ is the Wess-Zumino (WZ) term
\begin{equation}
\Gamma_{WZ} = \int_{\v x, w} \epsilon_{\mu \nu \rho} \tr \left [\left ( \tilde O^{-1}
\partial_\mu \tilde  O\right ) \left (\tilde O^{-1} \partial_\nu \tilde O \right
)\left (\tilde  O^{-1} \partial_\rho \tilde O\right )\right ],
\label{eq:WZW}
\end{equation}
where $\epsilon_{\mu\nu\rho}$ denotes the Levi-Civita symbol.
The definition of the WZ term involves an auxiliary coordinate $w \in \left [0,
1 \right ]$ and smooth fields $\tilde O \left (\v x , w \right )$ satisfying
$\tilde
O \left ( \v x, w = 0 \right ) = {\rm const} $ and $\tilde O \left (\v x , w
=1\right
) = O \left (\v x \right )$.
As a result the compactified two-dimensional coordinate space $\mathbb{R}^2 \cup
\left \lbrace \infty \right  \rbrace \simeq \mathbb{S}^2$ is promoted to the
solid 3-ball $\mathbb{B}^3$ (i.e., the ``filled'' sphere). 

\subsection{Free NL$\sigma$M of class AII}
\label{sec:freeNLSM}

\subsubsection{Disorder, frequency, and the chemical potential}
\label{sec:freqboson}

The action (\ref{eq:S-WZNW}) is the bosonized counterpart of the second
(proportional to velocity) term of the microscopic action \eqref{eq:Sofeta}.
Let us now consider the first term in Eq.~\eqref{eq:Sofeta} which carries
information about the chemical potentials, frequency and random potential.

Bosonization of the terms with frequency and the chemical potential in the microscopic action \eqref{eq:Sofeta} yields
\begin{align}
\delta S 
= 
  2 \int_{\v x} \tr \left [\left (i\hat \epsilon +\mu\right )\eta_{\uparrow}\eta_{\downarrow}^T\right ] 
\leftrightarrow 
 - 2\lambda  \int_{\v x}  \tr \left (\hat \epsilon - i\mu\right)O .
\label{eq:disordermassterms-frmu}
\end{align}

Upon disorder averaging and bosonization, the term with random potential
provides the following contribution to the field theory:
\begin{align}
\delta S_\textrm{dis} &= 
-  \frac{1}{\pi \nu \tau} \int_{\v x}\left (\tr
\eta_\uparrow \eta_\downarrow^T\right )^2 
+ \frac{1}{\pi \nu \tau} \int_{\v x}\tr (\eta_{\uparrow}
\eta_{\downarrow}^T)^2 \notag\\
&\leftrightarrow 
 \frac{\lambda^2}{\pi\nu\tau}\int_{\v x}  \left (\tr O\right )^2
 \notag \\& 
 +\frac{\lambda^2}{2\pi \nu \tau} \int_{\v x}  \tr \left (O^T-O\right )^T\left
(O^T-O\right ).
\label{eq:disordermassterms}
\end{align}

As we see, disorder induces mass terms for
$O$-matrices. Both mass terms in Eq. \eqref{eq:disordermassterms} are strictly non-negative.
Therefore, they are minimized by arbitrary traceless symmetric orthogonal matrix. It is convenient to choose the specific saddle-point solution as
\begin{equation}
O = \Lambda .
\label{eq:Lamda}
\end{equation}
This saddle-point solution coincides with the SCBA. Indeed, Eq. \eqref{eq:SCBA} can be written as
\begin{align}
\frac{i}{2\tau} \Lambda {\otimes \textbf 1}_\sigma&= \frac{2}{\pi \nu \tau}   \left \langle \left (\begin{array}{cc}
-\eta_\downarrow \eta_\uparrow^T & -\eta_\downarrow \eta_\downarrow^T \\
 \eta_\uparrow \eta_\uparrow^T &  \eta_\uparrow \eta_\downarrow^T
\end{array} \right )\right\rangle_{\text{SCBA}} \notag \\
& \leftrightarrow \frac{2}{\pi \nu \tau} \left \langle \left (\begin{array}{cc}
i\lambda O^T & \frac{-1}{8\pi v_F} O^T \partial_- O \\
\frac{1}{8\pi v_F} O \partial_+ O^T & i\lambda O
\end{array} \right )\right \rangle .
\end{align}
It is solved by the saddle-point solution \eqref{eq:Lamda} provided the auxiliary UV energy scale $\lambda$ introduced in
Eq.~(\ref{eq:Bosonrules}) is related to the density of states (i.e., to the chemical
potential),
\begin{equation}
\lambda = \frac{\pi \nu}{4} = \frac{\vert \mu \vert}{8 v_F^2}. \label{eq:lambdamu}
\end{equation}
We will rederive this relation from a different viewpoint below, see
Sec.~\ref{sec:Densityresponse}.

Equation (\ref{eq:Lamda}) is not the only solution of the saddle point
equation. It is easy to see that rotations
\begin{equation}
O\rightarrow O^T_{\text{soft}} O O_{\text{soft}},\; O_{\text{soft}}\in \mathbf G=
\mathbf{O}\left (2_\tau \times 2N_M\times N_R\right )
\end{equation}
leave the mass term unaffected. On the other hand, the saddle-point $O=\Lambda$
is invariant under rotations from a smaller group,  $O_{\text{soft}}\in \mathbf{K}=\mathbf
O\left (2_\tau \times N_M\times N_R\right )\times \mathbf O\left ( 2_\tau \times
N_M\times N_R\right )$. This can be understood as a breakdown of symmetry
$\mathbf G\to \mathbf K$.
 We thus obtain a non-trivial manifold of saddle-points
annihilating the mass term. Allowing for a slow variation of $O_{\text{soft}}$
and restricting other terms in the action to this manifold, we will obtain the
NL$\sigma$M action.

\subsubsection{Free NL$\sigma$M with $\mathbb Z_2$ topological term}
\label{sec:Z2term}

As we have just discussed, we keep only the soft modes
\begin{equation}
Q = O^T_{\text{soft}} \Lambda O_{\text{soft}}  \text{ with } 
O_{\text{soft}} \in
\mathbf{G} .  \label{eq:defQ}
\end{equation}
The subscript $_{\text{soft}}$ will be omitted in the remainder. The NL$\sigma$M
manifold ${\cal M}=\mathbf G/ \mathbf K$. 
We also rename the coupling constants according to the conventional notation of
diffusive NL$\sigma$Ms and restore the  surface index $s$,
\begin{equation}
S^{\rm free} = \sum_s \int_{\v x} \frac{\sigma_s}{16} \tr \left (\nabla
Q_s\right )^2 - 2 \pi T z_s \tr \left [\hat \eta Q_s\right ]+ i S^{(\theta)}_s.
\label{eq:freeNLSM}
\end{equation}
As will become clear from linear response theory (Sec. \ref{sec:Kubo}),
$\sigma_s$ measures the DC conductivity of surface $s$ (in units $e^2/h$). Its
bare value is the Drude conductance depending on the chemical potential $\mu_s$, as can be directly verified, see Appendix
\ref{sec:appendixBosonization}. The coupling constants $z_s$ determine the
renormalization of the specific heat.

The non-trivial second homotopy group of the NL$\sigma$M manifold $\pi_2 ({\cal
M})= \mathbb Z_2$ allows for topological excitations (instantons), similarly
to the QHE theory. A crucial difference is that in the QHE case the second
homotopy group is $\mathbb Z$, so that any integer topological charge (number
of instantons) is allowed. Contrary to this, in the present case
any configuration of an even number of instantons can be continuously deformed
to the trivial, constant vacuum configuration. Therefore, the theta term
$S^{(\theta)}_s$ appearing in \eqref{eq:freeNLSM} only distinguishes between an
even ($S^{(\theta)}_s = 0 \; \text{mod} \; 2\pi$) and odd ($S^{(\theta)}_s = \pi
\;\text{mod}\; 2\pi$) number of instantons.

Such a $\mathbb Z_2$ theta term $S^{(\theta)}$ does not appear in the case of
usual metals with strong spin-orbit coupling; it results from the
Dirac-fermion nature of carriers and is a hallmark of topologically protected
metals (in our case, the surface of a topological insulator).
The topological term flips the sign of the
instanton effects (as compared to the case of a
usual metal with spin-orbit interaction) from localizing to
delocalizing.
Thus, the theta term translates the protection against Anderson
localization into the NL$\sigma$M approach.

We are now going to show that $S^{(\theta)}_s$ is nothing but
the WZ term (obtained from non-Abelian bosonization) restricted to the smaller
symmetry group:
\begin{equation}
S^{(\theta)}_s =  \frac{1}{24 \pi} \left . \Gamma_{WZ,s} \right \vert_{\tilde
O_s\left (\v x, w = 1\right ) = Q_s\left (\v  x\right ) = Q_s^T\left (\v x\right
)}.
\label{eq:thetaterm}
\end{equation}
Note that, since the second homotopy group of the NL$\sigma$M manifold is
non-trivial, the definition of the WZ term requires that away from $w =
1$ the extended fields can take values in the big orthogonal group $\mathbf G$.

To show that Eq.~(\ref{eq:thetaterm}) is indeed the $\mathbb
Z_2$ theta-term, we proceed in the same way as was recently done for symmetry
class CII.\cite{KOPM2012} First of all, it is straightforward to check that
$S^{(\theta)}_s$ is invariant under small variations of the sigma-model field,
$Q_s \rightarrow Q'_s = Q_s + \delta Q_s$ ($Q_s'^2 = \mathbf{1} = Q_s^2$). Thus,
$S^{(\theta)}_s$ only depends on the topology of the field configuration.
This immediately implies that it is zero in the topologically trivial sector.
In order to proof that $S^{(\theta)}_s$ also returns the correct value
$S^{(\theta)}_s = \pi \; (\text{mod} \; 2\pi)$ in the topologically  non-trivial
sector, it is sufficient to insert a single instanton into $S^{(\theta)}_s$.
Instantons are field configurations that per definition can not be continuously
deformed into the vacuum configuration. Introducing the third dimension and
allowing the field to take values in the entire orthogonal group we can
continuously shrink the instanton in the $w=1$ sphere to the constant at $w =
0$. A necessary condition for this untwisting to happen is that
for some subinterval of $\left (0,1\right )$ the field leaves the NL$\sigma$M
manifold for the larger orthogonal group.
A direct calculation shows that the group volume covered while untwisting
indeed yields the value $iS^{(\theta)}_s = i\pi$, see Appendix
\ref{sec:Instanton}).

There have been alternative derivations of the $\mathbb Z_2$ term before
\cite{RyuZ2, OGM2007}. Viewing this theta term as a symmetry-broken WZ-term,
Eq.~(\ref{eq:thetaterm}), yields a
local expression for it  and implies the following advantages. First, this form
is very useful for understanding the crossover between 3D topological insulators
of class DIII and AII. Second and more importantly, an analysis of response of
the system to an external electric field requires coupling of the diffusive
matter fields to $\mathbf{U}(1)$ gauge potentials. In particular, one should
gauge the topological term, which can be done in a standard way by using a
local expression for it. We will show in Section \ref{sec:EdotB} that such a
procedure yields the correct linear response theory for the anomalous quantum
Hall effect of Dirac fermions. 

In addition to a non-trivial second homotopy group $\pi_2$, the sigma model
manifold of the class AII possesses also a non-trivial first homotopy group,
$\pi_1({\cal M}) = \mathbb Z_2$. For this reason, the RG flow in 2D systems
of class AII (as well as in other classes with a non-trivial $\pi_1$ group,
namely AIII, BDI, CII, and DIII) is affected by vortices, as was shown in
Ref.~\onlinecite{KOPM2012}. In the case of AII (and DIII) class these are $\mathbb
Z_2$ vortices,\cite{KOPM2012} i.e., a vortex is identical to an anti-vortex.
In a recent work \cite{FuKaneVortices} it was argued that such vortices are
crucial for establishing localization in the class AII. {Conversely, the robustness of a non-localized state on the
surface of a {weak} topological insulator and of the critical state separating 2D trivial and topological insulator were explained by vanishing of the 
corresponding fugacity.}

On the surface of a strong 3D TI, the effect of vortices is erased by the
$\mathbb Z_2$ topological term, in the same way as argued
previously\cite{KOPM2012} for the case of the symmetry class CII.
Specifically, due to the $\mathbb Z_2$ theta term, the vortices acquire an
internal degree of freedom which, upon averaging, annihilates the
contribution of vortices  to renormalization. For this reason, the vortices
{need not} be taken into account in the present context.

\subsection{Interacting NL$\sigma$M}
\label{sec:interactingNLSM}

In the previous subsection we have derived the diffusive non-linear sigma model
for non-interacting particles. The next step is to include the electron-electron
interactions.

\subsubsection{Interacting Fermi gas}

We concentrate first on the case of a weak Coulomb interaction ($\alpha \ll
1$).
At length scales larger than the screening length the interaction is effectively
pointlike:
\begin{equation}
S_{\rm int} = \frac{T}{2} \sum_{m,\alpha;ss'} \int_{\v x} \tr {\left (I_m^\alpha
\psi_s \bar\psi_s\right )} U^q_{ss'}{\left (I_{-m}^\alpha \psi_{s'}
\bar\psi_{s'}\right )}
\end{equation}
where $ U^q_{ss'}$ is the ``overscreened'' Coulomb interaction matrix i.e., the $q
\rightarrow 0$ limit of Eq. \eqref{eq:coupledsurf} (for its generalization in case of an
asymmetric dielectric environment, see Appendix \ref{sec:Elstat}). We use the
bosonization rule
\begin{align}
\tr I_m^\alpha \psi_s \bar \psi_s & = \tr I_m^\alpha \left (1-\tau_y\right )\eta_{s,\uparrow} \eta_{s,\downarrow}^T -\tr I_m^\alpha\left (1-\tau_y\right ) \eta_{s,\downarrow} \eta_{s,\uparrow}^T \notag \\
&\leftrightarrow i \lambda \left [ \tr I_m^\alpha\left (1-\tau_y\right ) \left
(O_s + O_s^T\right )\right ].
\end{align}
When disorder is introduced, the matrices $O$ become restricted to the
sigma-model manifold ${\cal M}$, and we obtain
\begin{equation}
S_{\rm int} {=} - \lambda^2 8T \sum_{m,\alpha;ss'} \int_{\v x} {\tr\left
[J_{-m}^\alpha Q_s\right ]} U^q_{ss'} {\tr\left [J_{m}^\alpha Q_{s'}\right ]}.
\label{eq:bosonizedIA}
\end{equation}
Here we have defined $J_n^\alpha = I_n^\alpha \frac{1+\tau_y}{2}$.
As has been already emphasized, we want to treat the general case of strong
interactions up to $\alpha \sim 1$. Therefore, in the
following (and in more detail in Appendix \ref{sec:cleanFL}), we present the
Fermi liquid (FL) treatment of strongly interacting surface states of a thin 3D TI film.

\subsubsection{Effective spinless theory}

\begin{figure}
\includegraphics[scale=.25]{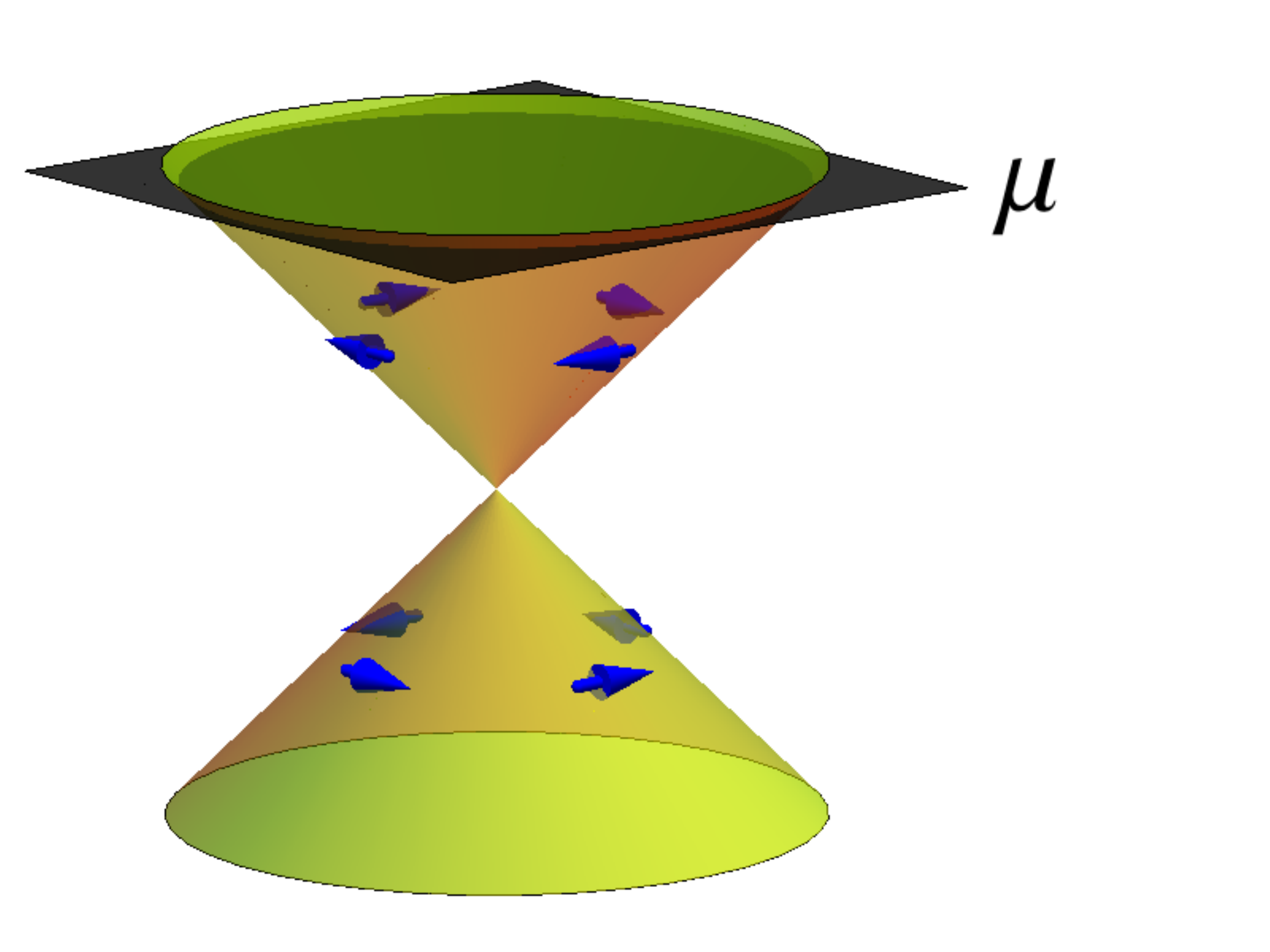}
\caption{Schematic representation of the Dirac cone and the strong Rashba spin
orbit coupling.
If the chemical potential (black plane) is large compared to the typical energy
scale $E$ (e.g., temperature), only one kind of helical states can take part in
the dynamics.
}
\label{fig:cone}
\end{figure}

One of the most striking peculiarities of the surface states of 3D topological
insulators is their Rashba-like kinetic term. As a consequence, spin and
momentum are locked in a manner visualized in Fig.~\ref{fig:cone}. Such
states are called helical;  one associates helicity eigenvalues $+1$ ($-1$)
with states with positive (respectively, negative) kinetic energy.
As has been stated above, we will be interested in the low energy regime $E \ll
\vert \mu_{1,2} \vert$. Hence, at each of the surfaces only one type of helical  states
represents dynamical low energy degrees of freedom, while the other one is suppressed by a
mass $\approx 2 \vert \mu_{1,2} \vert$. Therefore, we project onto the appropriate helicity
eigenstate of each surface using the following projection operator
\begin{equation}
\mathcal{P}_{s} = \vert \mu_s, \v p \rangle \langle \mu_s, \v p \vert \text{
with } \vert \mu_s, \v p \rangle = \frac{1 }{\sqrt{2}} \left (\begin{array}{c}
1 \\
i \text{sgn}\mu_s \; e^{i \phi( \v p )}
\end{array} \right ), 
\end{equation}
where we have defined the polar angle $\phi$ of the momentum, $p_x \equiv \vert
\v p \vert \cos \phi$ and $p_y \equiv \vert \v p \vert \sin \phi$.
The clean single-particle action becomes effectively spinless:
\begin{equation}
S^{(s)}_0	=  -\sum_s \int_{\v p} \bar{\zeta}_{s} \left ( \v p \right ) \left
[i \hat \epsilon + \text{sgn}\left(\mu_s\right) \left( \vert \mu_s \vert - v_F^s
\vert \v p \vert \right ) \right ] \zeta_{s} \left ( \v p  \right ),
\label{eq:spinlessSkin}
\end{equation}
where $\zeta_{s}$, $\bar \zeta_{s}$ are
the fields associated with the helicity eigenstates, $\zeta_{s} = \langle
\mu_s, \v p \vert \psi_{s}$ and $\bar \zeta_{s} = \bar \psi_{s,\sigma}\vert
\mu_s, \v p \rangle$.

\subsubsection{Scattering channels}
\label{sec:scattchannels}

In the presence of a Fermi surface, the electron-electron interaction at low
energies decouples into separate scattering channels defined by small
energy-momentum transfer and by the tensor structure in the surface space:
\begin{equation}
S_{\rm int} = - \frac{T}{2} \int_{P_1, P_2, K} \sum_{\alpha} \left
[\mathcal{O}^{IA}_{0+1} + \mathcal{O}^{IA}_2 +  \mathcal{O}^{IA}_c\right ]
\label{eq:realIA}
\end{equation}
with
\begin{eqnarray}
\mathcal{O}^{IA}_{0+1} &=&  \sum_{s_1s_2}\left [\bar \zeta^\alpha_{s_1} \left
(P_1\right )  \zeta^\alpha_{s_1} \left (P_1 + K\right )\right ] \notag \\ &&
\times \Gamma^{0+1,q}_{s_1,s_2;\hat p_1, \hat p_2} \left [\bar \zeta^\alpha_{s_2}
\left (P_2\right ) \zeta^\alpha_{s_2} \left (P_2 - K\right )\right ], 
\end{eqnarray}
\begin{eqnarray}
\mathcal{O}^{IA}_2 &=&  \sum_{s_1s_2}\left [\bar \zeta^\alpha_{s_1} \left
(P_2\right )  \zeta^\alpha_{s_1} \left (P_1 + K\right )\right ] \notag \\ &&
\times \Gamma^{2,q}_{s_1,s_2;\hat p_1, \hat p_2} \left [\bar \zeta^\alpha_{s_2}
\left (P_1\right ) \zeta^\alpha_{s_2} \left (P_2 - K\right )\right ], 
\end{eqnarray}
and
\begin{eqnarray}
\mathcal{O}^{IA}_c &=& \sum_{s_1s_2} \left [\bar \zeta^\alpha_{s_1} \left
(P_2\right )  \zeta^\alpha_{s_1} \left (-P_1 + K\right )\right ] \notag \\ &&
\times \Gamma^{c,q}_{s_1,s_2;\hat p_1, \hat p_2} \left [\bar \zeta^\alpha_{s_2}
\left (-P_2+K\right ) \zeta^\alpha_{s_2} \left (P_1\right )\right ]. 
\end{eqnarray}
Here the
capital letters denote 2+1 momenta. The smallness of $K = \left (\omega_m, \v
q\right )$ means that the following conditions hold $\left (\omega_m, \vert \v q \vert \right) \ll \left (
\vert \mu_{s} \vert,p^{(s)}_{F}\right )$ for both $s = 1,2$. We emphasize that all
``Dirac factors'' of 3D surface electrons are included in the angular dependence
of the scattering amplitudes (subscripts $\Gamma_{\hat p_1,\hat p_2}$).

\begin{figure}
\includegraphics[scale=.4]{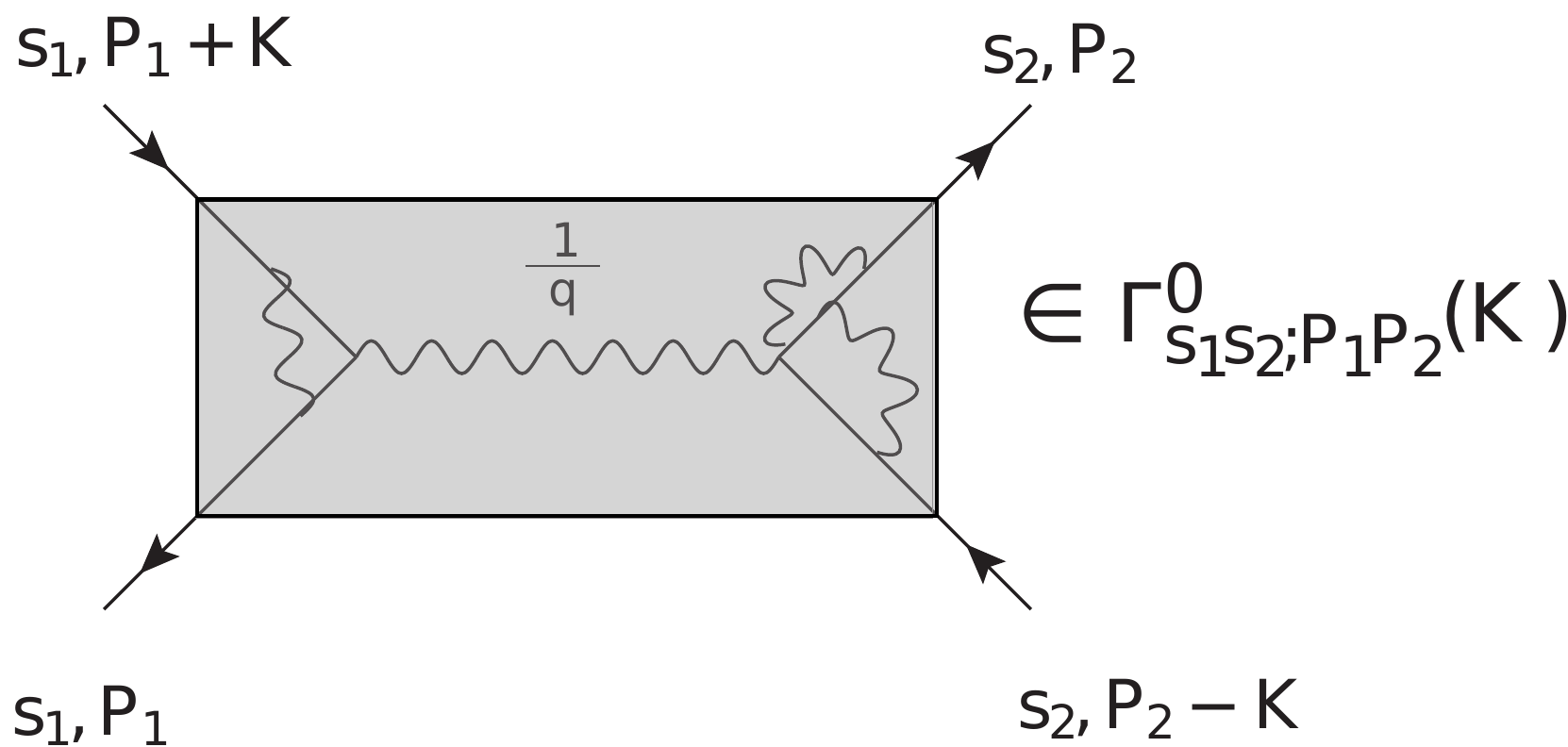}
\caption{An example of contribution to a one Coulomb-line reducible small angle scattering
amplitude. Independently of $\text{sgn}\left (\mu_s\right )$, ingoing arrows
denote fields $\zeta_s$, outgoing arrows $\bar \zeta_s$.}
\label{fig:Gamma0}
\end{figure}

\begin{figure}
\includegraphics[scale=.4]{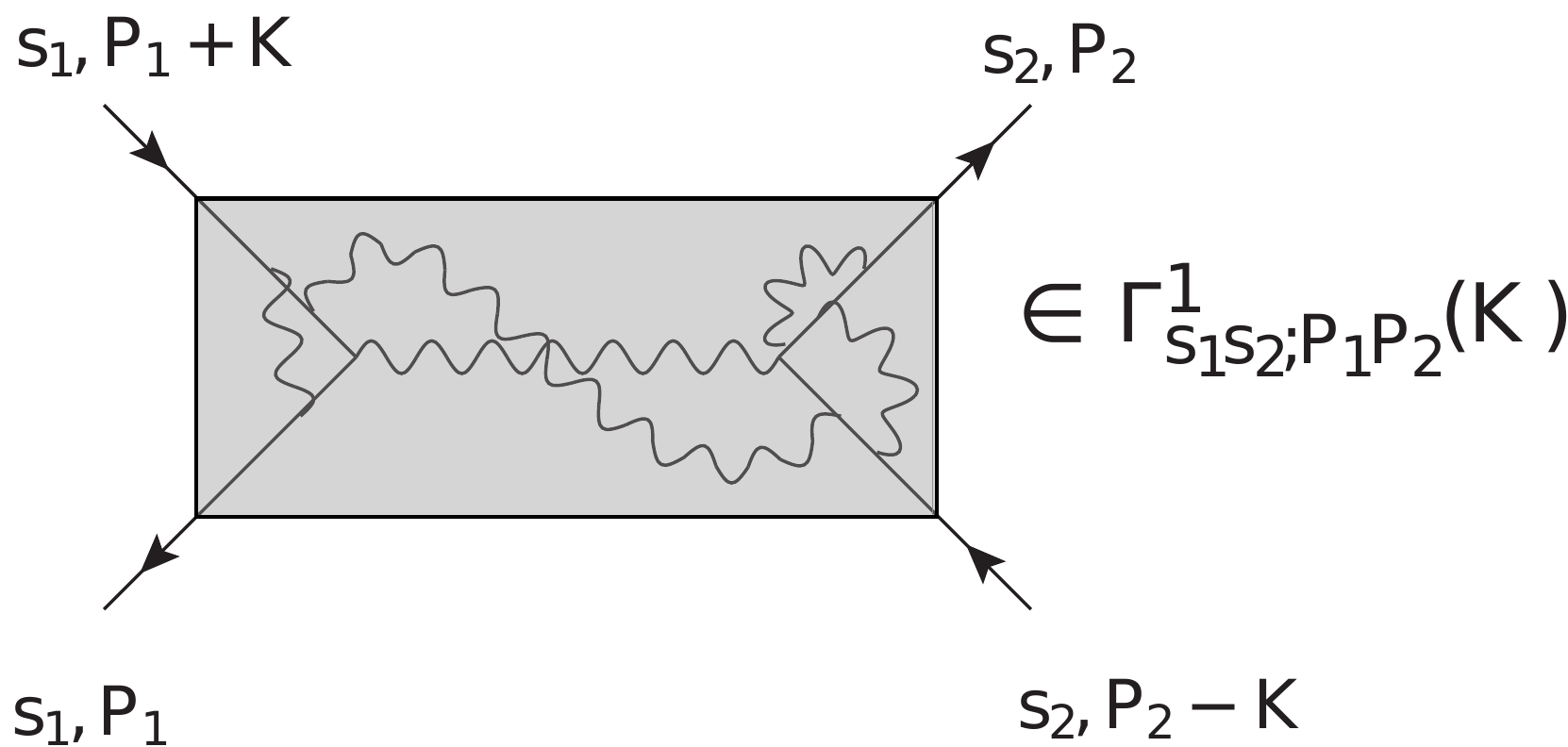}
\caption{An example of contribution to a one-Coulomb-line irreducible small-angle scattering
amplitude.}
\label{fig:Gamma1}
\end{figure}

\begin{figure}
\includegraphics[scale=.4]{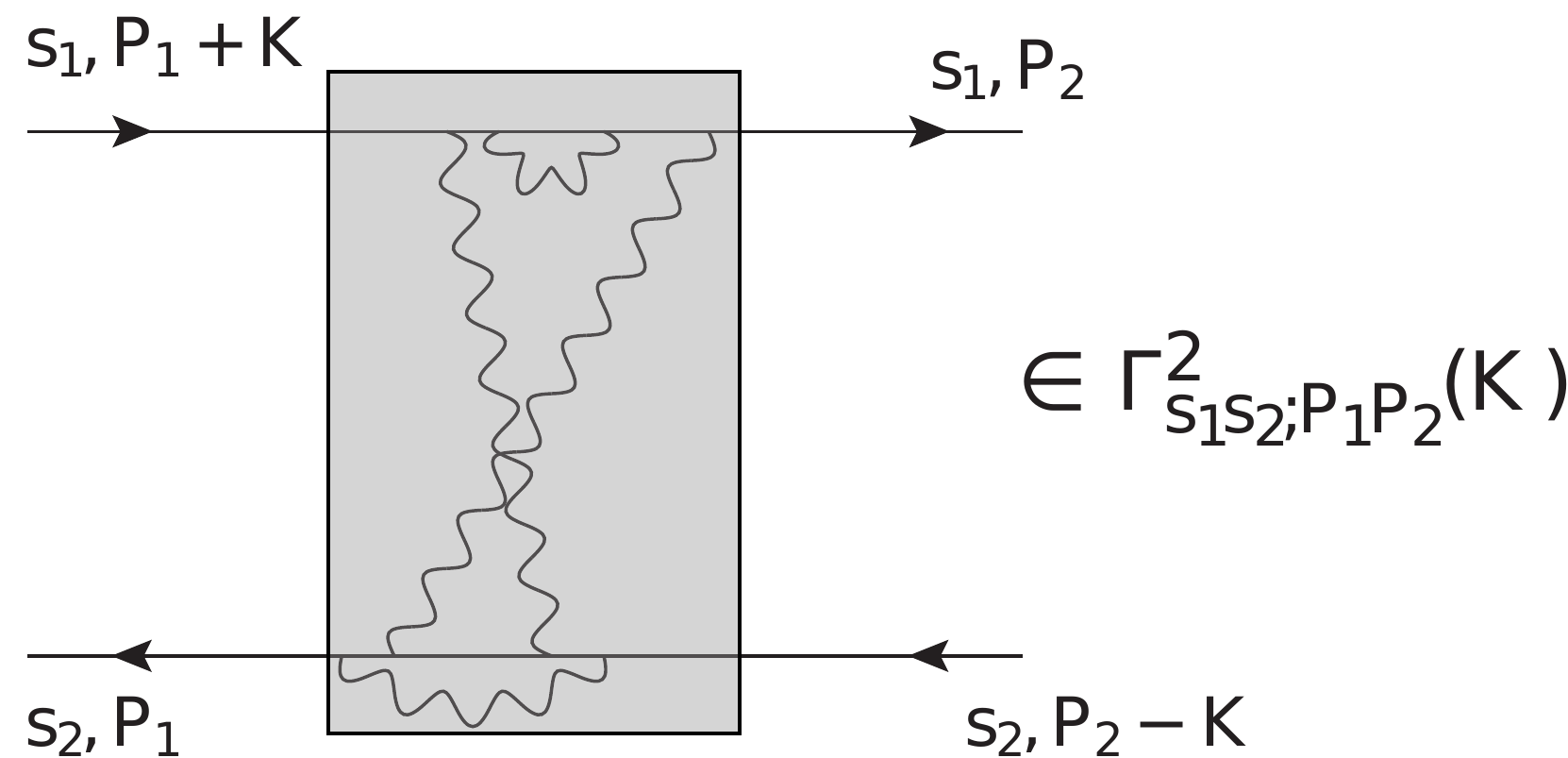}
\caption{An example of contribution to a large-angle scattering amplitude.}
\label{fig:Gamma2}
\end{figure}

\begin{figure}
\includegraphics[scale=.4]{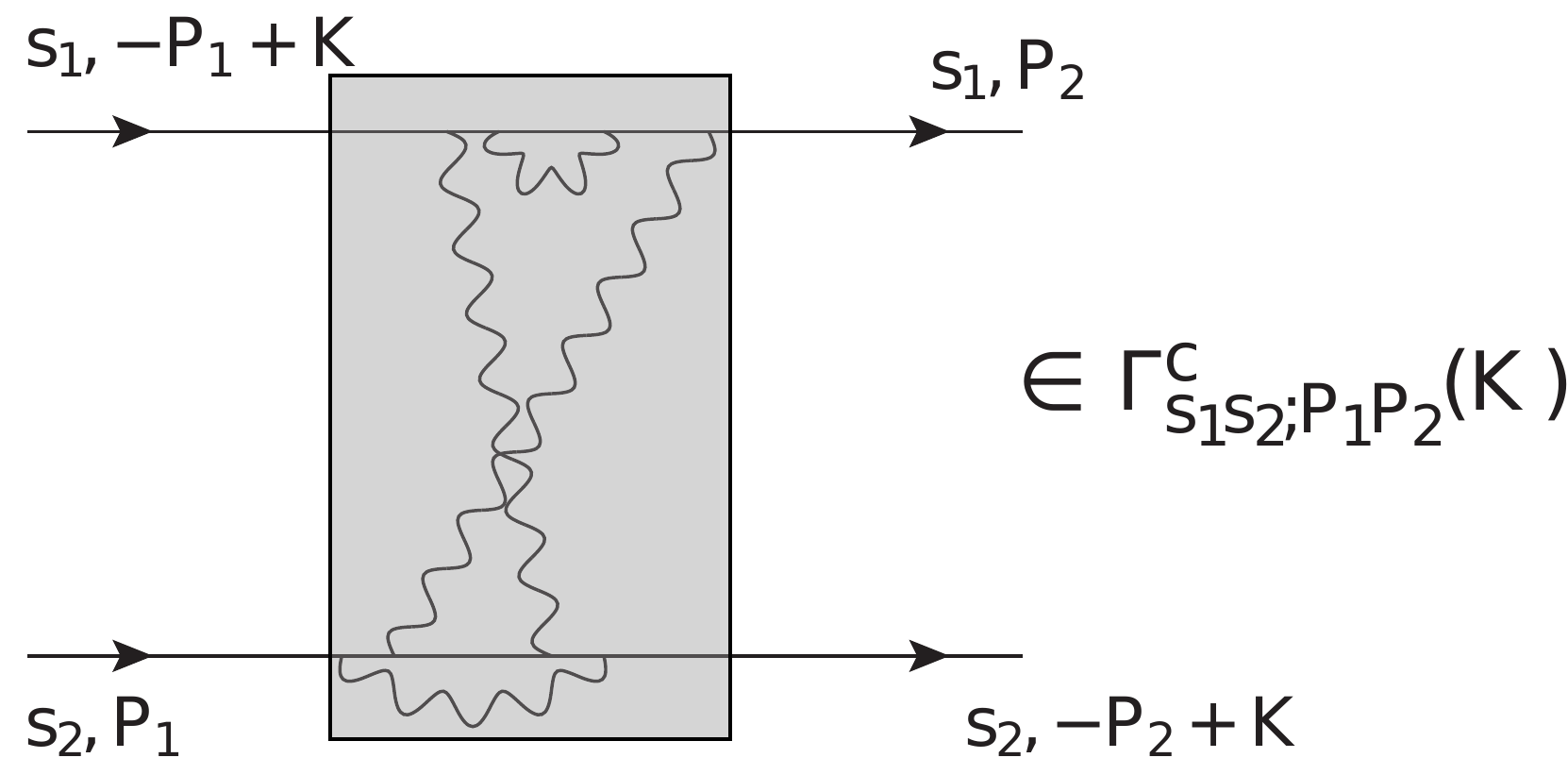}
\caption{An example of contribution to a scattering amplitude in the Cooper channel.}
\label{fig:Gammac}
\end{figure}

We refer to the three scattering channels as small angle scattering channel
($\Gamma^{0+1}$), large angle scattering channel ($\Gamma^2$), and the Cooper channel
($\Gamma^c$). The quantities entering Eq.~\eqref{eq:realIA} are the static
limit of the corresponding scattering amplitude, $\Gamma\left (\omega_m = 0, \v q
\right )$. They already include static screening and do not acquire any
tree-level corrections due to disorder. \cite{Finkelstein1990, Finkelstein2010}
Exemplary diagrams are given in figures \ref{fig:Gamma0} -- \ref{fig:Gammac}.
There, the small angle scattering amplitude is subdivided into its one Coulomb
line reducible part ($\Gamma^0$) and irreducible part ($\Gamma^1$) such that 
\begin{equation}
\Gamma^{0+1}=\Gamma^{0}+\Gamma^{1}.
\end{equation}
The irreducible part $\Gamma^1$
also includes the short range interaction induced by the finite thickness of the 3D TI film (see Appendix \ref{sec:Elstat} and \ref{sec:BareGamma}).

For the short-range interaction amplitudes ($\Gamma^1$, $\Gamma^2$, $\Gamma^c$),
the static limit coincides with the ``q-limit'' $\Gamma^q = \lim_{q\rightarrow 0
}\Gamma\left (\omega_m = 0 , \v q \right )$, see also Appendix
\ref{sec:cleanFL}. It should be kept in mind that for the one-Coulomb-line-reducible part $\Gamma^{0}$ (it is long-ranged) the ``q-limit'' $\Gamma^{0,q}$ is only a valid
approximation if the mean free path $l$ exceeds the screening length. This applies to most realistic situations. (In the opposite
case
$\Gamma^0$ is parametrically small. {On top of this, the $q$-dependence of the Coulomb potential implies a strong scale dependence of both conductivity corrections and the interaction amplitude until the running scale reaches} the screening length at which $\Gamma^{0} \approx \Gamma^{0,q}$ is again
justified.)

We conclude this section with a side remark concerning the topological exciton
condensation. \cite{TEC} In order to find the conventional pole structure of the
FL Green's functions for the case $\text{sgn}(\mu_s) =-1$ one needs to transpose the
bilinear form in action \eqref{eq:spinlessSkin} and swap the notation $\zeta_s
\left (\epsilon_n\right )\leftrightarrow \bar \zeta_s \left (-\epsilon_n\right
)$. If $\text{sgn}(\mu_1 \mu_2) =-1$, this interchange of notations obviously happens in only
one surface. In this case, the large-angle scattering amplitude $\Gamma^{2}_{12}$ and the
Cooper-channel amplitude $\Gamma^c_{12}$ are interchanged. Even though this procedure illustrates the analogy between exciton condensation (divergence in $\Gamma^{2}_{12}$) and Cooper instability (divergence in $\Gamma^c_{12}$), in the
following we
choose to keep our original notation of $\zeta_s$ and $\bar \zeta_s$ also in the case of $\mu_s <0$.

\subsubsection{Clean Fermi liquid theory}
\label{sec:cleanFLMaintext}

A systematic treatment of the scattering amplitudes involves the field-theory of
the FL\cite{Nozieres_Luttinger_1962, AbrikosovGorkovDzyaloshinski,
LandauLifshitz9} (see Appendix \ref{sec:cleanFL}.) It is valid down to energy
scales $\sim\tau^{-1}_{1,2}$ and therefore constitutes the starting point for
the effective diffusive theory at lower energies, $T \ll \tau^{-1}_{1,2}$.

In contrast to the Green's function of the free theory, in the FL the
exact electronic propagator contains both a singular and a regular part. The
singular part (``quasiparticle pole'') includes a renormalized dispersion
relation and its residue is no more equal to unity but rather is $a_s \in \left
(0,1\right )$. As usual in the context of disordered FLs,\cite{Finkelstein1990} we absorb the quasiparticle residue by rescaling
the fermionic fields and redefining the scattering amplitude.

The conservation of the particle number separately in each of the two
surfaces leads to the following Ward identities:
\begin{equation}
\Pi^\omega_{s_1,s_2} \equiv \lim_{\omega_m \rightarrow 0} \Pi_{s_1,s_2} \left
(\omega_m,\v q = 0 \right ) = 0
\label{eq:Piomega}
\end{equation}	
and
\begin{equation}
 \Pi^q_{s_1,s_2}  \equiv \lim_{\vert \v q \vert \rightarrow 0} \Pi_{s_1,s_2}
\left (\omega_m = 0,\v q \right ) =  -\frac{\partial N_{s_1}}{\partial
\mu_{s_2}}.
\label{eq:Piq}
\end{equation}
Since these identities reflect the gauge invariance, they can not be altered
during the RG procedure. Thus, the static polarization operator is always given
by the compressibility $\partial N_{s_1}/\partial \mu_{s_2}$.

The FL theory in a restricted sense contains only
short range interactions $\Gamma^1$, $\Gamma^2$ and $\Gamma^c$. For electrons in
metals, one has also to include the long-range Coulomb interaction. Following
Ref. \onlinecite{Nozieres_Luttinger_1962}, the associated scattering amplitude
$\Gamma^0$ is obtained by means of static RPA-screening of Coulomb interaction
with the help of the FL renormalized polarization operator and
triangular vertices (see Fig.~\ref{fig:Gamma0def}). In Appendix
\ref{sec:cleanFL} we explicitly perform the formal FL treatment. This determines the interaction amplitudes at ballistic scales. They will serve as bare coupling constants of the diffusive NL$\sigma$M (see Sec. \ref{sec:barevalues}).
We now turn our attention to the disordered FL. This will allow us to find out which of the interaction channels give
rise to soft modes within our problem.

\begin{figure}
\includegraphics[scale=.5]{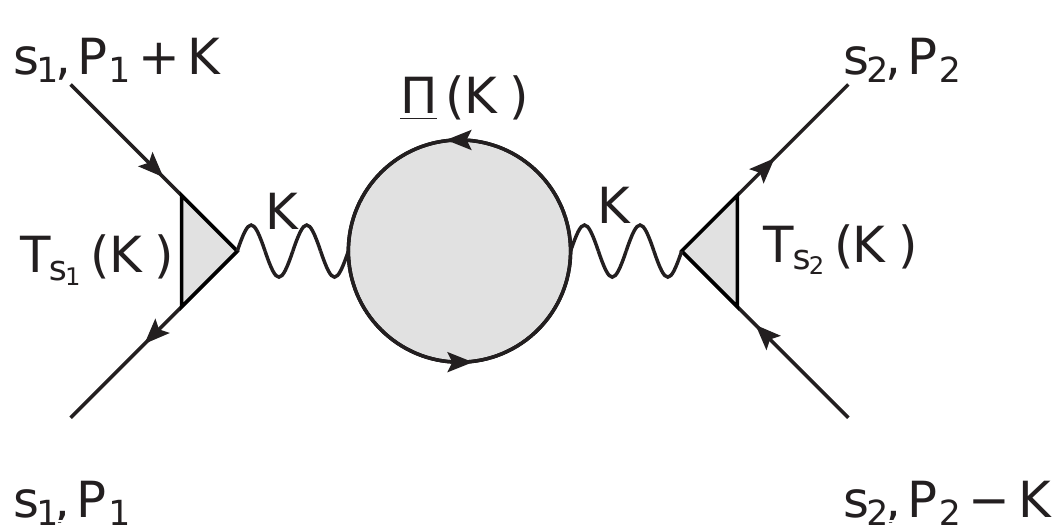}
\caption{A diagram contributing to $\Gamma^0$.}
\label{fig:Gamma0def}
\end{figure}

\subsubsection{Diffusive Fermi liquid theory}

The full amplitudes $\Gamma^{0+1}\left (K\right )$, $\Gamma^2\left (K\right )$
and $\Gamma^c\left (K\right )$ contain, among others, diagrams describing
multiple particle-hole (in the Cooper channel, particle-particle) scattering
(see Appendix \ref{sec:cleanFL}). The very idea of dirty FL lies in
replacing the dynamic part of these particle-hole (particle-particle) sections
by their diffusive counterpart.\cite{Finkelstein1990, Finkelstein2010} In
particular, only the zeroth angular harmonic of the scattering amplitudes
survives in the diffusive limit.

The scattering amplitude $\Gamma^2_{12}$ (as well as $\Gamma^c_{12}$) contains
only particle-hole (respectively, particle-particle) sections consisting of
modes from
opposite surfaces of the topological insulator. Since we assume the disorder to
be uncorrelated between the surfaces, these modes will not become diffusive
and are hence not of interest for the present investigation. We therefore do not
consider $\Gamma^2_{12}$ and $\Gamma^c_{12}$ any longer. As one can see from figures \ref{fig:Gamma0} - \ref{fig:Gamma2}, the large angle scattering amplitudes $\Gamma^2_{11}$ and
$\Gamma^2_{22}$ cannot be distinguished from the small angle scattering amplitudes $\Gamma^{0+1}_{11}$ and
$\Gamma^{0+1}_{22}$, respectively. We incorporate the effect of $\Gamma^2_{11}$ and
$\Gamma^2_{22}$ into the ``singlet channel'', which has the following matrix structure in the surface space
\begin{equation}
\underline \Gamma^{\rho} = \left (\begin{array}{cc}
\Gamma^{0+1-2}_{11} & \Gamma^{0+1}_{12} \\
\Gamma^{0+1}_{12} & \Gamma^{0+1-2}_{22}
\end{array} \right ). \label{eq:Gammasinglet}
\end{equation}
Here we used
\begin{equation}
\Gamma^{0+1-2}=\Gamma^{0+1}-\Gamma^{2}.
\end{equation}
The intrasurface Cooper channel interaction {$\Gamma^c_{ss}$} will be also neglected. Its bare value is repulsive for the Coulomb
interaction, so that the Cooper renormalization on ballistic scales $1/\tau \ll
E \ll \vert \mu \vert$ renders it small on the UV scale of the diffusive theory (i.e., at
the mean free path). Within the diffusive RG of a single 3D TI surface it
quickly becomes of the order of $1/\sqrt{\sigma}$ and thus negligible (see Ref. \onlinecite{Finkelstein1990} and
supplementary material of Ref. \onlinecite{OGM2010}).
{Consequently we drop the Cooper channel amplitude and do not consider the superconductive instability in this work.\footnote{The superconducting instability in a disordered system of Dirac fermions on a 3D TI surface (in the absence of density-density interactions) was recently addressed in Ref. \onlinecite{NandkishoreSondhi2013}.} For the opposite case of attraction in the Cooper channel Coulomb interaction suppresses the transition temperature $T_c$ .\cite{Finkelstein1987} The difference between Coulomb and short-range repulsive interaction was addressed in Ref. \onlinecite{BurmistrovGornyiMirlin12}.}

\subsubsection{Bosonization of Fermi Liquid}

The non-Abelian bosonization relies on the Dirac nature of the 2D
electrons
and on the associated non-Abelian  anomaly. On the other hand, for $\alpha \sim
1$ the spectrum of the system gets strongly renormalized by interaction.
An appropriate description in such a situation is the FL
theory which is restricted to fermionic excitations close to the Fermi level.
So, one can ask whether the result of non-Abelian bosonization remains
applicable for $\alpha \sim 1$. The answer is yes, for the follwing reasons.
All terms of the bosonized theory except for the $\mathbb Z_2$ theta term are
determined by fermionic excitations close to the  Fermi energy. Therefore, they
equally hold for the FL if the coupling constants are appropriately
redefined in terms of the corresponding FL parameters.

On the other hand, the $\mathbb Z_2$ theta term is a consequence of
the chiral anomaly and thus the only term determined by energies far from $\mu$.
However, it is well known that anomalies in quantum field theories are
insensitive to interactions. Hence, the $\mathbb Z_2$ term in the diffusive
NL$\sigma$M persists even for $\alpha \sim 1$. This follows also from
the key property of the FL state: its spectrum is
adiabatically connected to the free spectrum. This implies that
topological implications remain unchanged.
To summarize, the only difference between the NL$\sigma$M for the
weakly interacting
Fermi gas ($\alpha \ll 1$) and the FL ($\alpha \sim 1$) is
the replacement of the interaction strength by the appropriate FL
constant,
\begin{equation}
\underline U^q \rightarrow - \underline \Gamma^\rho \notag
\end{equation}
in Eq.~\eqref{eq:bosonizedIA}.

\subsubsection{Bare value of scattering amplitudes}
\label{sec:barevalues}

According to the formal FL treatment (Appendix
\ref{sec:appendixCoulomb}), the singlet-channel interaction  amplitude is given
by
\begin{equation}
\underline \nu \underline \Gamma^{\rho} \underline \nu = - \underline \nu -
\frac{\det  \underline \Pi^q}{\Pi^q_{11}  +\Pi^q_{22}+2\Pi^q_{12}} \left
(\begin{array}{cc}
1 & -1 \\
-1 & 1
\end{array} \right ), \label{eq:Coulombcond}
\end{equation}
where $(\underline \nu)_{ss'} = \nu_s \delta_{ss'}$ and
\begin{equation}
\underline \Pi^q = - \underline \nu - \underline \nu \left (\begin{array}{cc}
\Gamma^{1-2}_{11} & \Gamma^{1}_{12} \\
\Gamma^{1}_{12} & \Gamma^{1-2}_{22}
\end{array} \right ) \underline \nu. \label{eq:barePolop}
\end{equation}
Here $\Gamma^{1-2}=\Gamma^{1}-\Gamma^{2}$.
The remarkably simple matrix structure of $\underline \nu + \underline \nu
\underline \Gamma^{\rho} \underline \nu$ is actually due to the presence of the
long-range Coulomb interaction. This fact will be explained by means
of $\mathcal
F$-invariance in section \ref{sec:Finv}. It has very important consequences for
the RG flow in the diffusive regime, see Sec.~\ref{sec:Densityresponse}.

\subsubsection{Action of NL$\sigma$M}

We are now in a position to present the full action of the diffusive interacting
NL$\sigma$M for the problem under consideration:
\begin{equation}
\label{eq:TOTALNLSM}
S = \sum_s \left [S^{({\rm kin})}_s + i S^{(\theta)}_s\right ] +
S^{(\eta\,+\,{\rm int})}.
\end{equation}
It contains the kinetic term
\begin{equation}
S^{({\rm kin})}_s =  \frac{\sigma_s}{16} \int_{\v x} \tr \left (\nabla Q_s\right
)^2
\end{equation}
and the ${\mathbb Z}_2$ theta term
\begin{equation}
S^{(\theta)}_s =  \frac{1}{24 \pi} \left . \Gamma_s \right \vert_{\tilde
O_s\left (\v x, w = 1\right ) = Q_s\left (\v x\right ) = Q_s^T\left (\v x\right
)}
\label{eq:Stheta}
\end{equation}
for each of the surfaces, as well as the frequency and interaction terms,
\begin{eqnarray}
S^{(\eta\,+\,{\rm int})} &=& -\pi T \left [\sum_s 2z_s \tr \hat \eta Q_s \right . \notag
\\
& - & \left . \sum_{ss';n,\alpha} \tr \left [J_n^\alpha Q_s\right ]
\Gamma_{ss'}\tr \left [J_{-n}^\alpha Q_{s'}\right ]\right ] .
\label{eq:IANLSM}
\end{eqnarray}
Here we have introduced the notation
\begin{equation}
\Gamma_{ss'} = \frac{8}{\pi} \lambda_s \Gamma^\rho_{ss'}\lambda_{s'}.
\label{eq:defGammaNLSM}
\end{equation}

\subsection{Inclusion of scalar and vector potentials into the NL$\sigma$M}

In this subsection, we investigate consequences of the gauge invariance for the
interacting NL$\sigma$M.

\subsubsection{Electromagnetic gauge invariance}
\label{sec:elmaggauge}

We include the scalar potential $\Phi_s$ and the vector potential
$A_{\mu,s}$ for surface $s$ in the microscopic action
\eqref{eq:Highenergyaction}
by means of covariant derivatives.  This makes the action
gauge-invariant, i.e., unchanged under
local $\mathbf{U}(1)$-rotations of the fermionic fields $\psi$ and $\bar \psi$
accompanied by the corresponding gauge transformation of the potentials.
Note that locality implies independent rotations on the top and bottom surfaces of the TI film.

The rotations of $\psi$-fields imply the following rotation of bispinors:
\begin{equation}
\eta_s\left (\v x\right ) \rightarrow W_s \eta_s \left (\v x\right ),
\label{eq:etarotation}
\end{equation}
where
\begin{equation}
W_s = \left [e^{-i \hat \chi_s^T} \frac{1 +\tau_y}{2} + e^{i \hat \chi_s}
\frac{1 -\tau_y}{2} \right ]
\label{eq:deofofW}
\end{equation}
and we use the following convention for hatted matrices: $\hat a \equiv
\sum_{n,\alpha} a_n^\alpha I_n^\alpha$.
Let us recall that the $\eta_s$ fields are considered as
vectors in the Matsubara space. Upon introducing replica indices in the theory,
the $\mathbf{U}(1)$ rotation angles and correspondingly the gauge potentials
get replicated as well.

\subsubsection{$\mathcal{F}$-algebra and $\mathcal{F}$-invariance}
\label{sec:Finv}

As a direct consequence of \eqref{eq:etarotation}, $Q$-matrices transform under
a gauge transformation $\chi_s$ in the following way: 
\begin{equation}
Q_s \rightarrow W_s Q_s W_s^T. \label{eq:Wrot}
\end{equation}
Under such rotations, in the limit $N_M^\prime,  N_M \to \infty$,
$N_M/N_M^\prime \to 0$, the frequency term acquires the correction 
\cite{MishandlingI}
\begin{equation}
\delta_\chi \tr \hat \eta Q_s = 2 \sum_{n, \alpha} \left [in \chi_{s,n}^\alpha
\tr J_{-n}^\alpha Q_s - n^2  \chi_{s,n}^\alpha \chi_{s,-n}^\alpha\right ],
\label{eq:gaugetraforQ}
\end{equation}
while the factors entering the interaction term vary as follows:
\begin{equation}
\delta_\chi \tr J_n^\alpha Q_s = -i2 n \chi_{s,n}^\alpha.
\end{equation}
As explained in Sec. \ref{sec:symmetries}, the presence of the Coulomb
interaction
implies invariance of the fermionic action \eqref{eq:Highenergyaction} under
a simultaneous rotation in both surfaces by the same spatially constant
(``global'') but
time-dependent $\mathbf{U}(1)$-phase even \textit{without} inclusion of gauge
potentials (``$\mathcal F$-invariance''). This symmetry has to be
preserved on NL$\sigma$M level, implying that
\begin{equation}
\left (\underline{z} + \underline{\Gamma}\right )\left (\begin{array}{c}
1 \\
1
\end{array} \right ) = 0 .
 \label{eq:Finv0}
\end{equation}
Here $(\underline z)_{ss'} = z_s \delta_{ss'}$. Since the intersurface
interaction is symmetric, $\Gamma_{12}=\Gamma_{21}$, Eq. \eqref{eq:Finv0} 
yields
\begin{equation}
\underline{z} + \underline{\Gamma} =  \text{const.} \times \left
(\begin{array}{cc}
1 & -1 \\
-1 & 1
\end{array} \right ) \label{eq:Finv}.
\end{equation}
This relation is consistent with Eq. \eqref{eq:Coulombcond}. However, contrary
to Eq. \eqref{eq:Coulombcond}, the relation \eqref{eq:Finv} is {manifestly}
imposed by the symmetry (``$\mathcal{F}$-invariance'') of the action 
\eqref{eq:TOTALNLSM}. It should therefore remain intact under RG flow.

\subsubsection{Gauging the NL$\sigma$M and linear-response theory}
\label{sec:Kubo}

Generally, the requirement of gauge invariance prescribes the correct
coupling to the scalar and vector potentials in the action of the NL$\sigma$M,
Eq. \eqref{eq:TOTALNLSM}.
In particular, in the kinetic term one has to replace $\partial_\mu Q_s
\rightarrow D_{\mu,s} Q_s$ with the long derivative $D_\mu$ of the form
\begin{equation}
D_{\mu,s} Q_s \equiv \partial_\mu Q_s + \sum_{n,\alpha} i A^\alpha_{\mu,s, -n}
\left [J_n^\alpha - \left (J_n^\alpha\right )^T,Q_s\right ].
\end{equation}
For simplicity, the electron charge is absorbed into the vector potential here and in the following subsection.

As the theory is non-local in the imaginary time, the inclusion of the scalar
potential is non-linear. The corresponding term that should be added to the
NL$\sigma$M
\eqref{eq:TOTALNLSM} reads
\begin{eqnarray}
S^{\Phi} &=& -2 \sum_{n\alpha,ss'} {\Phi_{n,s}^\alpha} {\left (\underline
z+\underline \Gamma\right )_{ss'}} \tr J_{n}^{\alpha}Q_s \notag \\ &+&
\frac{1}{\pi T} \sum_{n\alpha,ss'} {\Phi_{n,s}^\alpha} {\left (\underline
z+\underline \Gamma\right )_{ss'}} \Phi_{-n, s}^\alpha .
\end{eqnarray}
The inclusion of the scalar and vector potentials allow us to express the
density-density correlation function and the conductivity in terms of the
matrix fields $Q_s$ by means of the linear-response theory.
In particular, a double differentiation of the partition function with respect
to the scalar
potential yields the density-density response,
\begin{eqnarray}
\Pi^{\rm RPA}_{ss'}\left (\omega_n, \v q\right ) &=& -\frac{2}{\pi} \left
(z+\Gamma\right )_{ss'} \notag \\
 && + 4 T\sum_{s_1,s_2}  \left (z+\Gamma\right )_{ss_1} \left \langle \tr
J_{n}^{\alpha}Q_{s_1}\left (\v q\right )\right . \times \notag \\
 && \times \left . \tr J_{-n}^{\alpha}Q_{s_2}\left (- \v q\right )\right \rangle
\left (z+\Gamma\right )_{s_2s'}  .
\label{eq:Kubodensity}
\end{eqnarray}
Here $\left \langle ... \right \rangle$ denotes average with respect to the action
\eqref{eq:TOTALNLSM}. The superscript $^{\rm RPA}$ emphasizes that the quantity appearing in the total density-density response includes RPA resummation. It is thus one-Coulomb-line-reducible and only its irreducible part
corresponds to the polarization operator.

In the same spirit, we obtain the expression for the conductivity (in units
of $e^2/h$) at a finite, positive frequency $\omega_n$:
\begin{equation}
\sigma'_{ss'}\left (\omega_n\right ) = B^{(s)}_1 \delta_{ss'} +  B^{(ss')}_2.
\label{eq:Kubocurrent}
\end{equation}
Here we introduced two correlators:
\begin{equation}
B^{(s)}_1 = \frac{\sigma_s }{8n} \left \langle \tr \left [J_n^\alpha - \left
(J_n^\alpha\right )^T,Q_s\right ]\left [J_{- n}^\alpha - \left
(J_{-n}^\alpha\right )^T,Q_s\right ]\right \rangle \label{eq:defB1}
\end{equation}
and
\begin{eqnarray}
\hspace*{-0.5cm} B^{(ss')}_2 &=& \frac{\sigma_s\sigma_{s'}}{128 n} \int_{\v x -
\v x'} \sum_{\mu = x,y} \notag \\
&& \left \langle \tr\left \lbrace \left [J_n^\alpha - \left (J_n^\alpha\right
)^T,Q_s\right ] \partial_\mu Q_s \right  \rbrace_{\v x} \times \right . \notag
\\
&& \times \left . \tr\left \lbrace \left [J_{-n}^\alpha - \left
(J_{-n}^\alpha\right )^T,Q_{s'}\right ] \partial_\mu  Q_{s'} \right \rbrace_{\v
x'} \right \rangle. \label{eq:defB2}
\end{eqnarray}
Substituting the saddle-point value $Q_s = \Lambda$, we obtain the classical value
$\sigma'_{ss'} \left (\omega_n\right ) = \sigma_{s}\delta_{ss'}$. Hence the
dimensionless coupling constant of the NL$\sigma$M has been identified with the
physical conductivity in units of $e^2/h$.

\subsubsection{Gauging the theta term and anomalous quantum Hall effect}
\label{sec:EdotB}

The local expression of the $\mathbb Z_2$ theta term, i.e, the WZW-term, Eq.
\eqref{eq:Stheta}, also allows of inclusion of gauge
potentials. \cite{PolyakovWiegman, DiVecchia, Faddeev,
Gerasimov, Nekrasov, Smilga} However, the
situation is more subtle here. Specifically, it turns out that the contribution
of non-singular gauge potentials to the topological term $S^{(\theta)}$
vanishes. We explicitly show this in Appendix \ref{sec:appendixBosonization}.

The situation changes when the time-reversal symmetry is broken (at least,
in some spatial domain at the surface) by a random or/and unform magnetic
field.
Subjected to a strong magnetic field, 3D TI surface states display the
characteristic quantum Hall effect of Dirac electrons \cite{BruehneHgTe,NomuraQHE}
 with quantized transverse conductance
\begin{equation}
\sigma_{xy} = g \left (n \pm \frac{1}{2}\right )\frac{e^2}{h}, \; n \in \mathbb
Z,
\label{eq:anomalous_QHE}
\end{equation}
where $g$ is the degeneracy of Dirac electrons, e.g., $g = 2$ for two 3D TI
surfaces. {It is intimately linked to the topological magnetoelectric effect.\cite{QiZhangTME,QiZhangMonopole,EssinMooreVanderbilt09,PesinMacDonald12}}
Theoretically, the anomalous quantum Hall effect was explained and discussed in a previous
work by three of the authors. \cite{OGMQHE} We will explain in the following how to understand it in the
framework of the
linear response theory within the NL$\sigma$M. As it turns out, the crucial
point is that gauge potentials drop from $S^{(\theta)}$.

We first briefly recall the NL$\sigma$M field theory  describing the
ordinary integer
QHE (i.e., for electrons with quadratic dispersion). It contains Pruisken's
theta
term, \cite{Pruisken} which assumes the following form upon inclusion of the vector potential: \cite{MishandlingI} 
\begin{subequations}
\begin{align}
S^{\text{QHE}}=&\frac{ \vartheta}{16\pi} \int_{\v x} \epsilon_{\mu\nu} \tr Q_U
\partial_\mu Q_U \partial_\nu Q_U
\label{eq:Pruiskenterm}\\
&+ \frac{ i \vartheta}{4\pi} \int_{\v x} \epsilon_{\mu\nu} \tr \partial_\mu \hat
A_\nu Q_U  \label{eq:FQ} \\
&+ \frac{\vartheta}{4\pi} \int_{\v x}
\epsilon_{\mu\nu}
\sum_{n,\alpha} n A_{\mu,n}^\alpha A_{\nu,-n}^\alpha.
\label{eq:ChernI}
\end{align}\label{eq:QHE}
\end{subequations}
Here $Q_U = U^{-1}\Lambda U$ with  $U \in \mathbf{U}\left (2 N_M \times
N_R\right )$, $\epsilon_{\mu\nu} = -\epsilon_{\nu\mu}$ is the 2D antisymmetric symbol ($\epsilon_{xy} \stackrel{\text{def.}}{=} 1$), and $\vartheta$ is the theta angle of the Pruisken's NL$\sigma$M. We
emphasize, that the last two terms (Eqs.~\eqref{eq:FQ} and \eqref{eq:ChernI})
determine the effective electromagnetic response and thus prescribe the relation
between the physical observable $\sigma_{xy}$ (in units of $e^2/h$) and the theta angle
$\vartheta$. In particular, $\vartheta/2\pi$ is identified as the bare value of the Hall conductance.\cite{PruiskenBurmistrovCPN-1}

Let us now turn to a single Dirac surface state. As has been discussed above,
all gauge potentials drop from $S^{(\theta)}$. Let us first add a random magnetic
field (keeping zero average magnetic field) to the gauged NL$\sigma$M.
This implies a breakdown of the symmetry:
\begin{equation}
\mathcal{M} \rightarrow \frac{\mathbf{U}\left (2
N_M N_R\right )}{\mathbf{U}\left ( N_M  N_R\right )\times\mathbf{U}\left (N_M
N_R\right )}.
\end{equation}
The $\mathbb Z_2$ theta term becomes the Pruisken's theta
term \cite{BocquetSerbanZirnbauer} (recall $\theta = \pi\, \text{mod}\, 2\pi$)
\begin{equation}
S^{(\theta)}_{U}=\frac{\theta}{16\pi} \int_{\v x} \epsilon_{\mu\nu} \tr Q_U
\partial_\mu Q_U \partial_\nu Q_U.  \label{eq:Pruiskentermcriticality}\\
\end{equation}
We emphasize that {together with the gauged kinetic term} $S^{(\theta)}_{U}$ is the complete gauged theory, no extra terms of
the type 
\eqref{eq:FQ} and \eqref{eq:ChernI} appear. Being topological, the Pruisken's theta
term is invariant under smooth $\mathbf U\left (1\right )$ rotations. Recall
that exactly the terms  \eqref{eq:FQ} and \eqref{eq:ChernI} provided a link
between $\vartheta$ and $\sigma_{xy}$ in the conventional (non-Dirac) QHE
setting. Their absence in Eq. \eqref{eq:Pruiskentermcriticality} is thus
physically very natural: without a net magnetic field the Hall conductivity is
zero.

We consider now the case when the average magnetic field is non-zero. The
action of the NL$\sigma$M  describing a Dirac fermion is then given by a sum of
Eqs.~\eqref{eq:QHE} and \eqref{eq:Pruiskentermcriticality}. 
The renormalization of the action of the NL$\sigma$M is governed by the full theta angle $\vartheta + \theta$. On the other hand, only $\vartheta$ is related with the bare value of $\sigma_{xy}$. Then standard arguments for the quantization of the Hall conductivity \cite{PruiskenBurmi} leads to the result
(\ref{eq:anomalous_QHE}) for the anomalous QHE.

\section{One-loop RG}
\label{sec:1loopRG}

In the preceding section we have derived the diffusive {NL$\sigma$M},
Eqs. \eqref{eq:TOTALNLSM}. We will now investigate its behavior under
renormalization. This will allow us, in particular, to deduce the scale
dependence of the conductivity. The most important steps of the calculation are
presented in the main text; further details can be found in Appendix
\ref{sec:RGderiv}.

We calculate the renormalization of the NL$\sigma$M parameters within the
linear-response formalism (rather than the background-field method). This is
favorable since it implies a more direct physical interpretation of the
NL$\sigma$M coupling constants. Furthermore, this way one can in principle
treat simultaneously different infrared regulators, such as temperature or
frequency. However,
for the sake of  clarity of presentation we restrict ourselves to a purely
field-theoretical regularization scheme and add a mass term to the action
\begin{equation}
S_{L} = -\sum_{s=1,2}\frac{\sigma_s L^{-2}}{8} \int_{\v x} \tr \Lambda Q_s.
\label{eq:Massterm}
\end{equation}
The connection between the running length scale $L$ and the physical regulators
temperature or frequency was analyzed in Ref. [\onlinecite{MishandlingII}].
Roughly speaking, in the presence of a single infrared scale $E$, e.g. when calculating DC conductance at finite
temperature and assuming an infinite sample, one can replace $L$ by
$L_E$ in the results.

We will calculate all UV-divergent contributions in the dimensional regularization scheme.
This allows us to preserve the local
$\text{O}\left (2_\tau \times N_M \times N_R\right ) \times \text{O}\left
(2_\tau \times N_M \times N_R\right )$-symmetry of the $Q$-matrix
\eqref{eq:defQ} and to ensure the renormalizability of the theory.

\subsection{Diffusive propagators}
\label{sec:Propagators}

We employ the exponential parametrization of the matrix fields $Q_s = \Lambda
\exp W_s$. The antisymmetric fields
$$W_s = \left
(\begin{array}{cc}
0 & q_s \\
-q_s^T & 0
\end{array} \right )
$$
anticommute with $\Lambda$.
Further, we define a set of real matrices in the particle-hole space: $\tilde
\tau_\mu \equiv 2^{-1/2} \left (\mathbf{1}, \tau_x, i \tau_y, \tau_z \right
)$. This allows us to introduce the fields $q^{(\mu)} \equiv \tr^{\tau} q \tilde
\tau_\mu^T$, where $\tr^\tau$ is the trace in the particle-hole space only.
With these definitions at hand, we expand the action, Eqs.~\eqref{eq:TOTALNLSM}
and \eqref{eq:Massterm}, to quadratic order in $q^{(\mu)}$ and obtain the
NL$\sigma$M propagators that describe the diffusive motion in the particle-hole
(diffusons) and particle-particle (cooperons) channels.

The fields $q^{(1)}$ and $q^{(3)}$ describe cooperons. Their propagator is
unaffected by interaction (since we have discarded the interaction in the
Cooper channel),

\begin{eqnarray}
\left \langle \left [q_s^{(\mu)}(\v p)\right ]_{m_1m_2}^{\alpha_1\alpha_2} \left
[q_{s'}^{(\nu)}(-\v p)\right ]_{n_1n_2}^{\beta_1\beta_2} \right \rangle = \frac{4}{\sigma_s} D_s
\left (\omega_{n_{12}},\v p\right )  \delta_{ss'} \notag \\
\hfill\times \delta_{\mu\nu} \delta_{n_1m_1}\delta_{n_2m_2}\delta_{\alpha_1\beta_1}
\delta_{\alpha_2\beta_2}\bigl (\delta_{\mu 1} + \delta_{\mu 3}  \bigr ) , \hspace{0.3cm}{}\,
\end{eqnarray}
where
\begin{equation}
\bigl [ D_s\left (\omega_{n_{12}},\v p\right ) \bigr ]^{-1}= \v
p^2 + L^{-2} + \frac{4z_s}{\sigma_s} \omega_{n_{12}}.
\label{eq:usualpropagator}
\end{equation}
The Matsubara indices $n_1$, $m_1$ are non-negative, while the indices
$n_2$, $m_2$ are negative; we have also defined $n_{12} \equiv n_1-n_2 > 0$ and
$m_{12} \equiv m_1-m_2 > 0$.

Next, we consider the diffusons $q^{(0)}$ and $q^{(2)}$. Their Green's function,
written as a matrix in surface space, is
\begin{eqnarray}
&\left \langle \left [q^{(\mu)}_s(\v p)\right ]_{m_1m_2}^{\alpha_1\alpha_2}
\left [q^{(\nu)}_{s'}(-\v p)\right ]_{n_1n_2}^{\beta_1\beta_2} \right  \rangle
= \displaystyle \frac{4}{\sigma_s} D_s
\left (\omega_{n_{12}},\v p\right ) 
\delta_{\mu\nu} \notag \\
& \times\delta_{n_{12},m_{12}}\delta_{\alpha_1\beta_1}
\delta_{\alpha_2\beta_2}\bigl (\delta_{\mu 0} + \delta_{\mu 2}  \bigr ) \notag \\
&\times 
 \left [\delta_{n_1m_1} \delta_{ss'}
- \displaystyle\frac{8\pi T}{\sigma_{s'}} \delta^{\alpha_1\alpha_2} \left (\underline{\Gamma  D^c}\left (\omega_{n_{12}},\v p\right )
 \right )_{ss'}\right ]. 
\label{eq:totaldiffusonpropagator}
\end{eqnarray}
Here we have introduced 
\begin{equation}
\bigl [D^c\left  (\omega_{n_{12}},\v p\right )\bigr ]^{-1}_{ss'} = D_s^{-1}\left (\omega_{n_{12}},\v p\right )\delta_{ss'}  + \frac{4 \omega_{n_{12}}}{\sigma_s}
\Gamma_{ss'} .
\label{eq:AApropagator}
\end{equation}

\subsection{RG invariants}
\label{sec:Densityresponse}

The bare action contains, aside from the mass $L^{-1}$, seven running coupling
constants: $\sigma_1$, $\sigma_2$, $z_1$, $z_2$, $\Gamma_{11}$, $\Gamma_{22}$
and $\Gamma_{12}$. We are now going to show that three linear combinations of
them are conserved under RG. To this end we evaluate the density-density
response \eqref{eq:Kubodensity} at the tree level:

\begin{equation}
\underline{\Pi}^{\rm RPA}\left (\omega, \v p\right ) = -\frac{2}{\pi} \left
[\underline{z} + \underline{\Gamma}\right ]\bigl (1-4\omega \underline{\sigma^{-1} D^c}\left (\omega, \v p\right )\left
[\underline{z} + \underline{\Gamma}\right ]\bigr )
\label{eq:Pired}
\end{equation}
where $(\underline{\sigma})_{ss'}=\sigma_s\delta_{ss'}$. There is no need for infrared
regularization here and we therefore omit the mass term \eqref{eq:Massterm}.

On the other hand, the density-density response function can be obtained from
the fermionic formulation of the theory, see Appendix \ref{sec:Piredappendix}:

\begin{equation}
\underline{\Pi}^{\rm RPA} = \left
[\underline{\Pi}^q - \underline{\nu\Gamma}^0 \underline \nu\right ]\left (1+
\omega \underline \Delta^{\Gamma}\left (\omega, \v p\right )\left [\underline{\Pi}^q -
\underline{\nu\Gamma}^0 \underline \nu\right ]\right ),
\label{eq:fermionicPired}
\end{equation}
where
\begin{equation}
\Delta^{\Gamma} \left (\omega, \v p\right ) = \left
[\underline{\nu D} \v p ^2 + \omega\left ( \underline {\nu} + \underline{\nu}
\underline{\Gamma}^{\rho,q} \underline{\nu}\right ) \right ]^{-1} .
\end{equation}

The equality of {Eqs.} \eqref{eq:Pired} and \eqref{eq:fermionicPired} 
{relates two functions of momentum and frequency. In the static limit, we find}
the following {constraint} connecting the  NL$\sigma$M coupling constants with
physical FL parameters:

\begin{eqnarray}
\frac{2}{\pi}\left (\underline{z} + \underline{\Gamma}\right )  =  -
\underline{\Pi^q} +  \underline{\nu}\underline{\Gamma^0}\underline{\nu}.
\label{eq:constraint1}
\end{eqnarray}

{Next, from comparison of momentum dependence in Eqs. \eqref{eq:Pired} and \eqref{eq:fermionicPired},} we find {the Einstein relation:} $\sigma_s = 2\pi\nu_s D_s$. Accordingly, $\sigma$ measures the
conductance in units of $e^2/h$, consistently with what has been found
in Secs.~\ref{sec:semiclassical} and \ref{sec:Kubo}.

In view of gauge invariance (Sec.
\ref{sec:cleanFLMaintext}), the static polarization operator
entering Eq.~(\ref{eq:constraint1}) is nothing but the
compressibility
$$\Pi^{ss',q} = - \frac{\partial N_s}{\partial \mu_{s'}}.$$
Its value is not renormalized because it can be expressed as a derivative of a
physical observable with respect to the chemical potentials. On ballistic
scales the chemical potential enters
logarithmically divergent corrections only as the UV cutoff of the
integrals. In the diffusive regime, the UV cutoff is provided by the scattering
rates $\tau^{-1}_s \ll \vert \mu_s \vert$. Therefore, diffusive contributions to the
derivative with respect to the chemical potential
vanish.\cite{Finkelstein1990} Since $\underline \nu \underline \Gamma^0
\underline \nu$ only depends on $\underline \Pi^q$ (see Appendix
\ref{sec:appendixCoulomb}) it is not renormalized as well.
Therefore, the right-hand side of
\eqref{eq:constraint1} is not renormalized and hence neither is its
left-hand-side, i.e., $\underline z + \underline \Gamma$.
This matrix constraint yields three RG invariants: $z_1 + \Gamma_{11}$,
$z_2 + \Gamma_{22}$, and $\Gamma_{12}$. Thus,  only four out of seven
NL$\sigma$M parameters are independent running coupling constants. We
emphasize that, in contrast to Eq. \eqref{eq:Finv}, this reasoning is valid also
in the absence of long-range interaction.

Finally, let us evaluate {Eq.} \eqref{eq:constraint1} on the bare level. Expressing the static polarization operator as $\underline \Pi^q = - \underline \nu -
\underline{\nu}\underline{\Gamma^{1-2}}\underline{\nu}$ {and using} the definition of $z_s$ in Sec. \ref{sec:freqboson}
one can {find the following relations for} the bare values
\begin{equation}
\frac{4\lambda_s}{\pi} \equiv \frac{2}{\pi} z_s = \nu_s .
\label{eq:relrel0}
\end{equation}
Equivalently, the same relationship between $\lambda_s$ and $\nu_s$
can be obtained by comparing the bare definition of $\underline \Gamma$ [Eq.
\eqref{eq:defGammaNLSM}] with the right hand side of \eqref{eq:constraint1}.
{The relation \eqref{eq:relrel0}} has been foreseen earlier on the basis of SCBA, see Eq.
\eqref{eq:lambdamu}. In conclusion, the SCBA and the density response
independently show that the  UV cutoff scale for the bosonization is
automatically set by the chemical potential (which is also very natural from
the physical point of view).

\subsection{Renormalization of conductivities}

\subsubsection{{Correlator $B_1$}}

We will first analyze the correlator $B^{(s)}_1$, Eq. \eqref{eq:defB1}. The
one-loop correction is determined by the expansion to second order in
$q^{(\mu)}$. The tensor structure in particle-hole space implies that the
diffuson contribution ($\mu = 0,2$) vanishes. The classical
value together with the cooperon contribution ($\mu = 1,3$) is

\begin{equation}
B^{(s)}_1 = \sigma_s +2 \int_{\v p }D_s(\omega_n,\v p)  .
\label{eq:usualHLN1}
\end{equation}
We evaluate this term in the announced regularization scheme:
\begin{eqnarray}
B^{(s)}_1 &=& \sigma_s + 2 \text{I}^{(2+\epsilon)}_1\\ 
	&=& \sigma_s + \frac{1}{2\pi} \left [-\frac{2}{\epsilon} + 2\ln {L}/{l}
+ \text{const.}\right ] .
\label{eq:usualHLN}
\end{eqnarray}
For dimensional reasons we have introduced the reference length scale $l$,
which for the present diffusive problem is set by the mean free path $l =
\max_{s=1,2}l_s$. We have further evaluated the following standard
dimensionless integral
\begin{eqnarray}
\text I^{(D)}_1 &\equiv & l^{D-2}\int \frac{d^D p}{\left (2 \pi\right )^D}
\frac{1}{\v p^2 + L^{-2}} \notag \\
&=& \frac{ \left (\frac{l^2}{L^2}\right )^{\frac{D}{2}-1}}{\left (4\pi\right
)^{\frac{D}{2}}} \Gamma \left (1- \frac{D}{2}\right ) \notag \\
&\stackrel{D=2+\epsilon}{=}& \frac{1}{4\pi}\left [-\frac{2}{\epsilon} + 2\ln
{L}/{l} + \ln 4\pi - \boldsymbol \gamma +  \mathcal{O}\left (\epsilon\right )\right ],
\notag 
\end{eqnarray}
where $\boldsymbol \gamma \approx 0,577 $ is the Euler-Mascheroni constant. 

The logarithmic term in Eq.~\eqref{eq:usualHLN} is nothing but the
well-known weak-antilocalization effect.\cite{HLN}

\subsubsection{{Correlator $B_2$}}

Next we turn our attention to $B^{(ss')}_2$, Eq.~\eqref{eq:defB2}. Because of
the presence of gradients {it} does not contribute neither at classical nor at
tree level. Furthermore, due to the absence of the Cooper channel and the
uncorrelated disorder on the {top and bottom} surfaces, there are no quantum corrections to
the transconductance $\sigma_{12}$.
The correlator $B^{(ss')}_2$ can be recast into the form (see Appendix \ref{sec:RGderiv})

\begin{eqnarray}
B_2^{(ss')} &=& \frac{16\delta_{ss'}}{n\sigma_s}\int_{\v p} \v p^2  \sum_{\omega_m>0} \omega_m \notag \\
&&\times \Bigl [ \left ( \underline{D\Gamma D^c}\right )_{ss} \left (\omega_{m},
{\v p}\right )D_s \left (\omega_{m+n}, {\v p}\right ) \notag \\ 
&& - \left ( \underline{D\Gamma D^c}\right )_{ss} \left (\omega_{m+n}, {\v
p}\right )D_s \left (\omega_{m+2n}, {\v p}\right ) \Bigr ] .
\label{eq:AAcorr1}
\end{eqnarray}

For its evaluation it is instructive to separate
contributions stemming from intrasurface interaction $\Gamma_{ss}$ and
intersurface interaction $\Gamma_{12}$.  This leads to 

\begin{eqnarray}
B_2^{ss'} &=& -4 \delta_{ss'} \left (\underbrace{1 -
\frac{1+\gamma_{ss}}{\gamma_{ss}} \ln \left ( 1 + \gamma_{ss}\right
)}_{\text{single surface}} \right .\notag \\
& +& \left .\underbrace{\left (1+\gamma_{ss}\right )\left (\frac{\ln \left ( 1 +
\gamma_{ss}\right )}{\gamma_{ss}} - \frac{\ln \left ( 1 + \tilde
\gamma_{ss}\right )}{\tilde \gamma_{ss}} \right )}_{\text{intersurface
interaction}} \right ) I_2^{(2+\epsilon)} \notag \\
&=& -\frac{\delta_{ss'}}{\pi}\left (1 - \frac{1+\gamma_{ss}}{\tilde \gamma_{ss}} \ln \left
( 1 + \tilde \gamma_{ss}\right )\right )\times \notag \\
&& \times \left [- \frac{2}{\epsilon} + 2 \ln L/l + \text{const}\right ] .
\label{eq:AAcorr}
\end{eqnarray}

We have introduced $\gamma_{ss} = \Gamma_{ss}/z_s$, $\tilde{\gamma}_{11}
= \gamma_{11} + (\sigma_1/\sigma_2)  (1+\gamma_{11} )$ and
$\tilde{\gamma}_{22}= \gamma_{22} + (\sigma_2 / \sigma_1)
(1+\gamma_{22} )$. Note that in the limit of $z_2 + \Gamma_{22} = 0$
[which corresponds to $\Gamma_{12} =0$ in view of \eqref{eq:Finv}] we recover
the well-known conductivity corrections to $\sigma_{11}$ for a single surface
(see also Sec. \ref{sec:surfacebulk}).
Further, in Eq.~(\ref{eq:AAcorr}) we have evaluated the second standard
diverging integral
\begin{eqnarray}
\text I^{(D)}_2 &\equiv & l^{D-2}\int \frac{d^D p}{\left (2 \pi\right )^D}
\frac{\v p^2}{\left (\v p^2 + L^{-2}\right )^2} \notag \\
&=& \frac{ \left (\frac{l^2}{L^2}\right )^{\frac{D}{2}-1}}{\left (4\pi\right
)^{\frac{D}{2}}} \frac{D}{2}\Gamma \left (1- \frac{D}{2}\right ) \notag \\
&\stackrel{D=2+\epsilon}{=}& \frac{1}{4\pi}\left [-\frac{2}{\epsilon} + 2\ln
{L}/{l} + \ln 4\pi -1 - \boldsymbol \gamma +  \mathcal{O}\left (\epsilon\right )\right ].
\notag
\end{eqnarray}

\subsection{Renormalization of the interaction amplitudes}

The renormalization of the interaction amplitudes, or equivalently, of Finkelstein
parameters $z_s$, is intimately linked to the renormalization of the specific
heat. \cite{CastellaniDiCastro86} This is because the scale (e.g., temperature)
dependence of
the total thermodynamic potential $\Omega$ is governed by the scale dependence of $z_s$. In the
present case of coupled surfaces we can only extract the correction to the
sum $z_1 + z_2$ from the (one-loop) correction to the {total} {thermodynamic potential}: \cite{MishandlingII}

\begin{equation}
z'_1 +z'_2 = \frac{1}{2\pi \tr \eta \Lambda} \frac{\partial}{\partial T}
\frac{\Omega}{T} .
\label{eq:z12def}
\end{equation}
At the classical level Eq. \eqref{eq:z12def} yields the relation $z'_1 +z'_2 = z_1+z_2$. 
Evaluating the quantum corrections in Eq. \eqref{eq:z12def}, we find
\begin{equation}
\left (z'_1 +z'_2\right ) = \left (z_1 +z_2\right ) + 2 \sum_{s=1,2} \Gamma_{ss}
\int_{\v p} D_s\left (0, \v p\right ) .
\end{equation}
As the correction is a sum of contributions from the two
opposite surfaces, it is natural to assume that the parameters $z_s$ are
renormalized $separately$ (and without intersurface interaction effects):

\begin{eqnarray}
 z_s' 	&=&z_s + 2  \Gamma_{ss}
\int_{\v p} D_s\left (0, \v p\right ) \notag \\
		&=&z_s + 2 \frac{\Gamma_{ss}}{\sigma_s} \text{I}^{(2+\epsilon)}_1
\notag \\
		&=& z_s + \frac{1}{2\pi} \frac{\Gamma_{ss}}{\sigma_s} \left [-\frac{2}{\epsilon}
+ 2 \ln L/l + \text{const}\right ].  \label{eq:Renormz}
\end{eqnarray}

We have directly proven this assumption of separate $z_s$ renormalization by
the background field method. \footnote{Instead of considering the
renormalization of $z_s$ one can equivalently consider the renormalization of
$\Gamma_{ss}$. It is governed by the interaction term $S^{int}$ in Eq.
\eqref{eq:IANLSM}. Within the background field method two types of contributions
can arise. First, there is $\left \langle S^{int}\right
\rangle_{\text{fast}}$. This term does not involve a frequency integration.
Because disorder is uncorrelated between the surfaces, $\Gamma_{11}$ and
$\Gamma_{22}$ are renormalized separately. This is described by Eq.
\eqref{eq:Renormz}. All possible further contributions at this order would arise from $\left \langle \left
(S^{int}\right ) ^2\right \rangle_{\text{fast}}$. This term generates so-called
ring diagrams.\cite{Finkelstein1990} We have explicitly checked 
that the ring diagrams vanish in one-loop approximation.}

\subsection{The one-loop RG equations}

Applying the minimal subtraction scheme to Eqs.~\eqref{eq:usualHLN}, \eqref{eq:AAcorr} and \eqref{eq:Renormz},
{we derive the one-loop perturbative RG equations}:
\begin{subequations}
\begin{align}
\T \B \frac{d\sigma_1}{dy} & = - \frac{2}{\pi} F\left
(\gamma_{11},\frac{\sigma_1}{\sigma_2}\right ),\\
\T \B \frac{d\sigma_2}{dy} & = - \frac{2}{\pi} F\left
(\gamma_{22},\frac{\sigma_2}{\sigma_1}\right ), \\
\T \B \frac{d\gamma_{11}}{dy} &= - \frac{\gamma_{11} \left (1+\gamma_{11}\right
)}{\pi \sigma_1}, \label{eq:RGgamma11} \\
\T \B \frac{d\gamma_{22}}{dy} &= - \frac{\gamma_{22} \left (1+\gamma_{22}\right
)}{\pi \sigma_2},
\label{eq:RGgamma22}
\end{align}
\label{eq:RGeqs}
\end{subequations}
where {$y = \ln L/l$, $\gamma_{ss} = \Gamma_{ss}/z_s$, $l =
\max_{s=1,2}l_s$ and
\begin{equation}
F\left (\gamma,x\right ) = \frac{1}{2}  - \frac{1+\gamma}{x \left
[1+\gamma\left (1+ \frac{1}{x}\right )\right ]} \LN\left [\left (1+x\right
)\left (1 + \gamma\right )\right ].
\end{equation}
We recall that
$\Gamma_{12}$, $z_1 + \Gamma_{11}$ and $z_2 + \Gamma_{22}$ are not
renormalized. {We mention that the mass $L^{-1}$ acquires a quantum correction
\cite{MishandlingII} but it does not affect the one-loop  renormalization of the
other parameters $\sigma_s$, $z_s$ and $\Gamma_{ss'}$.}   

For an alternative presentation of the RG equations \eqref{eq:RGeqs} we introduce the total
conductivity $\sigma = \sigma_1 + \sigma_2$ and the ratio of the conductivities
of the two surfaces $t = \sigma_1/\sigma_2$. In terms of these parameters the RG
equations take the following form:

\begin{widetext}
\begin{subequations}
\begin{align}
\T \B \frac{d\sigma}{dy} & = - \frac{2}{\pi} \left \lbrace  1 - \frac{1}{t}
\frac{1+\gamma_{11}}{1+\gamma_{11}\left (1+ \frac{1}{t}\right )}  \LN\left
[\left (1+{t}\right )\left (1 + \gamma_{11}\right )\right ]  - {t}
\frac{1+\gamma_{22}}{1+\gamma_{22}\left (1+ {t}\right )} \LN\left [\left
(1+\frac{1}{t}\right )\left (1 + \gamma_{22}\right )\right ] \right  \rbrace,
\\
\T \B \frac{dt}{dy} & = - \frac{2}{\pi} \frac{1+t}{\sigma}\left \lbrace
\frac{1-t}{2}   - \frac{1}{t} \frac{1+\gamma_{11}}{1+\gamma_{11}\left (1+
\frac{1}{t}\right )}   \LN\left [\left (1+{t}\right )\left (1 +
\gamma_{11}\right )\right ] + {t^2} \frac{1+\gamma_{22}}{1+\gamma_{22}\left (1+
{t}\right )} \LN\left [\left (1+\frac{1}{t}\right )\left (1 +  \gamma_{22}\right
)\right ] \right \rbrace ,\\
\T \B \frac{d\gamma_{11}}{dy} &= - \left (1+\frac{1}{t}\right )\frac{\gamma_{11}
\left (1+\gamma_{11}\right )}{\pi  \sigma},  \label{eq:RGeqs:gamma11}\\
\T \B \frac{d\gamma_{22}}{dy} &= - \left (1+t\right )\frac{\gamma_{22} \left
(1+\gamma_{22}\right )}{\pi \sigma} .  \label{eq:RGeqs:gamma22}
\end{align}
\label{eq:RGeqssigmatot}
\end{subequations}
\end{widetext}

\section{Analysis of the RG equations}
\label{sec:AnalysisofRG}

It is worthwhile to remind the reader that the RG equations \eqref{eq:RGeqs}
describe the quantum corrections to conductivity due to the interplay of two
distinct effects. First, they contain weak-antilocalization corrections (WAL)
$\delta \sigma^{WAL}_s = (1/\pi) \ln L/l$ due to quantum
interference in a disordered system with the strong spin-orbit coupling.
Second, these are interaction-induced contributions of Altshuler-Aronov (AA)
type, including effects of both, long-range and short-range interactions. The
result \eqref{eq:RGeqs} was obtained perturbatively to leading order in
$1/\sigma_s \ll 1$ but it is exact in the singlet interaction
amplitudes. While these equations describe the experimentally most relevant
case of Coulomb interaction, in Appendix \ref{sec:shortrange} we also
present the RG equations for the case of short-range interaction.

Equations \eqref{eq:RGeqs} which determine the flow of the coupling constants
$\sigma_1, \sigma_2, \gamma_{11}$ and $\gamma_{22}$ imply a rich phase diagram
in the four-dimensional parameter space. Before discussing the general
four-dimensional RG flow we highlight the simpler case of two equal
surfaces. 

\subsection{Two equal surfaces}
\label{sec:twoequalsurfaces}

Equal surfaces are defined by $\sigma_1 = \sigma_2 = \sigma/2$,
$\gamma_{11} = \gamma_{22} = \gamma$ and, because of Eq. \eqref{eq:Finv},
$\gamma_{12} = -1 - \gamma$. It can be checked that the plane of identical surfaces 
is an attractive fixed plane of the four dimensional RG-flow (see Appendix
\ref{sec:stableequalsurfaces}). The RG equations for the
two coupling constants $\sigma$ and $\gamma$ are
\begin{subequations}
\begin{align}
\T \B \frac{d\sigma}{dy} &=  - \frac{2}{\pi} \left [ 1 -  \frac{2+2\gamma}{1+2\gamma}  \LN \left (2 + 2 \gamma\right )\right ] , \label{eq:equallayers1} \\
\T \B \frac{d\gamma}{dy} &= -  \frac{2\gamma \left (1+\gamma\right )}{\pi \sigma}. \label{eq:equallayers2}
\end{align}
\label{eq:equallayers}
\end{subequations}
Experimentally, the case of equal surfaces is realized if both surfaces are
characterized by the same mean free path and the same carrier density and,
furthermore, if the dielectric environment of the probe is symmetric
($\epsilon_1 = \epsilon_3$).

\subsubsection{Flow Diagram within the fixed plane}

\begin{figure}
\includegraphics[scale=.4]{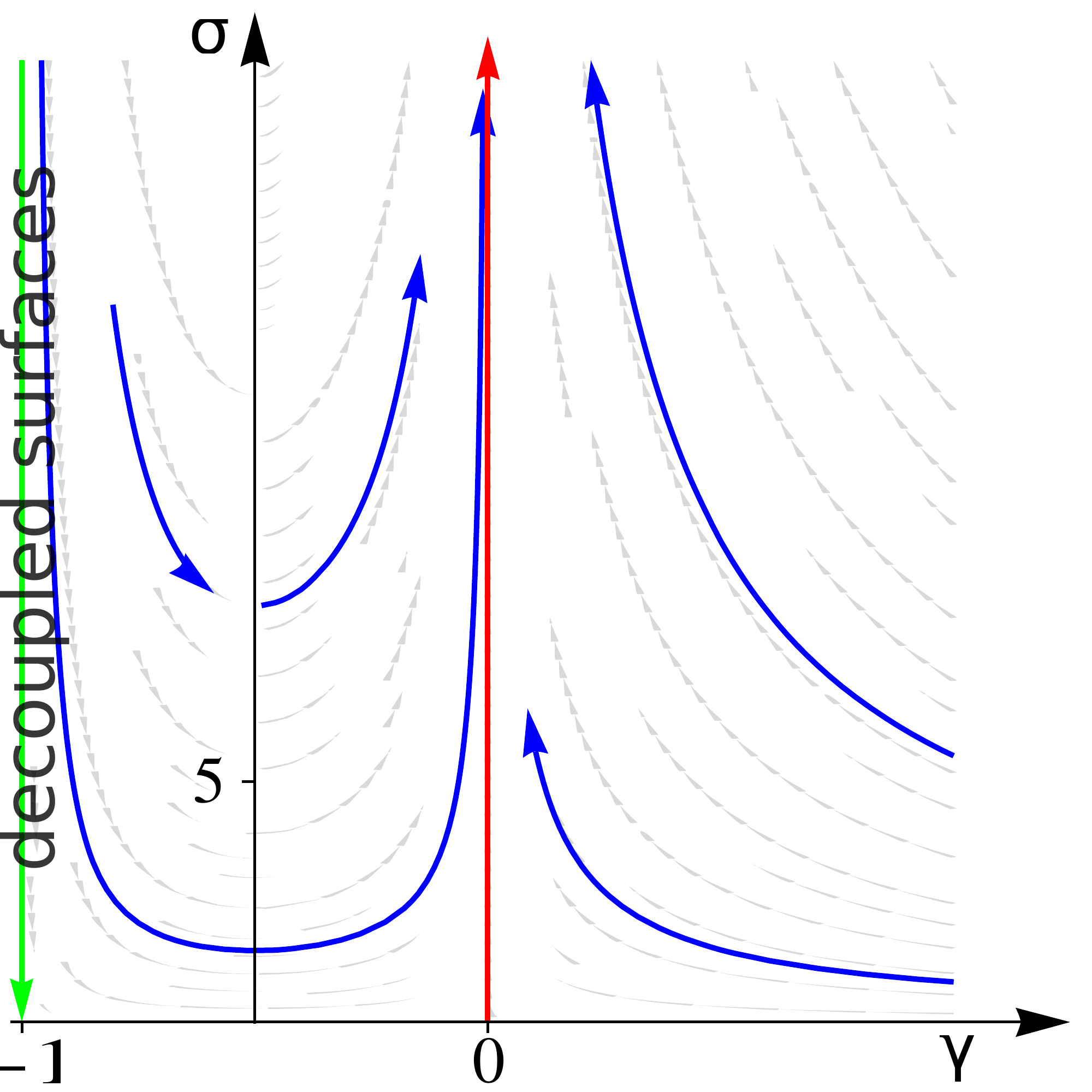}
\caption{RG flow for equal surfaces in the parameter space $\sigma$ (total
conductivity) and $\gamma$ (intra-surface interaction strength). Here and in all
following RG diagrams, arrows indicate the flow towards the infrared.}
\label{fig:RGflowequallayers}
\end{figure}

The RG flow within the $\sigma$--$\gamma$ plane is depicted in
Fig.~\ref{fig:RGflowequallayers}. The green vertical fixed line at $\gamma = -1$
corresponds to the case of two decoupled surfaces (recall $\gamma_{12} = -1
- \gamma$), and reproduces the result of Ref. \onlinecite{OGM2010} for a single
surface of 3D TI. In this limit the total correction to the  conductivity is
negative and obeys the universal law
\begin{equation}
\delta \sigma_{\gamma = -1} = 2 \times \frac{2}{\pi} \left
(\underbrace{1/2}_{\rm WAL} - \underbrace{1}_{\rm AA}\right ) \ln L/l=
-\frac{2}{\pi}
\ln L/l.
\end{equation}
The line of decoupled surfaces is repulsive, as can be seen from Eq.
\eqref{eq:equallayers2}. Flowing towards the infrared, the conductivity first
decreases before turning up again while the system approaches the second fixed
line at $\gamma = 0$. Note that on this line $\gamma_{12} = -1$: the intrasurface
interaction has died out, but the intersurface interaction is maximal. Here the
conductivity correction is positive indicating the flow into a metallic state:
\begin{equation}
\delta\sigma_{\gamma=0} = 2 \times \frac{2}{\pi} \left
(\underbrace{1/2}_{\rm WAL} -
\underbrace{\left [1 - \ln 2 \right  ]}_{\rm interaction}\right )\ln L /l .
\label{eq:equallayersgamma0}
\end{equation}
The flow on this fixed line is towards the perfect-metal point
$$
\left (1/\sigma^*, t^*,
\gamma_{11}^*,\gamma_{22}^* \right )= \left (0,1,0,0\right ),
$$
As discussed below, see Sec. \ref{sec:attractiveplane},
this is the only attractive fixed point even in the case of the general four
dimensional RG flow.
On the $\gamma = 0$ fixed line the intersurface interaction reduces the strength
of the WAL effect but it is not strong enough to reverse the
behavior.
The region $\gamma >0$ corresponds to attractive interaction in the singlet channel and is
shown on the flow diagram for the sake of completeness.

\subsubsection{Typical bare values and crossover scale}

Typically, before renormalization the intersurface interaction $\gamma_{12}$ is
weaker than or equal to the intrasurface interaction $\gamma$. This implies that
its bare value $\gamma_0$ takes values in the range between  $\gamma_0 = -1$
(decoupled surfaces, i.e. $\gamma_{12,0} = 0$) and $\gamma_0=-1/2 =
\gamma_{12,0}$. For small $\alpha$ we can approximate $\gamma_0$ by its RPA
value:
\begin{equation}
\gamma_0 = - \frac{1}{2} - \frac{1}{2} \frac{\kappa d}{1 + \kappa d}.
\label{eq:gammaRPA}
\end{equation}
Here $d$ is the system thickness and $\kappa = 2\pi\frac{e^2 }{\epsilon_2} \nu$
the inverse single surface screening length obtained for the general symmetric
situation: $\epsilon_1 = \epsilon_3 \neq \epsilon_2$, see Appendix
\ref{sec:Elstat}.
Note that at $\kappa d = 0$ the conductivity corrections due to WAL and AA
exactly compensate each other:
$$
\delta \sigma_{\gamma=-1/2} = \frac{2}{\pi} \left (2 \times
\underbrace{1/2}_{\rm WAL} - \underbrace{1}_{\rm AA}\right ) \ln L/l = 0,
$$
as can also be seen in Fig. \ref{fig:RGflowequallayers}.

 Typically $ \kappa d> 0$ or, as already explained on general grounds,
$-1 < \gamma_0 < - 1/2$. Then the most drastic consequence of intersurface
interaction is the non-monotonic temperature (or length) dependence: the
conductivity first decreases with lowering $T$ but eventually the sign of
$d\sigma/dT$ changes and the system is ultimately driven into the
metallic phase. It is natural to ask for the temperature scale, which is
associated with this sign change. The scale $y_*$ at which
the conductivity reaches its minimum can be extracted from
Eqs.~\eqref{eq:equallayers} and is expressed by the integral
\begin{equation}
y_* = - \frac{\pi \sigma_0}{2} \int_{\gamma_0}^{\gamma_*}
\frac{d\gamma'}{\gamma'}\frac{1+\gamma_0}{(1+\gamma')^2} \left
[\frac{\gamma'}{\gamma_0}\right ]^{1-2\ln 2} e^{2\left [f\left (\gamma'\right )
- f\left (\gamma_0\right )\right ]},
\label{eq:ystar}
\end{equation}
where $f\left (x\right ) = \text{Li}_2\left (-x\right ) - \text{Li}_2\left
(-\left (1+2x\right )\right )$, Li$_2$ is  the dilogarithm, and $\gamma_*
= -1/2$.

Numerical integration of \eqref{eq:ystar} yields the crossover length scale or
temperature $y_* = \ln L_*/l = 1/2 \ln T_0/T_*$. Its dependence on the bare
values $\sigma_0$ and $\gamma_0$ is plotted in Fig. \ref{fig:PlotT}. Using Eq.
\eqref{eq:gammaRPA} one can also investigate the dependence of $y_*$ on $\kappa d$
instead of $\gamma_0$ (see inset in Fig.\ref{fig:PlotT}).

\begin{figure}
\includegraphics[scale=.55]{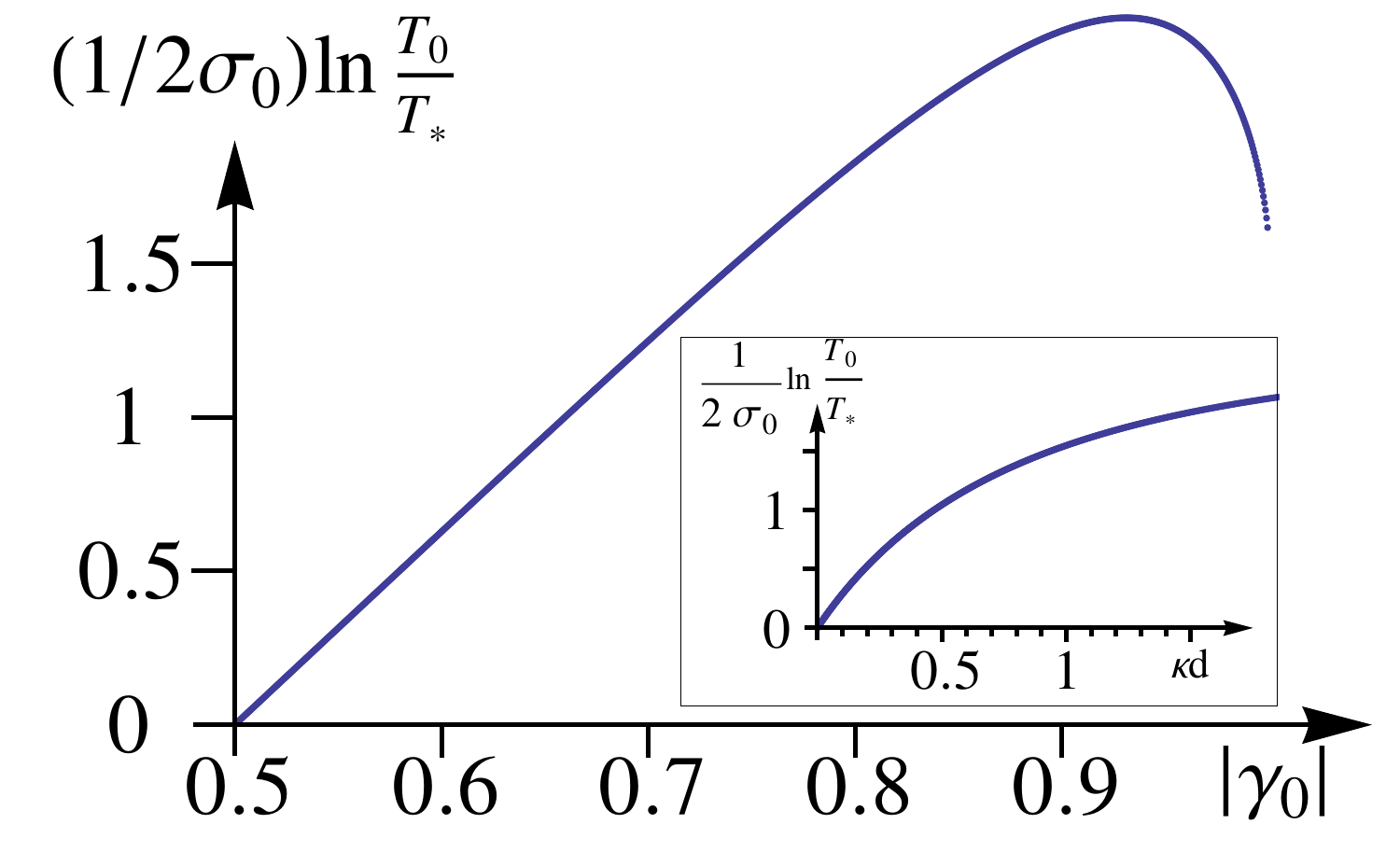}
\caption{Temperature scale associated with the minimum of $\sigma$ as a
function of the bare values $\sigma_0$ and $\gamma_0$. Inset: the same quantity as a
function of $\sigma_0$ and $\kappa d$.}
\label{fig:PlotT}
\end{figure}

\subsubsection{Role of topology: Dirac electrons vs. electrons with quadratic dispersion in the presence of
spin-orbit interaction}

The perturbative RG equations \eqref{eq:RGeqs} and \eqref{eq:equallayers} are
valid for $\sigma \gg 1$. Instanton effects are suppressed by $\exp(-2
\pi\sigma)$ in this region and we therefore neglected them.
{As has been discussed in Sec. \ref{sec:Z2term}, in the diffusive NL$\sigma$M of
Dirac electrons, the $\mathbb Z_2$ theta term reflects the topological
protection from Anderson localization. This term is absent in the case of
non-topological symplectic metals (NTSM) such as electrons with quadratic
dispersion subjected to strong spin-orbit coupling.\footnote{{The $\mathbb Z_2$
theta term is also absent for the critical state separating 2D trivial and
topological insulator. Such a state can be realized, in particular, on a
surface of a weak 3D topological insulator. Despite the absence of theta term
it is protected from Anderson localization due to topological reasons
\cite{RingelKrausStern12, OGM2012,
FuKaneVortices} (see Sec. \ref{sec:Z2term}) and hence do not fall into our
definition of non-topological symplectic metals.}}  The presence (respectively,
absence) of the topological term results in the opposite signs of the instanton
contribution in the two cases.}
However, as instantons are suppressed, our perturbative result is
equally applicable to the surfaces of a 3D TI and, for example, to a
double-quantum-well structure in a material with strong spin-orbit coupling.
Here we discuss non-perturbative differences between the two problems.

Let us start from the case of decoupled surfaces (green line, i.e. $\gamma =
-1$, in Fig.~\ref{fig:Comparison}). This limiting case has been analyzed
before \cite{OGM2010}. For {NTSM} localizing AA corrections overcome the
WAL effect and the system always flows towards localization (Fig.
\ref{fig:Comparison}, left). In contrast, for TI the topological protection implies
$d\sigma/dy > 0$ for small $\sigma$ and hence an attractive fixed point at
$\sigma \sim 1$ (Fig. \ref{fig:Comparison}, right).

As has been explained, the $\gamma=-1$ line is unstable with respect to the
intersurface interaction and the system eventually flows towards the
antilocalizing red line at $\gamma=0$. Let us now analyze this fixed line. The
fact that conductivity corrections \eqref{eq:equallayersgamma0} are
positive stems back to the (non-interacting) WAL effect. Its  contribution $2
\times (1/\pi) \ln L/l$ is independent of $\sigma$ only for $\sigma
\gg 1$. For {NTSM} it decreases with decreasing $\sigma$
and eventually becomes negative at the metal-insulator transition (MIT) point
$\sigma_{\rm MIT} \approx 2 \times 1.42\, e^2/h$. \cite{OhtsukiSlevinKramer2004, MarkosSchweitzer, EversMirlinRMP}
(As explained above, Sec. \ref{sec:Z2term}, in a recent
investigation\cite{FuKaneVortices} the crucial role of $\mathbb Z_2$ vortices
for this MIT was pointed out.)
Qualitatively, the picture of the MIT survives the presence of interactions,
which even enhance the tendency to localization. Therefore, for the double layer
system of {NTSM} we expect the antilocalizing 
RG flow on the
$\gamma=0$ line to turn localizing below some  $\sigma_{\rm MIT}\sim 1$. This MIT
point is indicated by a dot in the left panel of Fig.~\ref{fig:Comparison}.

In contrast, for the surfaces of a topological insulator the system is
topologically protected from Anderson localization, \cite{OGM2007}
i.e., the beta function $d\sigma/dy$ bends up when $\sigma
\rightarrow 0$. There is a numerical evidence \cite{bardarson, Nomura} that
in a non-interacting case this happens without any intermediate fixed points.
Again, the arguments are qualitatively unchanged by the presence of (pure
intersurface) interaction and this scenario is expected to hold also on the red
$\gamma=0$ line of the thin 3D TI film, see Fig.~\ref{fig:Comparison}, right.
(Strictly speaking, one cannot rule out a possibility that in the presence of
interaction there emerge intermediate fixed points but we assume the simplest
possible flow diagram consistent with large- and small-conductivity behavior.)

The interpolation between the limiting cases of decoupled surfaces and
maximally interacting surfaces produces the two phase diagrams shown in Fig.
\ref{fig:Comparison}.
For a {double layer of NTSM}, there is a separatrix connecting the weak-coupling,
decoupled layers fixed point $\left (\gamma, 1/\sigma\right ) = \left
(-1,0\right )$ with the critical MIT point $\left (\gamma, 1/\sigma\right ) \sim
\left (0,1\right )$ that we introduced above. (Strictly speaking, we cannot
exclude the possibility that this fixed point might lie slightly off the
$\gamma =0$ line.) Below the separatrix the conductivity renormalizes down to
$\sigma = 0$, i.e. the system is in the Anderson-localized phase. In
contrast, above the separatrix the characteristic non-monotonic conductivity
behavior leads to the metallic state. As the horizontal position in the phase
diagram is controlled by the parameter $\kappa d$,  we predict a
quantum phase transition between metal and insulator as a function of the
interlayer distance. On the other hand, in the case of the coupled top and bottom surfaces of a thin 3D TI film the flow is always towards the metallic phase. The critical point of decoupled
surfaces at $\gamma=-1$ with $\sigma \sim 1$ is unstable in the direction of $\gamma$.

It is worth recalling that in this paper we neglected the tunneling between the opposite surfaces of the 3D TI. If such a tunneling is included, it introduces a corresponding exponentially small temperature scale below which the two surfaces behave as a single-layer NTSM. This would imply a crossover to localizing behavior at such low temperatures.

\begin{figure}
\begin{minipage}[b]{0.5\linewidth}
\includegraphics[scale=0.2]{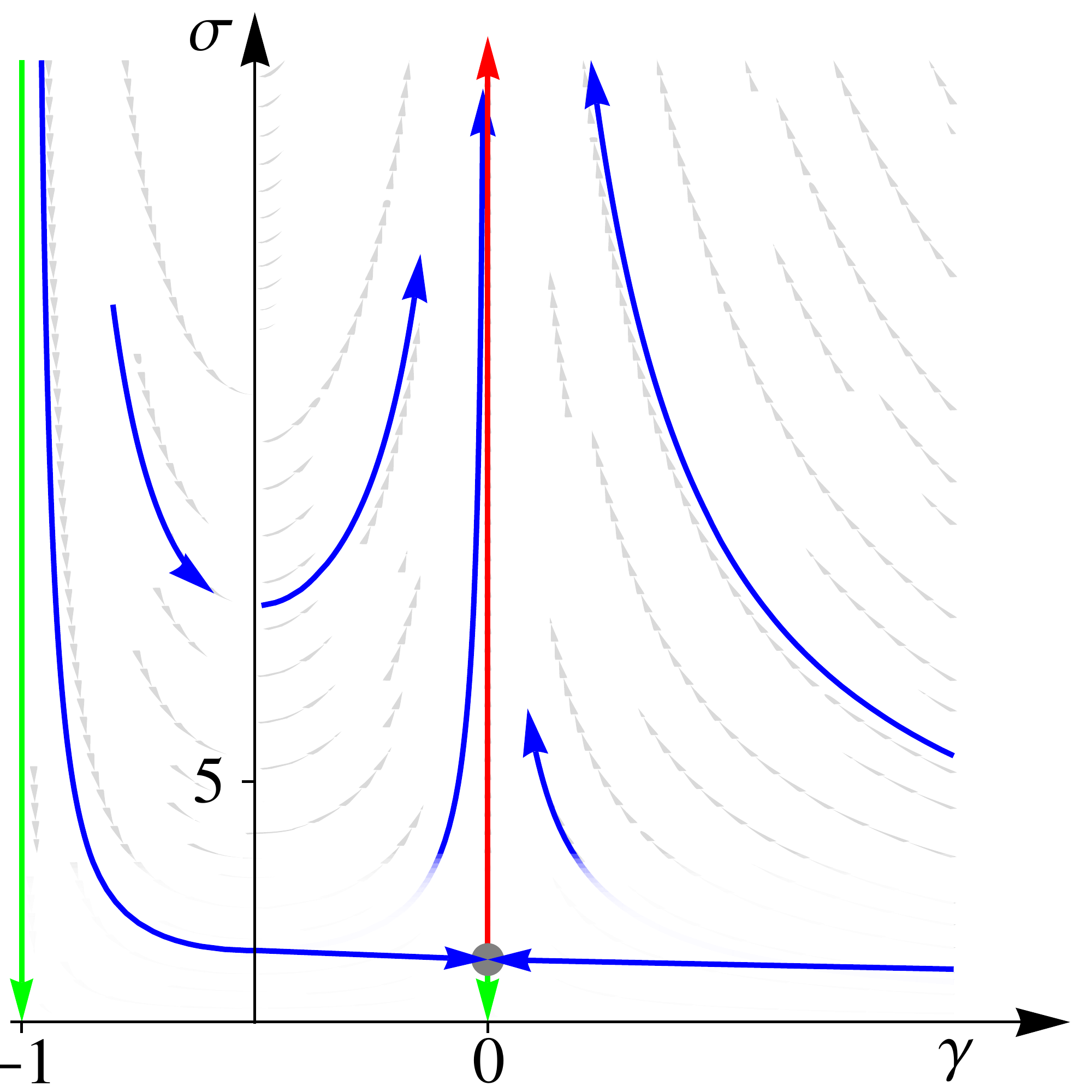}
\end{minipage}
\begin{minipage}[b]{0.5\linewidth}
\includegraphics[scale=0.2]{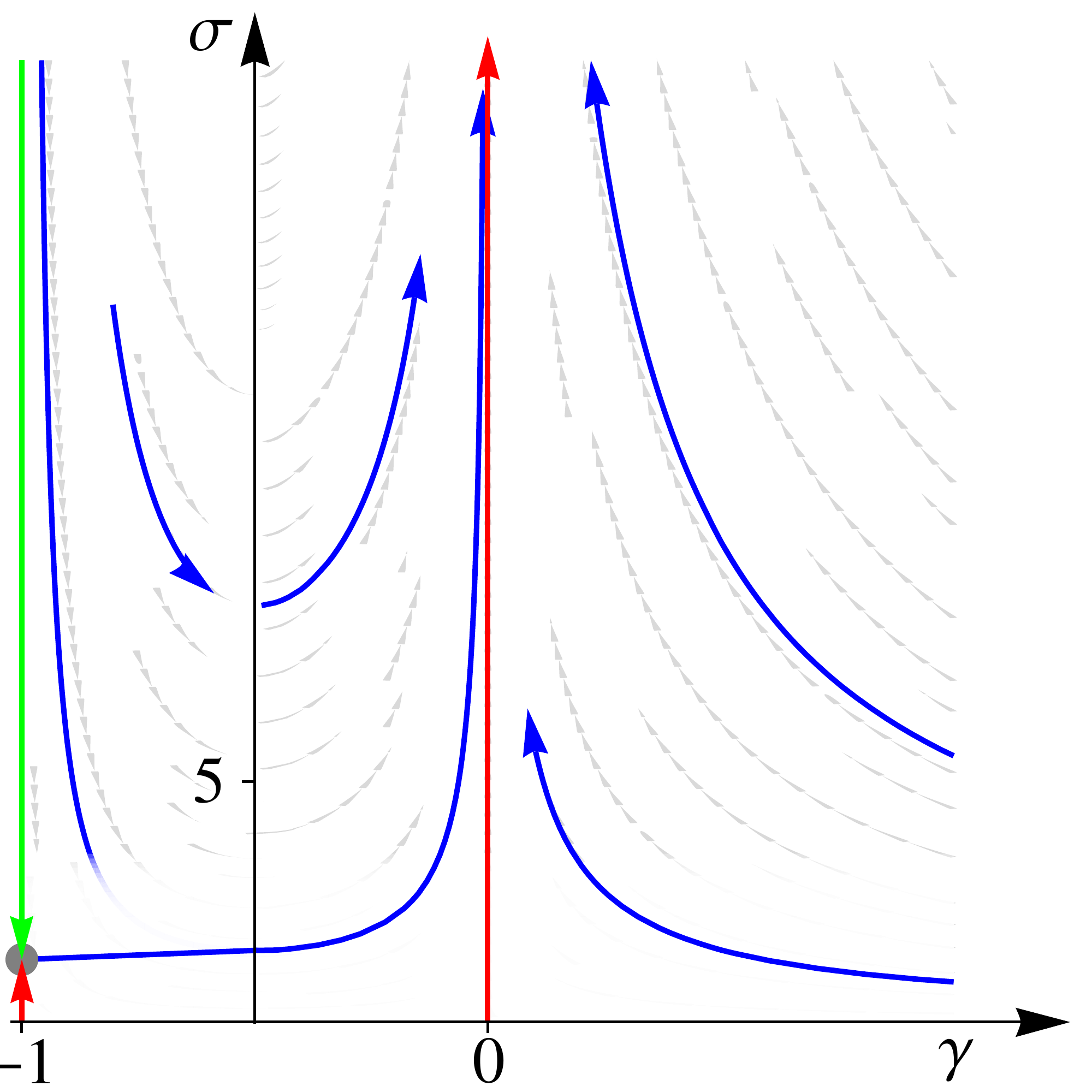}
\end{minipage}
\caption{Comparison between expected RG-flow for a double layer system of NTSM (left) and the coupled surfaces of a thin 3D TI film (right).}
\label{fig:Comparison}
\end{figure}

\subsection{General RG flow}

We now turn our attention to the complete analysis of RG equations
\eqref{eq:RGeqs} which, in general, describe the case of different
carrier density, disorder and interaction strength on the top and bottom surfaces of a 3D TI film.
The renormalization of interaction parameters $\gamma_{11}$ and $\gamma_{22}$,
Eqs.~\eqref{eq:RGeqs:gamma11} and \eqref{eq:RGeqs:gamma22},  determines four
fixed planes of the RG flow:
\begin{itemize}
\item \underline{$\gamma_{11} = -1 = \gamma_{22}$.} Repulsive fixed plane of two
decoupled surfaces with only intrasurface Coulomb interaction.
This problem has been studied in Ref.~\onlinecite{OGM2010}.
\item \underline{$\gamma_{11} = 0$, $\gamma_{22} = -1$ or vice versa}. Fixed
plane describing a 3D TI film with strongly  different surface population. This case
in analyzed in Sec. \ref{sec:surfacebulk} below.
\item \underline{$\gamma_{11}=0=\gamma_{22}$}. Attractive fixed plane.
Intrasurface interaction has died out and only intersurface interaction
survived. This case is analyzed in Sec. \ref{sec:attractiveplane} below.
\end{itemize}

Concerning the repulsive fixed planes, one should keep in mind that the
renormalization of interaction amplitudes is suppressed by the small factor
$1/\sigma$. Therefore even if the conditions on $\gamma_{11}$ and $\gamma_{22}$
are only approximately fulfilled the behavior in the fixed plane dictates the
RG flow in a large temperature/frequency window. RPA-estimates of the bare values of interaction amplitudes can be found in Appendix \ref{sec:BareGamma}.

We also remind the reader that the RG equations describing the model with
finite-range interaction (and thus the whole crossover between the
problem with Coulomb interaction and the non-interacting system)
is discussed in Appendix \ref{sec:shortrange}.

\subsubsection{Strongly different surface population}
\label{sec:surfacebulk}

We investigate here the fixed plane of Eqs.~\eqref{eq:RGeqs} in which
$\gamma_{11} = 0$ and $\gamma_{22} = -1$. (Clearly, the reversed situation
$\gamma_{11}=-1$ and $\gamma_{22}=0$ is completely analogous.) Both fixed planes
are ``saddle-planes'' of the RG flow, i.e., they are attractive in one of the
$\gamma$-directions and repulsive in the other. 

Before analyzing this fixed plane, it is worth explaining why
this limit is of significant interest for gate-controlled transport experiments,
in particular, those on Bi$_2$Se$_3$. As for this material the Fermi energy
is normally located in the bulk conduction band, an electrostatic gate is
conventionally used to tune the chemical potential into the bulk gap and hence
to bring the system into a topologically non-trivial regime. A situation as
depicted in Fig. \ref{fig:setupwithbulk} is then believed
to arise in a certain range of gate voltages: \cite{SteinbergPRB} one of the two
surfaces (here surface 1) is separated by a depletion region from
a relatively thick bulk-surface layer.

Recently, \cite{OGM2012, Glazman, Krueckl} disorder-induced interference
corrections for 3D TI bulk electrons have been investigated
theoretically: \footnote{It is worthwhile to repeat that in the diffusive regime
of typical experiments on thin films the 3D TI bulk electrons are subjected to 2D diffusive motion.} While at small length scales additional symmetries of the
Hamiltonian provide non-trivial localization behavior, at sufficiently large
scales the usual WAL effect sets in. The strong coupling between electron states
in the conducting part of the bulk and at surface 2 does not alter this
universal low-energy property. In
conclusion, at sufficiently large length scales the symplectic class
NL$\sigma$M, Eq.~\eqref{eq:TOTALNLSM}, is the adequate description of such a
system (under the assumption of  negligible tunneling between
surface 1 and the conducting part of the bulk).

Since the bulk-surface layer has a much higher carrier density than the carrier density on the spatially separated surface 1 we can expect that $\kappa_2\gg\kappa_1$. Provided $\kappa_1 d\ll 1$ the
electron-electron interaction on the spatially separated surface 1 is effectively
screened out such that $|\gamma_{11}|\approx(\kappa_1/\kappa_2)(1+2\kappa_2 d)\ll 1$ (see Eq. \eqref{eq:coupledsurf}). Conversely, the
effect of screening by electrons on the surface 1 is negligible
for Coulomb interaction of the bulk states: $1+\gamma_{22}\approx \kappa_1/\kappa_2\ll 1$.

\begin{figure}
\includegraphics[scale=.4]{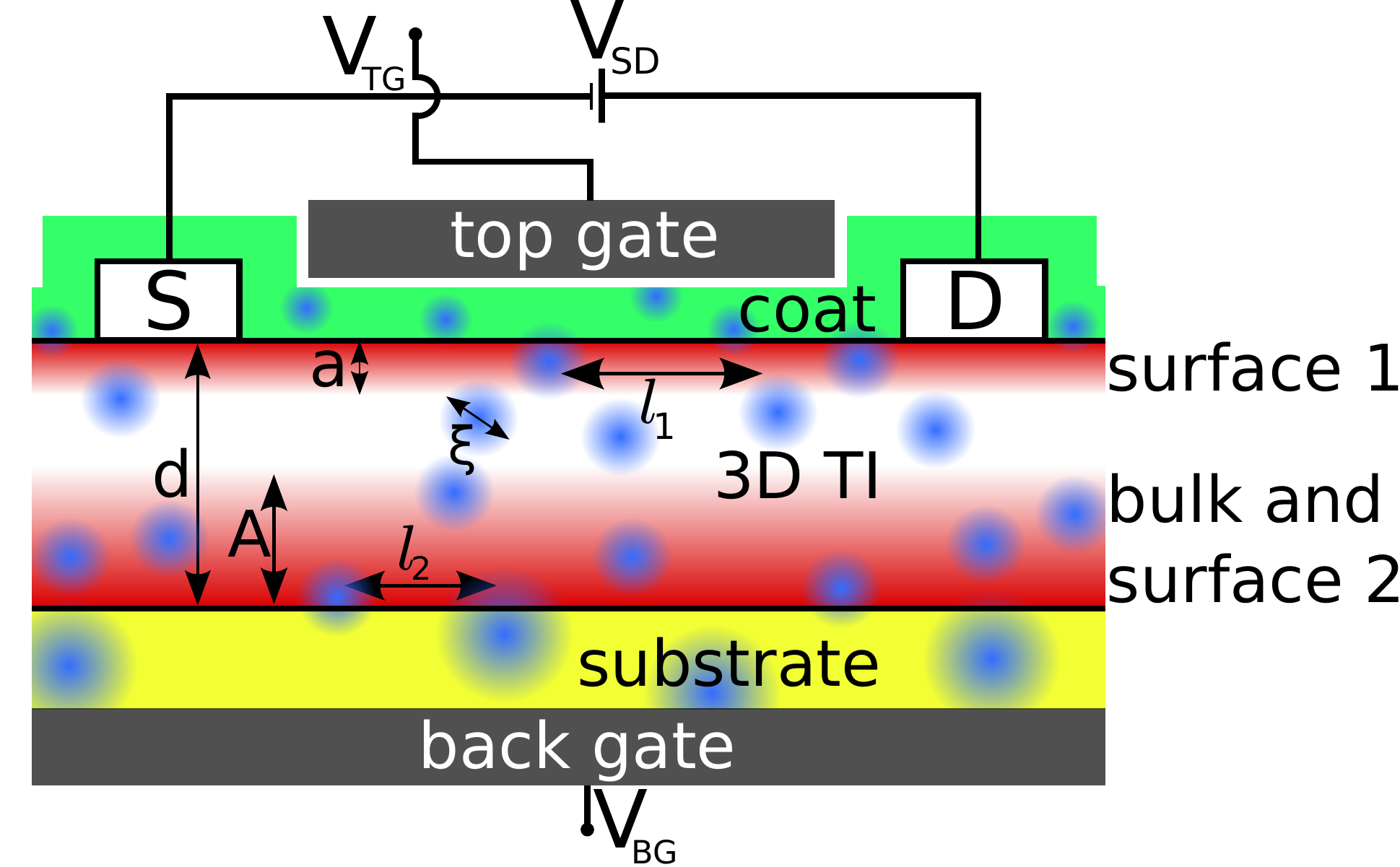}
\caption{Typical scenario for gate-controlled transport experiments: A
topologically protected surface separated from a  thick bulk-surface layer.}
\label{fig:setupwithbulk}
\end{figure}

Substituting $\gamma_{11}=0$ and $\gamma_{22}=-1$ into Eqs. \eqref{eq:RGeqssigmatot}, we find that the RG equations in this fixed plane are as follows
\begin{subequations}
\begin{align}
\T \B \frac{d\sigma}{dy} & = - \frac{2}{\pi} \left \lbrace  1 - \frac{1}{t}
\ln (1+{t}) \right \rbrace,  \label{eq:RGeqsunequal:sigma}\\
\T \B \frac{dt}{dy} & = - \frac{2}{\pi} \frac{1+t}{\sigma}\left \lbrace
\frac{1-t}{2}   - \frac{1}{t}  \ln (1+{t})\right \rbrace .
\label{eq:RGeqsunequal:t}
\end{align}
\label{eq:RGeqsunequal}
\end{subequations}
They can equivalently be written in terms of conductivities $\sigma_1$ and
$\sigma_2$:
\begin{subequations}
\begin{align}
\T \B \frac{d\sigma_1}{dy} & = - \frac{2}{\pi} \left \lbrace  \frac{1}{2} -
\frac{\sigma_2}{\sigma_1} \LN\left  [1+\frac{\sigma_1}{\sigma_2}\right ] \right
\rbrace, \label{eq:RGeqsunequalsigma12:sigma1} \\
\T \B \frac{d\sigma_2}{dy} & = - \frac{1}{\pi}.
\label{eq:RGeqsunequalsigma12:sigma2}
\end{align}
\label{eq:RGeqsunequalsigma12}
\end{subequations}

We emphasize that the limit $\gamma_{11} = 0$ and $\gamma_{22} = -1$ is very peculiar. Indeed, due to the relation
\eqref{eq:Finv}, this limit implies that the condition $z_1/z_2=0$ holds. Equations \eqref{eq:RGeqsunequal} and \eqref{eq:RGeqsunequalsigma12}
 are written under assumption that the ratio $t = \sigma_1/\sigma_2$ is finite in spite of the fact that $z_1/z_2=0$. In the experiment it corresponds to the case in which $\kappa_1/\kappa_2\ll 1$ but the ratio $D_1/D_2 \gg 1$ where $D_s=\sigma_s/4z_s$ is the diffusion coefficient.

Equations \eqref{eq:RGeqsunequalsigma12} become decoupled for $\sigma_1/\sigma_2 =
0$. Then, as expected, $\delta \sigma_1 =
\frac{1}{\pi}\ln L /l$ (WAL, no interaction on the surface 1) and $\delta \sigma_2 =
-\frac{1}{\pi} \ln L /l$ (WAL and AA due to Coulomb interaction on the surface 2).
However, the line $t=0$ is unstable. As one can see from Eq. \eqref{eq:RGeqsunequal:t}, due to the very same quantum
corrections the initially small parameter $t= \sigma_1 / \sigma_2$
increases under RG. The ultimate limit of the perturbative RG flow is $\sigma
\rightarrow 0$ and $t \rightarrow \infty$, see Fig.~\ref{fig:unnequallimit}.
The scale dependence of $\sigma_1$ is non-monotonous; the position of the
corresponding maximum is determined by zeros of the right-hand-side of
Eq.~\eqref{eq:RGeqsunequalsigma12:sigma1} shown by a green line in the right
panel of Fig.~\ref{fig:unnequallimit}.

As has been already emphasized, the perturbative RG equations are
sufficient only
in the regime of large $\sigma_{s}$. We now discuss the topological effects at
small values of conductivities. In the limit $\gamma_{11} = 0$, $\gamma_{22} = -1$ the
renormalization of $\sigma_2$ is {\it exactly} independent of the surface 1.
Indeed, {in the conductivity corrections,} the two surfaces influence each other only via mutual
RPA screening. In the NL$\sigma$M description the interaction amplitudes in the
full action \eqref{eq:TOTALNLSM} and hence in the propagators \eqref{eq:AApropagator} (diffusons and cooperons) fully account for this effect. 
Since the layer 2 includes a
single TI surface, we know that
$\sigma_2$ is topologically protected and flows towards $\sigma_2^*$ of the
order of the quantum of conductance (``interaction-induced
criticality'' \cite{OGM2010}). Before this happens, the flow of $\sigma_1$
becomes reversed from antilocalizing to localizing, see
Eq.~\eqref{eq:RGeqsunequalsigma12:sigma1}.
However, since the surface 1 is
also topologically protected, its states can not be strongly localized and
$\sigma \rightarrow \sigma_1^*>0$.\footnote{A priori $\sigma_1^*$
and $\sigma_2^*$ are different although we cannot exclude a possibility that
they might be equal.} Thus, both surfaces are at the quantum critical points with
conductivities of order $e^2/h$. The conclusion concerning the surface 1 is
particularly remarkable: even though $\gamma_{11}=0$, there is
``intersurface-interaction-induced criticality'' on the surface 1.

\begin{figure}
\begin{minipage}[b]{0.45\linewidth}
\includegraphics[scale=0.42]{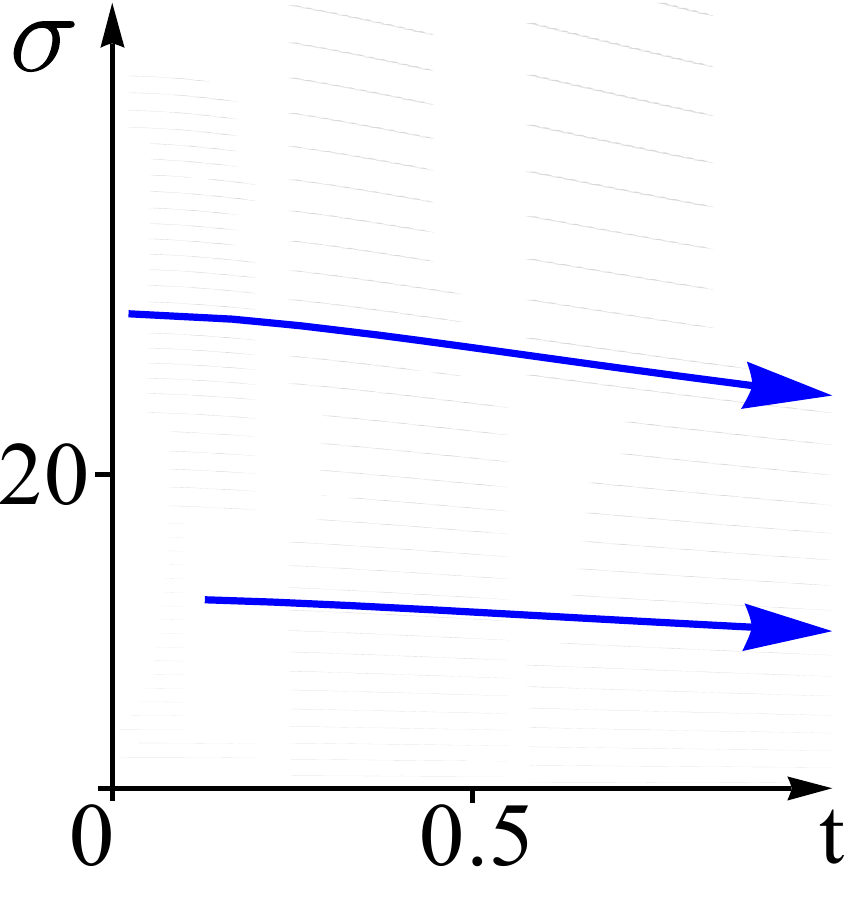}
\end{minipage}
\begin{minipage}[b]{0.45\linewidth}
\includegraphics[scale=0.49]{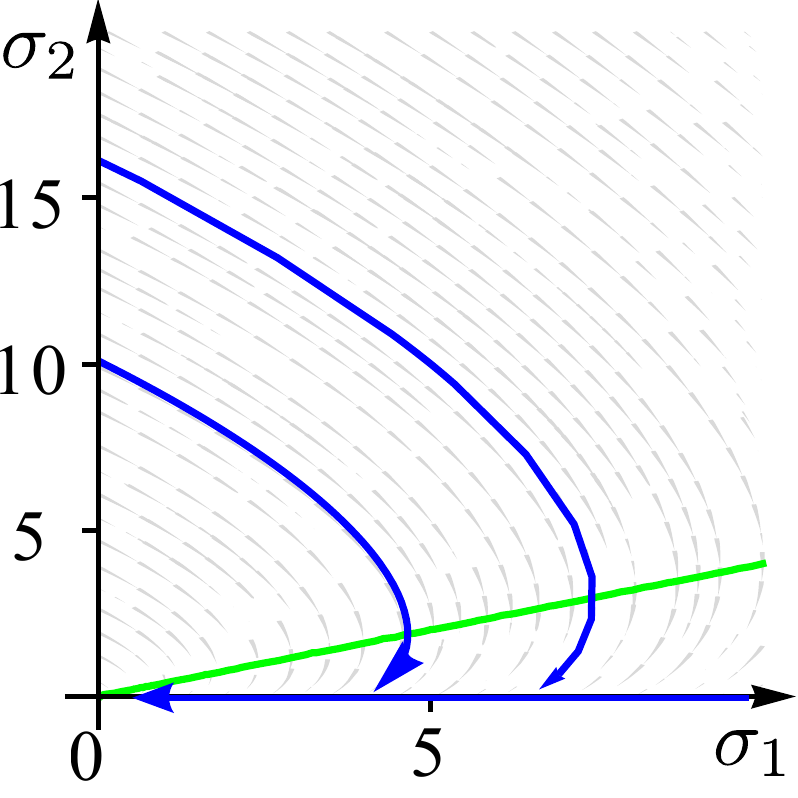}
\end{minipage}
\caption{Perturbative RG flow in the fixed plane $\gamma_{11} = 0$, $\gamma_{22}
= -1$. In the experimentally motivated scenario (Fig. \protect \ref{fig:setupwithbulk}),
the flow starts at $ t= \sigma_{1}/\sigma_{2} \ll 1$. The green line in the
right panel is a line of zeros of the right-hand-side of
Eq.~\eqref{eq:RGeqsunequalsigma12:sigma1}; it determines the maximum
in the RG flow of $\sigma_1$. }
\label{fig:unnequallimit}
\end{figure}

\subsubsection{Attractive fixed plane}
\label{sec:attractiveplane}

According to Eqs.~\eqref{eq:RGeqs:gamma11} and \eqref{eq:RGeqs:gamma22}, any
$\gamma_{ss} \notin \left \lbrace0,-1\right \rbrace$ is renormalized to zero.
The $\gamma_{11}=\gamma_{22}=0$ is thus an attractive fixed plane of the
general RG flow. The flow within this plane has the form
determined by the following RG equations
\begin{subequations}
\begin{align}
\T \B \frac{d\sigma}{dy} & = - \frac{2}{\pi} \left \lbrace  1 - \frac{1}{t}
\LN\left [1+{t}\right ]  - {t} \LN\left  [1+\frac{1}{t}\right ] \right \rbrace,
\\
\T \B \frac{dt}{dy} & = - \frac{2}{\pi} \frac{1+t}{\sigma}\left \lbrace
\frac{1-t}{2}   - \frac{1}{t}   \LN\left  [1+{t}\right ] + {t^2}  \LN\left
[1+\frac{1}{t}\right ] \right \rbrace ,
\label{eq:RGeqs:attractiveplane:t}
\end{align}
\label{eq:RGeqs:attractiveplane}
\end{subequations}
or, equivalently,
\begin{subequations}
\begin{align}
\T \B \frac{d\sigma_1}{dy} & =  - \frac{2}{\pi} \left \lbrace \frac{1}{2}  -
\frac{\sigma_2 }{\sigma_1} \LN\left  [1+\frac{\sigma_1}{\sigma_2}\right ] \right
\rbrace ,\\
\T \B \frac{d\sigma_2}{dy} & =  - \frac{2}{\pi} \left \lbrace \frac{1}{2}  -
\frac{\sigma_1}{\sigma_2 } \LN\left  [1+\frac{\sigma_2}{\sigma_1}\right ]\right
\rbrace .
\end{align}
\label{eq:RGeqs:attractiveplanesigma12}
\end{subequations}

\begin{figure}
\begin{minipage}[b]{0.5\linewidth}
\includegraphics[scale=0.44]{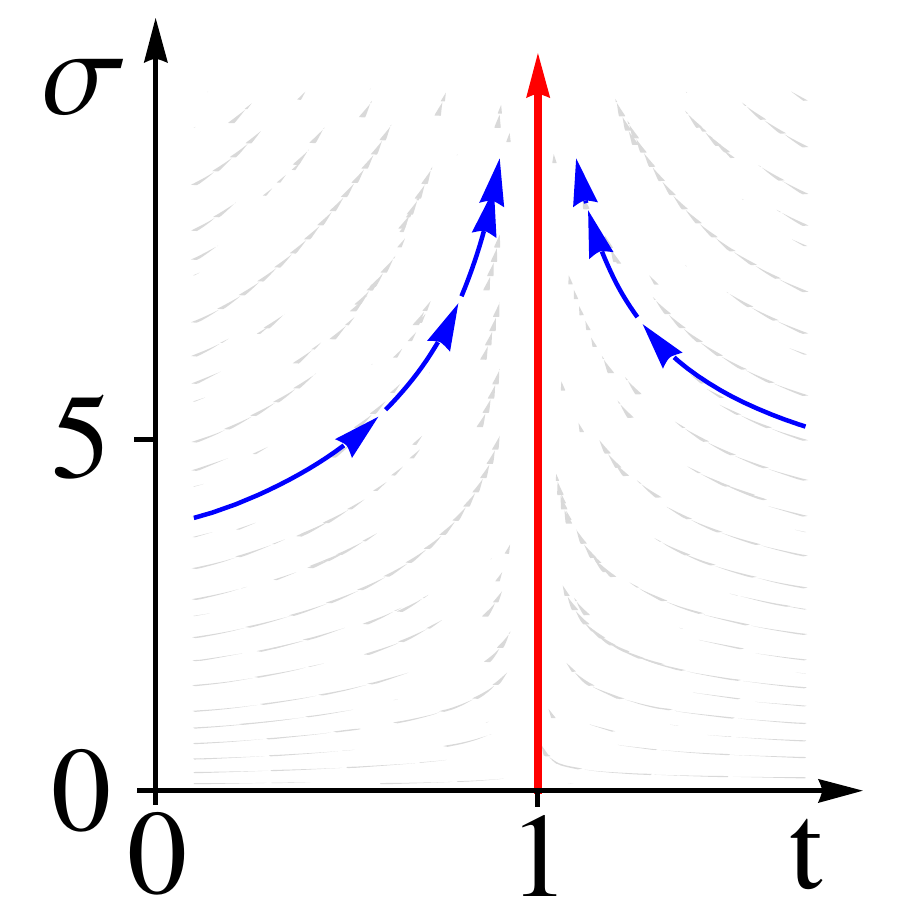}
\vspace*{.2cm}
\end{minipage}
\begin{minipage}[b]{0.5\linewidth}
\includegraphics[scale=0.46]{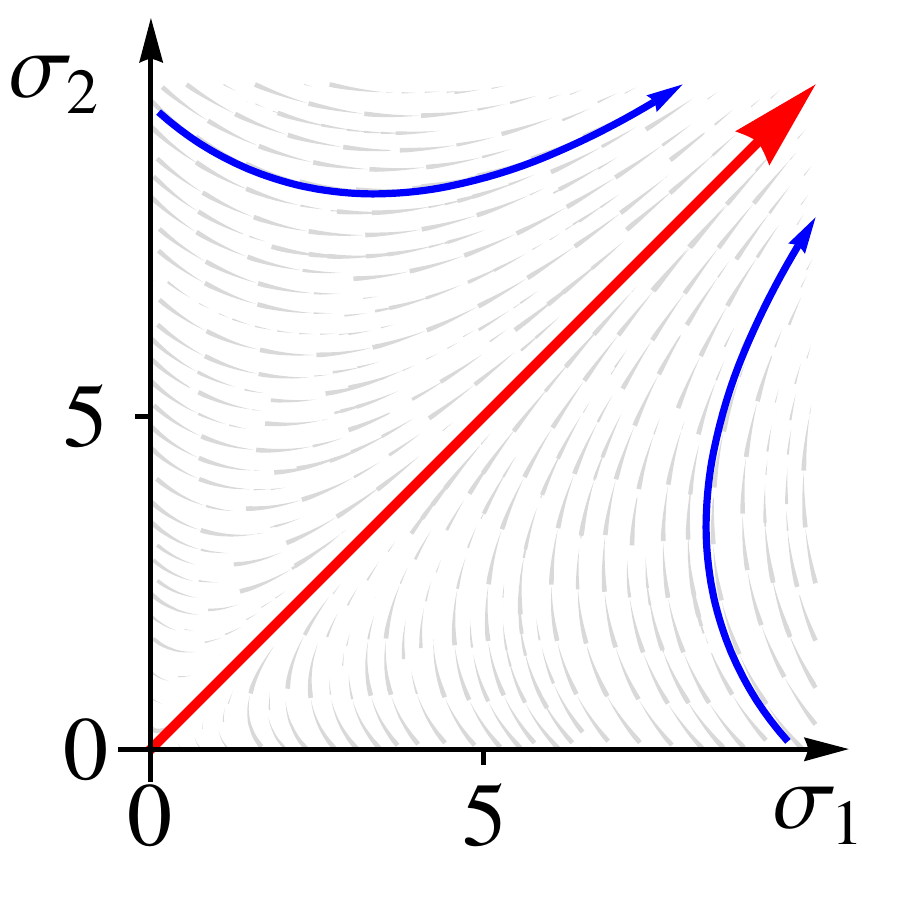}
\end{minipage}
\caption{The RG flow in the attractive fixed plane $\gamma_{11} = 0 =
\gamma_{22}$. The zero of Eq. \eqref{eq:RGeqs:attractiveplane:t} is displayed by
the red line.}
\label{fig:attractiveplane}
\end{figure}

Even though the single-surface conductivities $\sigma_s$ display
non-monotonic behavior within this plane, eventually all quantum corrections are
antilocalizing, see Fig.~\ref{fig:attractiveplane}. The ratio of conductivities
flows to the symmetric situation $t = \frac{\sigma_1}{\sigma_2} =1$, as has been
discussed in Sec.~\ref{sec:twoequalsurfaces}. We reiterate that at the
corresponding fixed line the WAL
effect is competing with a contribution of the opposite sign due to
intersurface interaction. While the WAL wins, the antilocalizing flow is slower
than for free electrons, see Eq. \eqref{eq:equallayersgamma0}.

\subsubsection{General RG flow}

After having analyzed the RG flow in various fixed planes,
we briefly discuss the general {RG} flow. According to
Eqs.~\eqref{eq:RGgamma11} and \eqref{eq:RGgamma22}, there is a single
attractive fixed point of the overall RG flow -- the metallic fixed point with
zero intrasurface interaction, $\sigma_1 = \sigma_2 \to \infty$ and $\gamma_{11}
= \gamma_{22} = 0$. On the other hand, for the values of $\gamma_{ss}$ close to $-1$ the corresponding conductivity $\sigma_s$ is first subjected to localizing quantum
corrections and will thus show a non-monotonic behavior towards
antilocalization. There also exists a range of initial parameters for the RG flow for which
the conductivity at one surface demonstrates monotonous antilocalizing behavior, while the conductivity in the other surface flows in the described non-monotonous manner.

\section{Discussion and experimental predictions}
\label{sec:experiment}

In the preceding Section we have performed a general analysis of the
renormalization {group} flow determined by the RG equations  \eqref{eq:RGeqs}. The
purpose of the present Section is to apply these results to specific
experimentally relevant materials.

\subsection{Parameters}

As explained in Sec.~\ref{sec:setup}, the RG equations \eqref{eq:RGeqs} apply
in the case of the following hierarchy of length scales:
\begin{subequations}
\begin{align}
 l &\ll  L_E,  \label{eq:diffusiveregimecond}\\
 d &\ll  l.  \label{eq:intersurfaceimportant}
\end{align}
In order to deal with $q$-independent interaction amplitudes, an additional
requirement occurs in the case $\kappa_s d \ll 1$ for both $s =1$ and $s=2$: 
\begin{equation}
 l_{\rm scr} \ll L_E.  \label{eq:AAexists}
\end{equation}
In view of condition \eqref{eq:diffusiveregimecond}, the constraint
\eqref{eq:AAexists} is fulfilled in the entire diffusive regime if  $l_{\rm scr}
\ll l$. 

Further, we have assumed that the intersurface tunneling is negligible; the
corresponding condition reads
\begin{align}
 a &\ll d. \label{eq:notunneling}
\end{align}
\end{subequations}
In this Section, we will concentrate on the case when the RG scale is set by
temperature, $L_E = l_T$.
We recall the definition of the length scales entering the above conditions: $l
= \max_{s=1,2} l_s$ is the larger mean free path,
$l_T = \min_{s=1,2} \sqrt{D_s/kT}$ the smaller thermal length, $d$ the sample
thickness, $a$ the penetration depth, $\kappa_s$ the inverse Thomas-Fermi screening
length for the surface $s$ and $l_{\rm scr}$ the total screening length for the
3D TI film.
The situation in which only one of the two surfaces is in the diffusive regime,
while the other one is in the ballistic regime (i.e. $T\tau_1 \ll 1$ and
$T\tau_2 \gg 2$ or vice versa) is also a conceivable and interesting
scenario. However, we do not address it in the present paper. 

The effect of intersurface interaction becomes prominent if the
sample thickness does not exceed too much at least one of the single surface
screening lengths $\kappa_s^{-1}$. As discussed above (Sec.
\ref{sec:AnalysisofRG}), this condition implies that the bare values of
interaction $\gamma_{11}$ and $\gamma_{22}$ are not too close to $-1$.

It is useful to present expressions for the length scales appearing in the
conditions \eqref{eq:diffusiveregimecond}-\eqref{eq:notunneling} in terms of
standard parameters characterizing samples in an experiment.
For simplicity, we assume $v_F^{(1)}=v_F^{(2)}$ and $\tau_1 = \tau_2 $ in these
formulas.

The densities of states (DOS) and inverse screening lengths for the top and
bottom surfaces are
\begin{equation}
\nu_s = \sqrt{\frac{n_s}{\pi v_F^2}},\qquad \kappa_s \equiv \frac{2\pi  e^2}{\epsilon_2} \nu_s = 2\pi
\alpha\sqrt{\frac{n_s}{\pi}} \label{eq:kappas1}
\end{equation}
where $n_s$ are the corresponding electron densities.
If the electron densities for each surface separately are not known, the total
density  $n_{\text{tot}} = n_1 + n_2$  can be used to estimate the {DOS} and the
screening lengths: 
\begin{equation}
\nu_1^2 + \nu_2^2 = \frac{n_{\text{tot}}}{\pi v_F^2},\qquad \kappa_1^2 +
\kappa_2^2 = \left (2\pi\alpha\right )^2 \frac{n_{\text{tot}}}{\pi}. 
\label{eq:kappas2}
\end{equation}

The mean free path can be expressed as
\begin{equation}
l = v_F \tau_{tr} = \frac{\sigma}{\pi v_F \left (\nu_1 + \nu_2\right )}.
\label{eq:MFpath}
\end{equation}
The thermal length in the diffusive regime is given by
\begin{equation}
l_T = \sqrt{\frac{D}{kT}} =  \sqrt{\frac{v_F l}{2 k T}}
= \sqrt{\frac{\sigma}{kT 2\pi \left ( \nu_1 + \nu_2\right ) }}
\end{equation}
Hence, the condition \eqref{eq:diffusiveregimecond} is fulfilled for
temperatures
\begin{equation}
kT \ll kT_{\text{Diff}},
\label{eq:Tmax}
\end{equation}
where
\begin{equation}
kT_{\text{Diff}}
= \frac{v_F}{2l}
= \frac{1}{\sigma  [e^2/h ]} \left (
\frac{v_F^2}{2}\right )2\pi (\nu_1 + \nu_2) 
\end{equation}
is the temperature scale at which the diffusion sets in.

In order to obtain $l_{\rm scr}$ entering Eq.~(\ref{eq:AAexists}), we have to
consider the full (inter- and
intrasurface) Coulomb interaction, see Appendix \ref{sec:Elstat}.
As explained in Sec. ~\ref{sec:IAimportant} it is only a meaningful quantity
provided $\kappa_s d \ll 1$. Taking into account the influence of the
surrounding dielectrics, we find 
\begin{equation}
l_{\rm scr} = \frac{\epsilon_1 + \epsilon_3}{2\epsilon_2} \frac{1}{\kappa_1 + \kappa_2}.
\label{eq:screeninglengthexp}
\end{equation}
When deriving Eq.~\eqref{eq:screeninglengthexp}, we assumed for simplicity that
$\epsilon_2 \lesssim \epsilon_1 + \epsilon_3$. Regarding the experimental setups discussed in Sec. \ref{sec:2materials}, this condition is well fulfilled for 
Bi$_2$Se$_3$ but only marginally for HgTe. Thus in the latter case
Eq.~\eqref{eq:screeninglengthexp} should be considered as a rough estimate. 

Finally, to check the validity of the condition \eqref{eq:notunneling}, one
needs to know the value of the penetration depth $a$. The latter can be
estimated from the condition
\begin{equation}
\frac{v_{F,\perp} p_\perp}{\Delta_{\text{bulk}}} \sim 1 ,
\end{equation}
where $p_\perp \sim 1/a$ denotes typical momenta perpendicular to the surface.
Provided $v_{F,\perp} \sim v_{F}$, it yields
\begin{equation}
a \sim
\frac{v_{F}}{\Delta_{\text{bulk}}}.
\label{eq:a_estimate}
\end{equation}

We are now going to consider two exemplary materials for 3D TIs: Bi$_2$Se$_3$
and strained HgTe. We shall estimate numerically all the
relevant parameters and present characteristic plots for temperature
dependence of conductivities.

\subsection{Exemplary 3D TI materials}
\label{sec:2materials}

\subsubsection{Bi$_2$Se$_3$}

%
\begin{table}
\begin{tabular}{|l|c|}
\hline Fermi velocity & $v_F \sim 5 \times 10^{5} $m/s \\
\hline Bulk gap  & $\Delta_{\text{bulk}} \sim 0.3$ eV \\
\hline Sample thickness   & $d \sim$ 10 nm \\
\hline Dielectric properties & \begin{tabular}{r l} Coat: & $\epsilon_1 \sim1$ \\
 3D TI (Bi$_2$Se$_3$): & $\epsilon_2 \sim 100$ \\
 Substrate (SrTiO$_3$): & $\epsilon_3 \sim10^3 - 10^4$ \end{tabular} \\
\hline Carrier density  & $n_{\text{tot}} \sim 3 \times 10^{12}$ cm$^{-2}$ \\
\hline Mobility  & $\mu_{el}\sim100 \dots 1000$ cm$^{2}/$V$\cdot$s \\
\hline Sheet resistance & $1/\sigma \sim 0.097 \, h/e^2$ at $T\sim 50$ mK \\
\hline \hline Effective coupling  &$\alpha \sim 4 \times 10^{-2}$\\
\hline Chemical potential & $\mu_1^2 + \mu_2^2 = \left (0.2\, \text{eV}\right )^2$ \\
\hline Penetration depth  &$a \sim 1$ nm \\
\hline Mean free path  &$l \sim  24 \dots 34$ nm\\
\hline Diff. temperature  & $T_\text{Diff} \sim 80 \dots 57$ K\\
\hline Screening length & $\kappa_1^2 + \kappa_2^2 \sim  \left (37 \text{ nm} \right )^{-2}$\\
\hline Scr. length (total) & $l_{\rm scr} \sim 132 \dots 186 \text{ nm}, \; \text{for} \; \epsilon_3 = 10^3$\\
\hline \hline Bare interaction (RPA) & \begin{tabular}{r l} top surface: & $\gamma_{11} \sim -0.6 \dots -1 $  \\
 bottom surface: &  $ \gamma_{22} \sim - 0.6 \dots 0$ \end{tabular}\\
\hline
\end{tabular}
\caption{Experimental values of sample parameters at the point 
of the minimal carrier density and associated length scales for transport
experiments on Bi$_2$Se$_3$ films of Refs. [\onlinecite{Chen,YQLi}]. The dots ``$\dots$'' separate values for the
symmetric ($n_1 = n_2$) and totally asymmetric ($n_1 = n_{\text{tot}}, n_2 = 0$)
cases. The bare interaction amplitudes are estimated in the random  phase
approximation (RPA).}
\label{tab:expvaluesBise}
\end{table}

Bi$_2$Se$_3$ is currently the most conventional material for experimental
realization of  the 3D TI phase. Typical experimental data (extracted from the
point of the minimal conductance in Refs. \onlinecite{Chen,YQLi}) is summarized
in
the upper part of Table \ref{tab:expvaluesBise}. Using Eqs.~\eqref{eq:kappas1}
-- \eqref{eq:a_estimate} we can estimate the hierarchy of length scales (lower
part of the same Table). One can see that all of the requirements of validity
for our theory are fulfilled for length scales above $l_{\rm
scr}$ [temperatures below $T_{\max} = 2.6 \dots 1.9 $K, see condition
\eqref{eq:AAexists}.] \footnote{The only assumption that can not be
directly verified on the basis of the quoted experimental data is the absence of
complete intersurface correlations of disorder. We remind the reader that in
the case of $\mu_1 = \mu_2$ a completely correlated disorder implies extra soft
modes \cite{BurmistrovGornyiTikhonov}. We see however no reason for such perfect
correlations between impurities at opposite surfaces of a {3D TI film}. } 

From the experimental data, the ratio of carrier densities is not known.
Therefore, we show in Fig. \ref{fig:BiSe} the expected temperature dependence of
total conductivity for various values of this ratio. Clearly, the behavior
strongly differs from the case of decoupled surfaces (dashed line). First,
the slope of $d\sigma/d\ln T$ is considerably smaller. Second, one observes a
clear curvature of the dependence $\sigma(\ln T)$ which is a manifestation of
the non-monotonicity. (For the parameters used in the plot the minimum of
$\sigma$ occurs at still lower temperatures.) This curvature is especially
pronounced for strongly different surfaces.

\begin{figure}
\includegraphics[scale=0.7]{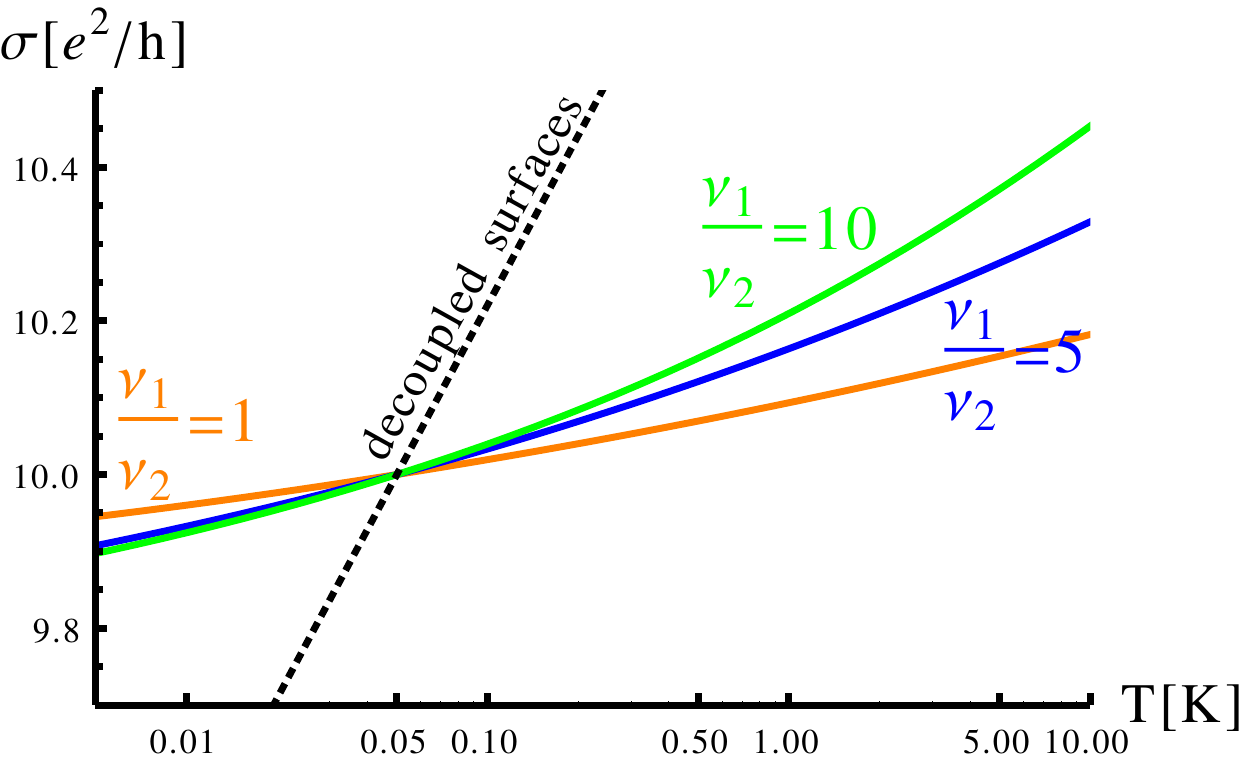}
\caption{Theoretical prediction for the temperature dependence of the total
conductivity in thin Bi$_2$Se$_3$ films.}
\label{fig:BiSe}
\end{figure}

It should be mentioned that the substrate used in Ref. \onlinecite{Chen} has a
strongly temperature-dependent dielectric function $\epsilon_3$ since SrTiO$_3$
approaches a ferroelectric transition at low temperatures. This could 
result in a temperature dependence of effective gate voltage and consequently of
carrier density. The resulting classical temperature dependence of
conductivity (and interaction constants) would mask the  quantum effects
described in our analysis. However, in the presence of the gating field, the
temperature dependence of $\epsilon_3$ saturates at low temperatures. This
motivates the presentation in Fig. \ref{fig:BiSe} where we assumed independent of temperature
$\epsilon_3 = 1000$.

\subsubsection{Strained HgTe}

\begin{table}
\begin{tabular}{|l|c|}
\hline Fermi velocity & $v_F \sim 5 \times 10^{5} $m/s \\
\hline Bulk gap & $\Delta_{\text{bulk}} \sim 0.022$ eV \\
\hline Sample thickness  & $d \sim$ 70 nm \\
\hline Dielectric properties & \begin{tabular}{r l} Coat:& $\epsilon_1 \sim1$ \\
  3D TI (HgTe): & $\epsilon_2 \sim 21$ \\
  Substrate (CdTe): & $\epsilon_3 \sim 10 $ \end{tabular}\\
\hline Carrier density & \begin{tabular}{r l} top surface: & $n\sim 4.8 \times
10^{11}$ cm$^{-2}$ \\
  bottom surface: & $n\sim 3.7 \times 10^{11}$ cm$^{-2}$  \end{tabular}\\
\hline Mobility & $\mu_{el}\sim 34 000$ cm$^2$/V$\cdot$s \\
\hline Sheet resistance & $1/\sigma \sim 0.04\, h/e^2$ at $T=50$ mK \\
\hline \hline Effective coupling & $\alpha \sim 0.21$\\
\hline Chemical potential & \begin{tabular}{r l} top surface: & $\mu_1 \sim 0.08$ eV \\
  bottom surface: & $\mu_2\sim 0.07$ eV \end{tabular}\\
\hline Penetration depth & $a \sim 15$ nm\\
\hline Mean free path & $l \sim 108$ nm \\
\hline Diff. temperature & $T_\text{Diff} \sim 18$ K\\
\hline Screening length & \begin{tabular}{r l} top surface: & $\kappa_1^{-1} \sim 19.53 $ nm \\
 bottom surface: &  $ \kappa_2^{-1} \sim 22.24$ nm \end{tabular}\\
\hline \hline Bare interaction (RPA) & \begin{tabular}{r l} top surface: &
$\gamma_{11} \sim -0.893 $  \\
 bottom surface: &  $ \gamma_{22} \sim - 0.878$ \end{tabular}\\
\hline
\end{tabular}
\caption{Typical experimental values for transport experiments on HgTe  films of Refs. [\onlinecite{BaarsSorger1972,
BruehneHgTe}].}
\label{tab:expvaluesHgTe}
\end{table}

Another very promising 3D TI material is strained HgTe. The presence
of Dirac-like surface states was experimentally confirmed by  the odd
series of QHE plateaus, as well as by ARPES \cite{BruehneHgTe}. While the
transport experiment indicates dominant surface conduction, the extracted
carrier density appears to be too large for a pure surface theory with linear
spectrum, yielding the value of the chemical potential $\mu$ larger than the gap
{$\Delta_{\rm bulk}$}, see Table \ref{tab:expvaluesHgTe}. (The role of the bulk
conduction band as well as the parabolic bending of the dispersion was also
discussed within an independent magneto-optical study by the same
experimental group. \cite{Hancock}) Thus, it remains to be clarified under what experimental
conditions the strained HgTe sample is in the true TI regime (i.e., the
bulk contribution to transport is negligible). Notwithstanding this point and
motivated by the excellent surface transport data, we apply our theory to the
HgTe experiment, see Fig. \ref{fig:HgTe}. In spite of the considerable thickness
of the probe, the effect of intersurface interaction is clearly visible: the
slope of $d\sigma/d\ln T$ is considerably smaller than it is expected for decoupled
surfaces.

\begin{figure}
\includegraphics[scale=0.8]{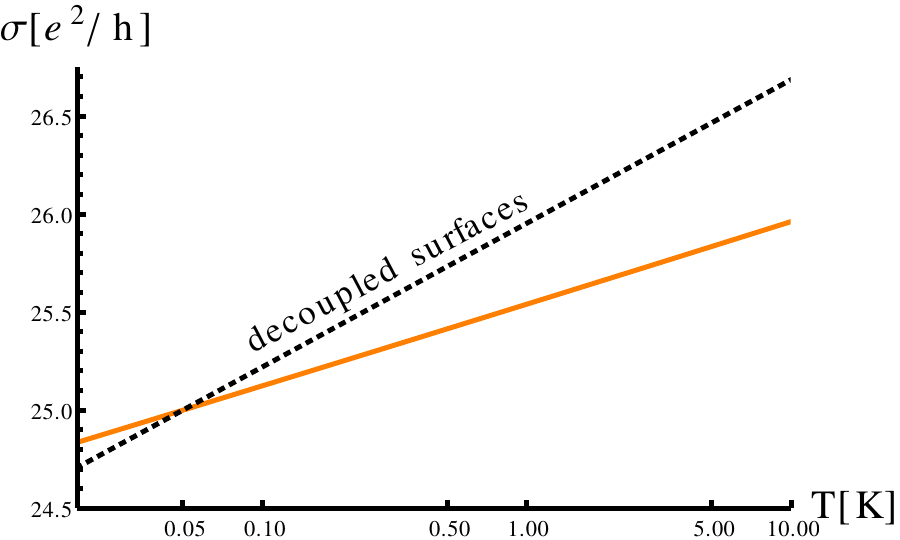}
\caption{Theoretical prediction for the temperature dependence of the total
conductivity in thin films of strained HgTe.}
\label{fig:HgTe}
\end{figure}

\subsection{Hallmarks of surface transport and interactions}

We briefly summarize now our most salient predictions for experimental
signatures of surface transport in 3D TI with an intersurface interactions.
\begin{itemize}
\item As already exploited in 3D TI experiments, \cite{Chen} the magnetoconductance formula \cite{HLN} for the total conductivity is
\begin{equation}
\delta \sigma \left (B\right ) =  - \frac{e^2}{2\pi h} \sum_{s=1,2}\left [ \psi \left
(\frac{1}{2}+\frac{B^{(s)}_{\phi}}{B}\right ) - \ln \left (\frac{B^{(s)}_{\phi}}{B}\right
)\right ] ,
\end{equation}
where the characteristic magnetic field $B^{(s)}_{\phi}=\hbar/(4eD_s^{(s)}\tau^{(s)}_\phi)$ is determined by the diffusion coefficient $D_s^{(s)}$ and the phase breaking time $\tau^{(s)}_\phi$ for the surface $s$. The function $\psi$ denotes the digamma function here.

\item The characteristic effect of intersurface
interaction is the non-monotonous temperature dependence of conductivity (see
Fig. \ref{fig:exemplarynonmonotonicity}, top).
It may happen that in experimentally accessible temperature window
this effect manifests itself only
as a deviation of the conductance slope 
\begin{equation}
\delta \sigma\left (T\right ) =  {\; \frac{e^2}{h}} c \ln T 
\end{equation} from the value $c=1/\pi$ characteristic
for two decoupled surfaces accompanied by some bending of the curve $\sigma(\ln
T)$, see Figs.~\ref{fig:BiSe} and \ref{fig:HgTe}.
The ultimate low-$T$ behavior of the coupled system  is always antilocalizing
and following the universal law 
\begin{equation}
\delta \sigma \left (T\right ) ={\; \frac{e^2}{\pi h}}
\left (1-2 \ln 2 \right ) \ln T  .
\end{equation}  However, depending on the
parameters, this asymptotics may become valid at very low temperatures
only.

\item The strength of intersurface interaction is governed by the parameters
$\kappa_1 d$ and $\kappa_2 d$, where $\kappa$ is the screening length.
Therefore, in contrast to usual, single surface conductivity corrections, the
predicted effect strongly depends on the carrier density (see Fig.
\ref{fig:exemplarynonmonotonicity}, bottom).
\end{itemize}
It is also possible to access the intersurface induced quantum corrections in
the frequency dependence of the AC conductivity (by the simple replacement $T
\rightarrow \omega$ in $\delta \sigma \left (T\right )$ if $\omega \gg T$).

\begin{figure}
\begin{minipage}{1\linewidth}
\includegraphics[scale=0.8]{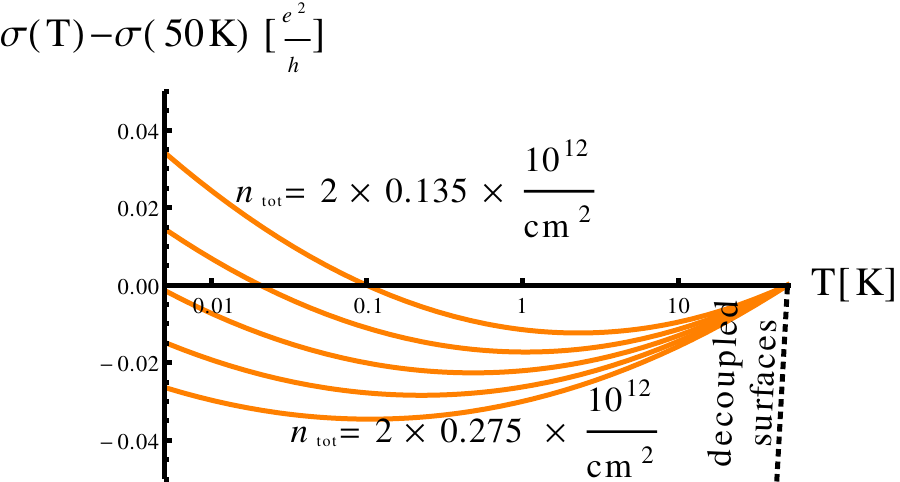}
\end{minipage}
\begin{minipage}{1\linewidth}
\includegraphics[scale=.85]{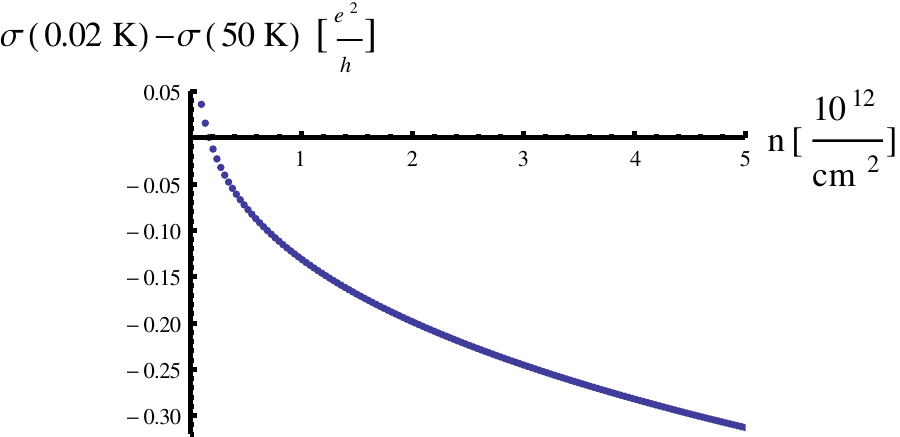}
\end{minipage}
\caption{{\it Top:} Conductivity corrections for low carrier
concentration. The total electron concentration in units of $10^{12} \text{cm}^{-2}$ is equal to $0.55, 0.48, 0.41, 0.34, 0.27$ 
from bottom to top.
The characteristic non-monotonous behavior is clearly seen; deviations
from the behavior of decoupled surfaces are very strong. {\it Bottom:}
Carrier-density dependence of conductivity corrections. The non-trivial
dependence is entirely due to the intersurface interaction: in the case of the
decoupled surfaces, the conductivity correction would be constant as a function
of density,  $\sigma\left (0.02 K\right ) - \sigma\left (50 K\right ) \approx -
2.49 e^2/h$.
We used the values of the parameters $d$, $v_F$ and $\alpha$ as in Table
\ref{tab:expvaluesBise}
for Bi$_2$Se$_3$. Further, we assumed the case of equal surfaces ($n_{tot} = 2 n$) and $T_{\text{Diff}} =
1/2\tau = 50 K$. }
\label{fig:exemplarynonmonotonicity}
\end{figure}

\section{Conclusions}
\label{sec:conclusions}

In this paper, we have investigated interference and interaction
effects in the surface state conductivity of 3D topological
insulator slabs. We have taken into account the electron-electron
interaction within the top and bottom surfaces of a slab and between
them. These two surfaces were in general assumed to be characterized by
different carrier densities and scattering rates, and by
asymmetric dielectric environment.

Our field-theoretical analysis was based on the interacting non-linear sigma
model approach describing the system at length scales above the mean
free path. We demonstrated how this effective theory can be obtained from the non-Abelian
bosonization. In particular, we have shown that upon inclusion of potential
disorder the Wess-Zumino term generates a local expression
for the $\mathbb Z_2$ theta term. The appearance of this topological term is the
hallmark of the Dirac surface states; it is absent in conventional 2D metals of
the same symmetry class. We have further analyzed the $\mathbf{U}(1)$-gauged
sigma model that describes a coupling to the external electromagnetic field.
This has allowed us to connect the physical linear-response characteristics of
the problem and the sigma-model {coupling constants}.  We have also analyzed the effect of
breaking of time-reversal symmetry, namely,  the anomalous quantum Hall effect
of Dirac electrons.

It is worth emphasizing that our theory treats the general situation of
potentially strong
interactions and thus went beyond perturbation theory. We have thus developed
the Fermi liquid theory of the strongly correlated double layer system in the
ballistic and diffusive regime.

We renormalized the interacting NL$\sigma$M of the two surfaces in the one-loop
approximation and obtained the RG equations, Eq.~\eqref{eq:RGeqs}. This way we
have determined the temperature (or else, frequency, or length scale) dependence
of the conductivities of both surfaces.
The RG  is controlled by a large conductivity, $k_F l \gg 1$.
Our calculations are exact in the singlet interaction amplitudes, while
contributions due to a repulsive Cooper interaction are parametrically small and
can be neglected.

Inspecting the RG equations,  we showed that intersurface
interaction is relevant in the RG sense and the limiting case of decoupled
surfaces is therefore unstable. The rich flow diagram has been  analyzed in
detail. For fully decoupled surfaces the system flows into an
intermediate-coupling fixed point (``interaction-induced criticality''). This
point is, however, unstable with respect to the intersurface coupling.
The flow is then towards a single attractive fixed point
which is ``supermetallic'', $\sigma \to \infty$, and at which even
originally different surfaces have the same transport properties, $\sigma_1 =
\sigma_2$, see Figs.  \ref{fig:RGflowequallayers} and
\ref{fig:attractiveplane}. Further, this fixed point is characterized by
vanishing intrasurface and finite intersurface interaction.
Typically, this fixed point is reached via a characteristic
non-monotonous temperature dependence of conductivity.

Our perturbative results are equally applicable to weak topological
insulator\cite{RingelKrausStern12,KobayashiOhtsukiImura13} thin films and to non-topological double
layer 
systems with spin-orbit interaction. {For the latter type of structures,} we
have also discussed the difference 
compared to the strong TI films which is in non-perturbative
topological effects, see a comparison of the flow diagrams in
Fig.~\ref{fig:Comparison}.
While in the TI case these effects lead to a topological
protection of the surface states from Anderson localization, a conventional
(non-topological) double layer system undergoes  a metal-insulator
transition which is tuned by the ratio of interlayer distance and screening
length.

Finally, we have estimated parameters and presented explicit predictions
for the temperature dependence of the conductivity for typical
experimental setups based on Bi$_2$Se$_3$ and strained HgTe materials.

Before closing, we discuss perspectives for further research. First,
experimental studies of temperature dependence of conductivity of 3D
topological insulators for different positions of chemical potentials would be
highly useful. A comparison of such experimental data with our theoretical
predictions would allow one to judge whether the system is in the truly
topological phase. Second, more work is needed on effects of local breaking of time-reversal
symmetry in TI slabs. 
{Third, it is known that Coulomb interaction in electronically decoupled
double-layer systems induces a finite but typically small
transconductance $\sigma_{12}$.\cite{SolomonDrag, GramilaDrag, CoulombDragCarrega, MichaelDrag}
However, the side walls of 3D TI films connect the two major surfaces, which
might be a serious obstacle for performing  Coulomb drag experiments. 
Fourth, {in view of recent experimental progress,\cite{WrayTopSC}} it would be
interesting to perform an RG analysis for a superconducting 
counterpart of the system that we have explored, namely, for surface states of a
3D topological superconductor with spin-orbit interaction (class DIII).\cite{FosterYuzbashyan2012}

\section{Acknowledgements}

We thank J. Smet, Y.Q. Li, A. Finkel'stein, M. Sch\"utt and U. Briskot for
useful discussions. The work was supported by BMBF, DAAD, DFG SPP 1666, RAS
programs ``Quantum mesoscopic and disordered systems'', ``Quantum Physics of Condensed Matter'', ``Fundamentals of nanotechnology and nano materials'', RFBR grants No.
11-02-12126 and No. 12-02-00579, Russian President grant No. MK-4337.2013.02,
and by the Russian Ministry of Education and Science under contract No. 8678.

\appendix

\section{Non-Abelian bosonization and the topological term}

In this appendix we include more detailed calculations concerning non-Abelian
bosonization, the gauged WZNW model and the topological term $S^{(\theta)}$. For
brevity, we omit the surface index $s$ in this appendix. 

\subsection{Gauged WZNW model}
\label{sec:appendixBosonization}
The Wiegmann-Polyakov formula\cite{PolyakovWiegman} allows the inclusion of
smooth $\mathfrak{o}\left (2_\tau \times 2N_M \times N_R\right )$ gauge fields
$A_\mu$. A generalization to potentially topological gauge potentials can be
found in Refs. \onlinecite{Faddeev,Gerasimov,Nekrasov,Smilga}. The gauged WZNW
model is {given as~\cite{Faddeev,Gerasimov,Nekrasov,Smilga}}
\begin{subequations}
\begin{align}
& S\left [O,A_\mu\right ] = -\frac{1}{16 \pi} \int\limits_{\v x} \tr \left (O^TD_\mu O\right )\left (O^TD_\mu O\right )  \\
	& + \frac{i\epsilon_{\mu\nu\rho}}{24\pi}  \int\limits_{\v x,w} \tr \left [\left (O^TD_\mu O\right )\left (O^TD_\nu O\right )\left (O^TD_\rho O\right )\right ]\label{eq:GammaOA} \\
	& - \frac{i \epsilon_{\mu\nu\rho} }{16 \pi} \int\limits_{\v x,w} \tr \left [F_{\mu \nu}\left (O^TD_\rho O + D_\rho O O^T\right )\right ] ,\label{eq:FmunuO}
\end{align}
\label{eq:gaugedWZNW}
\end{subequations}
where we introduced $D_\mu = \partial_\mu + \left [A_\mu,\cdot\right ]$ and
$F_{\mu\nu} = \left [D_\mu,D_{\nu}\right ]$. In the main text, we {were} mostly
interested in $\mathbf U\left (1\right )$ gauge fields $A_\mu = i \hat
A^T\frac{1+\tau_y}{2} - i \hat A\frac{1-\tau_y}{2}$. (In this appendix, the electron charge is absorbed into the vector potential.)

To obtain the Wiegmann-Polyakov formula, one can use the following identity
\cite{Smilga}
\begin{align}
\eqref{eq:GammaOA} &= \frac{i}{24\pi} \Gamma\left [O\right ] \notag \\
&- \frac{i}{8\pi} \int_{\v x, w} \epsilon_{\mu \nu \rho} \partial_\mu \tr \left [O A_\nu O^TA_\rho \right . \notag \\
&\hspace{2cm} \left .+ A_\nu\left (O^T\partial_\rho O + \partial_\rho OO^T\right )\right ] \notag \\
& + \frac{i\epsilon_{\mu\nu\rho}}{16 \pi} \int_{\v x,w}  \tr \left [F_{\mu \nu}\left (O^TD_\rho O + D_\rho O O^T\right )\right ] .
\label{eqAppAi1}
\end{align}
While the {last integral in Eq.~\eqref{eqAppAi1}} compensates the term \eqref{eq:FmunuO}, the total derivative term yields the Wiegmann-Polyakov formula provided $A_\mu$ is not singular:
\begin{align}
S\left [O,A_\mu\right ]&= S\left [O\right ] + \notag \\
	& + \frac{1}{8\pi} \int_{\v x} \tr \left [A_\mu\left (O\partial_\mu O^T + O^T\partial_\mu O\right ) \right  . \notag\\
	& \;  + A_\mu O^T A_\mu O - A_\mu^2 - i \epsilon_{\nu\rho} OA_\nu O^TA_\rho \notag\\
	& \;\left . -i \epsilon_{\nu\rho} A_\nu\left (O^T\partial_\rho O+ \partial_\rho O O^T\right )\right ] \label{eq:WiegmannPolyakov1}\\
	&= S\left [O\right ] + \notag \\
	& + \frac{1}{8\pi} \int_{\v x} \tr \left [A_- \left (O \partial_+ O^T\right ) + A_+ \left (O^T \partial_- O\right )\right . \notag \\
	&\left . + A_+ O^T A_- O - A_+A_-\right ] .
	\label{eq:WiegmannPolyakov}
\end{align}
Here we have introduced the (anti-)holomorphic combination of gauge potentials
$A_\pm = A_x \pm i A_y$. In the case of topological gauge potentials, the
integral over the total derivative yields also a contribution from the Dirac
string.

Equation \eqref{eq:WiegmannPolyakov} is a very powerful result. In particular,
it justifies a posteriori the bosonization rules
\eqref{eq:leftcurrent} and \eqref{eq:rightcurrent}. Also, it follows
immediately from expression \eqref{eq:WiegmannPolyakov1} 
that after disorder-induced symmetry breaking ($O \rightarrow Q = Q^T$) the
gauge-field-dependent contributions from the topological term vanish.  

Further, one can use Eq. \eqref{eq:WiegmannPolyakov} to determine the prefactor
of the kinetic term in the AII NL$\sigma$M, Eq. \eqref{eq:freeNLSM}. As
explained in the main text, soft rotations $O^T_{\text{soft}}OO_{\text{soft}}$
of the WZNW fields $O$ are not affected by disorder induced masses, Eq.
\eqref{eq:disordermassterms}. The effective action for topologically trivial
Goldstone modes contains
\begin{align}
S_{\rm eff,kin} \left [\Phi_\mu\right ] & = \frac{1}{8\pi} \int_{\v x}  \left
\langle \tr \left [\Phi_+O^T\Phi_-O - \Phi_+\Phi_-\right ]\right \rangle \notag
\\ 
&- \frac{1}{2} \left \langle \left ( \int_{\v x}  \tr \left [\Phi_+ j_-
+ \Phi_- j_+\right ] \right )^2\right \rangle, 
\label{eqAppi2}
\end{align}
where $j_\pm$ are the (bosonic) currents, $\left \langle \dots\right  \rangle$
denotes average with respect to the full bosonic theory (including the mass
terms) and $\Phi_\pm = O_{\text{soft}}\partial_\pm O_{\text{soft}}^T$. To the
leading order, the average can be calculated close to the saddle point.
Exploiting the equivalence of bosonic and fermionic theories one can equally
evaluate $\left \langle \dots \right \rangle$ using the fermionic fields at SCBA
level. At $\vert \mu \vert \tau \gg 1$ the major contribution comes from the
second line of Eq.\eqref{eqAppi2}, which, taking the vertex corrections into
account, yields the correct prefactor (i.e. the conductivity) of the kinetic
term {in Eq.~}\eqref{eq:freeNLSM}. 

\subsection{Instanton configuration}
\label{sec:Instanton}

We define the following four dimensional unit vector
\begin{eqnarray}
\underline{a} &\equiv & \left(a_0, a_1, a_2, a_3 \right) \notag \\
	&\equiv& \frac{1}{\vert \vec x- \vec x' \vert^2 + \lambda^2} \left(2
\lambda \left (\vec x- \vec x'\right ), \vert \vec x- \vec x' \vert^2  -
\lambda^2 \right) , \notag
\end{eqnarray}
where the 1+2 vector $\vec x - \vec x' \equiv \left ((1-w)/w, \v x - \v x'\right
)$ contains the extension parameter and the real-space  coordinates. It
describes a topological excitation at position $\left (1,\v x'\right )$ in a
three-dimensional base space.
With the help of {the vector} $\underline a$ we can define the following
extended field configuration 
\begin{equation}
\tilde O_{\rm inst} = \left (\begin{array}{cccc}
-a_0 i \tau_y + a_3 &  0 & a_1 + a_2 i \tau_y & 0 \\
0 & \v 1 & 0 & 0 \\
a_1 - a_2 i \tau_y & 0 & -a_0 i \tau_y - a_3 & 0 \\
0 & 0 & 0 & - \v 1
\end{array} \right ).
\end{equation}
For $a_0 = 0$, i.e. on the physical space $w = 1$, $\tilde O_{inst}$ is a
symmetric matrix and  characterizes the two-dimensional instanton. The choice of
the extension is arbitrary, but, as has been stressed in the main text, the
$\tilde O_{\rm inst}$ field has to leave the diffusive saddle-point manifold for
some subinterval $w \in I \subseteq \left (0,1\right )$. For $w \rightarrow 0 $
the extended field $\tilde O_{\rm inst}$ satisfies the boundary condition
$\tilde O \left (\v x, w=0\right ) = \Lambda = {\rm const}$. 

We are now in the position to insert the instanton configuration into the
WZNW-term. After tracing out the matrix degrees of freedom this leads to
\begin{eqnarray}
iS^{(\theta)} &=& \frac{-i}{6 \pi} \int_{x,w}  \epsilon_{\mu \nu \lambda} \Bigl 
(\epsilon_{abc} \; a_a \partial_ \mu a_b \partial _\nu a_c \partial _\lambda a_0
\nonumber \\
& - & \epsilon_{abd} \; a_a \partial_ \mu a_b \partial _\nu a_0 \partial
_\lambda a_d \nonumber \\
& + &\epsilon_{cda} \; a_a \partial_ \mu a_0 \partial _\nu a_c \partial _\lambda
a_d \nonumber \\
& - &\epsilon_{cdb} \; a_0 \partial_ \mu a_b \partial _\nu a_c \partial _\lambda
a_d \Bigr ) \nonumber \\
& = & i\pi .
\end{eqnarray}
Here the last line is obtained by a straightforward calculation.
We have thus shown that the topological term distinguishes between the trivial
and the non-trivial sectors as it acquires on them the values $0$ and $i \pi
\;(\text{mod }2\pi i)$, respectively.

\section{{Effect of dielectric environment} on Coulomb interaction}
\label{sec:Elstat}

\subsection{Electrostatic Potential and Single Particle Effects}
As has been stated above the experimental setup consists of a sandwich of (at
least) three {different} dielectrics (see figure \ref{fig:setup}). We define the
$z$-axis {to be} perpendicular to the two surfaces. The sandwich consists of the
coating material with a dielectric constant $\epsilon_1$ (for $d/2 < z$), the
topological insulator {film} with a dielectric constant $\epsilon_2$ (for $-d/2
\leqslant z \leqslant d/2$), and the substrate with a dielectric constant
$\epsilon_3$ (for $z < -d/2$). Taking these different dielectric properties into
account, {we} here present {the expression for the Coulomb potential which
generalizes Eq.~\eqref{eq:Coulombnoelstat}.} 

By the method of mirror charges, one can derive\cite{Profumo2010,
Katsnelson2011, CoulombDragCarrega} the electrostatic potential induced by a
single point charge $e$ located at $\left (\v x_0, z_0\right ) = \left
(0,0,z_0\right )$ inside the middle region  
of the sandwich ($z,z_0 \in \left [-\frac{d}{2},\frac{d}{2}\right ]$):
\begin{eqnarray}
\Phi\left (\v x, z, z_0\right ) & = &\frac{e}{\epsilon_2}	 \bigg[
\frac{1}{\sqrt{ \v x  ^2 + \left (z-z_0\right )^2}} \notag \\
								&				+			& r_{23}^{-1} \mathcal F\left (\v x, d+\left (z+z_0\right )\right ) \notag \\
								&				+			& r_{21}^{-1} \mathcal F\left (\v x, d - \left (z+z_0\right )\right ) \notag \\
								&				+			& \mathcal F\left (\v x, z-z_0\right ) \notag \\
								&				+			& \mathcal F\left (\v x, - \left (z-z_0\right )\right ) \bigg], \label{eq:generalpotential}
\end{eqnarray}
where 
\begin{equation}
\mathcal{F}\left ( \v x, z\right ) = \sum_{k=1}^{\infty} \frac{\left (r_{21}r_{23}\right )^k}{\sqrt{\v x ^2 + \left (z - 2dk\right )^2}}.
\end{equation}
and the ratios 
$$
r_{21} \equiv \frac{\epsilon_2 - \epsilon_1}{\epsilon_2 +
\epsilon_1}; \qquad 
r_{23} \equiv \frac{\epsilon_2 - \epsilon_3}{\epsilon_2 +
\epsilon_3}
$$ 
were introduced. If one of these ratios vanishes, the textbook limit
of two dielectric half-planes follows. Fourier transformation of the $\v x$
coordinates yields 
\begin{eqnarray}
\Phi\left (\v q, z, z_0\right ) & = &\frac{2\pi e}{q\epsilon_2}	
\bigg[e^{-\vert z-z_0\vert q} + \frac{e^{-2dq}}{1- r_{21}r_{23}e^{-2dq}} \times
\notag \\
								&				\times			& \Big ( r_{21} e^{(d+z+z_0)q} + r_{23} e^{(d-z-z_0)q} \notag \\
								&								& + 2r_{21}r_{23}\cosh((z-z_0)q)\Big )\bigg]. \label{eq:generalpotentialFT}
\end{eqnarray}

We consider now 3D TI surface states: the charges are located at a typical
distance $a \sim v_F/\Delta_{\rm bulk}$ (the penetration depth) from the
boundaries $z = \pm \frac{d}{2}$. The consequences of the general expression
\eqref{eq:generalpotentialFT} on the 3D TI surface states are twofold. 

First, there is a single particle effect, stemming from the interaction of the
charged particles with their own mirror charges. The associated electrostatic
energy is incorporated in the chemical potential in the main text and can be
expressed as  
\begin{eqnarray}
\Delta\mu_1 &=& \frac{e}{2}\Phi^{{\rm reg}}\left (0, \frac{d}{2}-a,
\frac{d}{2}-a\right ) \notag \\
	&=& \frac{e^2}{4\epsilon_2 }\left [\frac{r_{21}}{a} -
\frac{r_{21}+r_{21}^{-1} + 2}{ d} \ln \left (1 - r_{21}r_{23}\right )\right ] .
\nonumber \\
&&
\end{eqnarray}
The analogous shift of the chemical potential at the second surface $\Delta
\mu_2$ is easily obtained by interchanging $r_{21} \leftrightarrow r_{23}$. The
superscript $^{\rm reg}$ indicates that selfinteraction of the charges is
subtracted. In the second term we used the approximation $a \ll d$. The first
term, i.e. the interaction with the nearest mirror charge, is typically the
dominating contribution $\Delta \mu_1 \approx \alpha_2 r_{21}/4 \Delta_{\rm
bulk}$. 

Second, the electrostatic energy associated with two-particle interaction is the
quantity $\underline U_0$ entering $S_{\rm int}$ in Eq. \eqref{eq:S_Coulsimple}.
This leads to the interaction parameters analyzed below. 

\subsection{Interaction parameters}

The interaction parameters are obtained by placing a test charge into Eq.
\eqref{eq:generalpotentialFT}. 
We will present this effective Coulomb interaction in the surface space.
The terms induced by intersurface interaction contain a factor $\exp(-qd)$ ($q$
takes values between the IR and UV cutoffs, $q \in \left [L_E^{-1},l^{-1}\right
]$). As a result we have to distinguish between the following two cases. 

In the first case the momenta are large ($qd \gg 1$) 
throughout our RG-procedure if $dL^{-1} >1$ or for part of it if $d \in \left
[l,L\right ]$.
 Then the two surfaces {become decoupled and}
\begin{equation}
\underline U_0 = \frac{2\pi}{  q } \left (\begin{array}{cc}
\frac{2}{\epsilon_2 + \epsilon_1} & 0 \\
0 & \frac{2}{\epsilon_2 + \epsilon_3}
\end{array} \right ) .
\end{equation}
(Here and in all subsequent appendices we drop the electron charge, it is formally included into a redefinition of $\epsilon_{1},\epsilon_{2},\epsilon_{3}$.)

In the second case the momenta are small $qd \ll 1$. 
As we shall be interested in the low-energy theory, we keep only the Fourier
transformed terms which are not vanishing in the limit of small transfered
momentum $qd \rightarrow 0$. All others are irrelevant in the RG-sense. This way
we obtain the true long-range Coulomb part 
\begin{equation}
\underline U_{C} = \frac{2}{\epsilon_1 + \epsilon_3}\frac{2\pi  }{q} \left (\begin{array}{cc}
1 & 1 \\
1 & 1
\end{array} \right ). \label{eq:Ueff}
\end{equation}
As expected, it does no longer depend on $\epsilon_2$. The limit we considered
is the large-distance behavior in which the dominant part of the electric field
lines lives in the {dielectrics surrounding the film}. 

There are other contributions which do not vanish in the $qd \rightarrow 0$
limit. These are short range interaction amplitudes introduced by the finite
{thickness of the film}: 
\begin{eqnarray}
\underline F^{(d)} &=& - \frac{2\pi }{\epsilon_2}d  \left (\begin{array}{cc}
0 & 1 \\
1 & 0
\end{array} \right ) \notag \\
	& & - \frac{4\pi }{\epsilon_1 + \epsilon_3} d\left [ F_{\rm symm}\left
(\begin{array}{cc}
1 & 1 \\
1 & 1
\end{array} \right ) + \underline F_{M} \right ].
\end{eqnarray}
Here we have defined the scalar
\begin{equation}
F_{\rm symm} = \left (\epsilon_2- \epsilon_1\right )\left (\epsilon_2-
\epsilon_3\right ) \left \lbrace \frac{1}{2\epsilon^2_2} +
\frac{1}{\epsilon_2\left (\epsilon_1 + \epsilon_3\right )}\right \rbrace
\end{equation}
and the matrix
\begin{equation}
\underline F_{M} = \frac{1}{2\epsilon_2^2}\left (\begin{array}{cc}
\left (\epsilon_2+ \epsilon_1\right )\left (\epsilon_2- \epsilon_3\right ) & \epsilon^2_2 - \epsilon_1\epsilon_3 \\
\epsilon^2_2 - \epsilon_1\epsilon_3 & \left (\epsilon_2- \epsilon_1\right )\left (\epsilon_2+ \epsilon_3\right )
\end{array} \right ) ,
\end{equation}
which both vanish in the limit of $\epsilon_1 = \epsilon_2 = \epsilon_3$. In
summary, for coupled surface we can write $\underline U_0 = \underline U_C +
\underline F^{(d)}$. 

The derivation of the above equations includes some subtleties. First, we
derived the electric field configuration for a single point charge. Thus, in
particular, we did not consider the metallic surfaces between the dielectrics.
As in the theory of conventional metals, their effect will be incorporated in
the field theoretical description of the model (Appendix \ref{sec:cleanFL}). 
Second, we used the potential \eqref{eq:generalpotentialFT} derived for charged
particles at position $z, z_0$ and then moved them on the surface between the
dielectrics from inside {of the} {TI film} ($z_0 = \pm d/2 \mp a \approx \pm
d/2$ and equally for $z$). This requires that the (macroscopic) electrostatic
theory of continuous, homogeneous dielectrics can be applied to electrons
located at a distance $a$ from the boundary. This is justified, as we are
interested in the long-range behavior of the electric field. Furthermore, for
Bi$_2$Se$_3$ it is known that $a$ is of the order of a few nanometers
\cite{Linder, WeiZhang}, hence one order of magnitude larger than the atomic
scale.

\section{Clean Fermi liquid}
\label{sec:cleanFL}

In this appendix we present the formal resummation of scattering amplitudes
following references \onlinecite{Nozieres_Luttinger_1962,
AbrikosovGorkovDzyaloshinski, LandauLifshitz9}. We first consider the short
range (one-Coulomb-line-irreducible) part of the singlet channel (see also Eq.
\eqref{eq:realIA}) 
\begin{equation}
\Gamma^{1-2}_{ss'} = \Gamma^{1}_{ss'} - \Gamma^{2}_{ss} \delta_{ss'} \;,
\end{equation}
 and include the long-range, one-Coulomb-line-reducible, diagrams ($\Gamma^0$)
later on.

\subsection{Resummation of interaction amplitudes}

The first step is to single out the subset of particle-hole-section irreducible
diagrams {$I^{1-2}$}. The total interaction amplitude as a matrix in the surface
space and in 2+1-momentum space is given by {the Dyson equation} 
\begin{equation}
\underline \Gamma^{1-2} \left (K\right )= \underline I^{1-2} - \underline
I^{1-2} \underline R\left (K\right ) \underline \Gamma^{1-2} \left  (K\right )
\label{eq:DysonforGamma}
\end{equation}
(Matrix multiplication includes momentum integral $\int_\v p $and a Matsubara sum $T \sum_n$.) 

The matrix 
\begin{align}
\left [\underline R \left (K \right )\right ]_{PP',ss'} &=  \delta_{ss'}\delta_{PP'} R_{s,P}\left (K\right ), \\
R_{s,P} \left (K\right ) &\equiv  G_s\left (P\right )G_s\left (P+K\right ) 
\end{align}
describes particle-hole bubbles and in the singlet channel. This matrix is
diagonal in both 2+1 momentum and surface space: As we explained in the main
text, it is sufficient to keep only intrasurface bubbles in the assumed case of
uncorrelated disorder. In the presence of generic interaction, the quantity
$\underline R_{s,P} \left (K \right )$ can be represented as 
\begin{eqnarray}
R_{s,P} \left (K\right ) &= &  R_{s,P}^\omega +  \Delta_{{s,P}} \left (K\right ) \\
&= &  R_{s,P}^q + \tilde \Delta_{s,P} \left (K\right ). 
\end{eqnarray}
Here $R_{s,P}^\omega$ ($R_{s,P}^q$) are called regular (static) part of the
bubble. The $\omega$- and q-limits are defined in the main text ({see
Eqs.~}\eqref{eq:Piomega} and \eqref{eq:Piq}). The singular (dynamic) part of the
particle-hole bubble {is}  
\begin{eqnarray}
\Delta_{{s,P}} \left (K\right ) &=&   \beta \frac{-i \v v^F_s \cdot \v q}{\omega_m +i \v v^F_s \cdot \v q} \delta_P^{(s)}, \notag \\
\tilde \Delta_{s,P} \left (K\right ) &=&  \beta \frac{\omega_m}{\omega_m +i \v v^F_s \cdot \v q} \delta_P^{(s)} .\notag
\end{eqnarray}
(We have absorbed the Fermi liquid (FL) residues into a redefinition of the
scattering amplitudes.) From these definitions and
Eq.~\eqref{eq:DysonforGamma} we obtain the relations 
\begin{subequations}
\begin{equation}
\underline \Gamma^{1-2} \left (K\right ) = \underline \Gamma^{1-2,\omega} - \underline \Gamma^{1-2} \left (K\right ) \underline{\Delta}\left (K\right ) \underline \Gamma^{1-2,\omega}, \label{eq:singletsumomega}
\end{equation}
and
\begin{equation}
\underline \Gamma^{1-2} \left (K\right ) = \underline \Gamma^{1-2,q} - \underline \Gamma^{1-2} \left (K\right ) \underline{\tilde \Delta}\left (K\right ) \underline \Gamma^{1-2,q}.  \label{eq:singletsumq} \\
\end{equation}
\end{subequations}
This formal (re-)expression of the general scattering amplitude
will be used to calculate the polarization operator in the next subsection. 

\subsection{Definitions}
\label{sec:appendix:definitions}

In order to introduce the long-range Coulomb interaction and to describe its
screening we define the following quantities.
The bare triangular vertices are obtained in response to an external scalar potential $\phi^{(s)} \left (\omega_m, \v q\right )$:
\begin{equation}
 \underline{\v {v}}^{(1)}_0 = \left (1,0 \right ) \text{ and }  \underline{\v{v}}^{(2)}_0 = \left (0,1 \right ). 
\end{equation}
We used the approximation $\langle \mu_s, \v p \vert  \mu_s, \v p + \v q  \rangle \approx 1$. In our notation, bold, italic, underlined quantities are vectors in surface space.

The {triangular vertex $\underline{\v T}^{(s)}$ renormalized by interaction satisfies}
\begin{equation}
\underline{\v T}^{(s)} \left (K\right ) = \underline{\v v}_0^{(s)} - \underline{\v v}_0^{(s)}  \underline R \left (K\right ) \underline \Gamma^{1-2}\left (K\right ) . \label{eq:resumT}
\end{equation}
The polarization operator is a matrix in the surface space and can be written
as
\begin{eqnarray}
\hspace*{-0.5cm}
\Pi^{ss'} \left (K\right ) & = & \underline{\v v}_0^{(s)} \underline R\left (K\right )
[\underline{\v v}_0^{(s')}]^T \nonumber \\ 
& - & \underline{\v v}_0^{(s)} \underline R\left (K\right ) \underline
\Gamma^{1-2} \left (K\right )\underline R\left (K\right )  [\underline{\v v}_0^{(s')}]^T ,
\end{eqnarray}
which {transforms} into
\begin{subequations}
\begin{eqnarray}
\hspace*{-0.5cm}
\Pi^{ss'} \left (K\right )&=& \Pi^{{ss'},q} + \underline{\v T}^{(s),q} \underline
{\tilde{\Delta }}\left (K\right )  [\underline{\v T}^{(s'),q}]^T \notag \\
	&-&  \underline{\v T}^{(s),q} \underline {\tilde{\Delta }}\left (K\right )\underline
\Gamma^{1-2} \left (K\right )\underline {\tilde{\Delta }}\left (K\right )
[\underline{\v T}^{(s'),q}]^T  \label{eq:Pist+dynamic} \\ 
	&=& \Pi^{{ss'},\omega} + \underline{\v T}^{(s),\omega} \underline {{\Delta }}\left
(K\right )  [\underline{\v T}^{(s'),\omega}]^T \notag \\ 
&-& \underline{\v T}^{(s),\omega} \underline {{\Delta }}\left (K\right )\underline \Gamma^{1-2} \left (K\right )\underline {{\Delta }}\left (K\right ) 
[\underline{\v T}^{(s'),\omega}]^T.\hspace{0.5cm}\,{} \label{eq:Pireg+sing}
\end{eqnarray}
\end{subequations}
We will show below that these equations combined with Ward identities can be
used to derive the $\omega$ and $q$ limits of the polarization operator.

\subsection{Ward identities}

We will first investigate the Ward identities which are due to invariance under separate $\mathbf{U}(1)$ rotation of the fermionic fields. Following the standard procedure we obtain
\begin{equation}
\left ( \frac{\partial G^{-1}_1}{\partial p_0}, 0 \right ) = \underline{ \v T}^{(1), \omega} \text{ and }
\left ( 0 ,\frac{\partial G^{-1}_2}{\partial p_0} \right )  = \underline { \v T}^{(2), \omega}. \label{eq:TomegaWI}
\end{equation}
		
Next, we exploit that constant external fields can be reabsorbed into a redefinition of the chemical potentials. This leads to
\begin{equation}
\left ( \frac{\partial G^{-1}_1}{\partial \mu_s} , \frac{\partial G^{-1}_2}{\partial \mu_s} \right ) = \underline{\v T}^{(s), q} . \label{eq:TqWI}
\end{equation}
We insert this into the $\omega$- and $q$- limits of the polarization operator and obtain
\begin{equation}
			\Pi^{{ss'},\omega} = 0 \hspace*{.3cm} \text{ and } \hspace*{.3cm} \Pi^{ss',q} = - \frac{\partial N_s}{\partial \mu_{s'}} = -\frac{\partial N_{s'}}{\partial \mu_{s}}.\label{eq:PiWI}
\end{equation}

The Ward identities \eqref{eq:TqWI} and \eqref{eq:PiWI} have very profound
consequences. They relate the static triangular vertex and the static
polarization operator to derivatives of physical observables with respect to the
chemical potential. It is explained in the main text, that for this reason they
are not renormalized in the diffusive RG.~\cite{Finkelstein1990} 

\subsection{Screening of the Coulomb interaction}
\label{sec:appendixCoulomb}

We consider the singular part of the Coulomb interaction (see Eq. \eqref{eq:Ueff}), i.e. 
\begin{equation}
\underline U_0 = \frac{2\pi}{\epsilon_{\text{eff}}q} \left (\begin{array}{cc}
1 & 1 \\
1 & 1
\end{array} \right ) ,
\end{equation}
{where} $\epsilon_{\text{eff}} = (\epsilon_1 + \epsilon_3)/2$ for the most general situation of a dielectric sandwich structure. This matrix has zero determinant, $\det \underline U_0 = 0$.

The RPA-screened Coulomb interaction is defined {as}
\begin{equation}
 \underline U_{\rm scr} \left (\omega_m, \v q \right ) = \left (1-\underline U_0
\underline \Pi \right )^{-1} \underline U_{0}. \notag
\end{equation}

The static one-Coulomb-line-reducible singlet interaction amplitude is obtained
by attaching the (q-limit) triangular vertices to {$\underline U_{\rm scr} \left (\omega_m=0, \v q \right )$} from
both sides (see Fig.~\ref{fig:Gamma0def} in the main text). From the
definition in Sec. \ref{sec:appendix:definitions} we know {that} $\Pi^{ss',q}
= T^{s,q} \nu_{s'}$.
(Note that none of these three quantities is renormalized during RG.) Therefore,
we obtain
\begin{equation}
\underline \Gamma ^{0}	= - \underline{\left (\frac{1}{\nu}\right )} \underline
\Pi^q \underline U_{\rm scr} \left (\omega_m=0, \v q \right ) \underline \Pi^q \underline{\left
(\frac{1}{\nu}\right )}. \label{eq:defGamma0}
\end{equation}

By means of the orthogonal matrix 
\begin{equation}
O=\frac{1}{\sqrt2}
\begin{pmatrix}
1 & -1 \\
1 & 1 
\end{pmatrix}
\end{equation}
we can rotate {$\underline U_{\rm scr} \left (\omega_m=0, \v q \right )$} into
the basis where $\underline U_0$ is diagonal:
\begin{align}
& O^T \underline U_{\rm scr}{\left (\omega_m=0, \v q \right )} O = \left (1-\left (\begin{array}{cc}
\frac{4\pi}{\epsilon_{\text{eff}}q} & 0 \\
0 & 0
\end{array}\right )  O^T \underline \Pi^q  O\right )^{-1} \notag \\ 
& \times 
\left (\begin{array}{cc}
\frac{4\pi}{\epsilon_{\text{eff}}q} & 0 \\
0 & 0
\end{array}\right ) 
 = \frac{\frac{4\pi}{\epsilon_{\text{eff}}}}{q-\frac{2\pi}{\epsilon_{\text{eff}}} \left (\Pi^q_{11} +\Pi^q_{22}+2\Pi^q_{12}\right )} \left (\begin{array}{cc}
1 & 0 \\
0 & 0
\end{array}\right ) . 
\label{eq:effectiveU}
\end{align}
The denominator in the last line of Eq.~\eqref{eq:effectiveU} defines the
coupled surface screening length [analogously to Eqs.~\eqref{eq:coupledsurf},
\eqref{eq:screeninglength}].

In the considered parameter {range} we can take the $q$-limit {under the
following condition:} $\vert \frac{2\pi}{\epsilon_{\text{eff}}} \left
(\Pi^q_{11} +\Pi^q_{22}+2\Pi^q_{12}\right ) \vert\gg q$. Then we obtain 
\begin{equation}
 \underline U^q_{\rm scr} = - O \left (\begin{array}{cc}
\left[\hat e_1 ^T O^T \underline \Pi^q O \hat e_1\right]^{-1} & 0 \\
0 & 0
\end{array} \right ) O^T.
\end{equation}
The q-limit of {Eq.~}\eqref{eq:defGamma0} is
\begin{equation}
\underline \Gamma ^{0,q} =  \underline{\left (\frac{1}{\nu}\right )} \underline \Pi^q O \hat e_1 \otimes  \hat e_1^T O^T \underline \Pi^q \underline{\left (\frac{1}{\nu}\right )} \frac{1}{\hat e_1^T O^T \underline \Pi^q O \hat e_1}.  \label{eq:Gamma0}
\end{equation}
We multiply by $\underline \nu O \hat e_1$ from the right {side} and {find}
\begin{equation}
\left [ - \underline{\left (\frac{1}{\nu}\right )} \underline \Pi^q + \underline \Gamma ^{0,q} \underline \nu \right ] O \hat e_1 = 0 . \label{eq:Coulombconditionappendix}
\end{equation}
This {matrix equation} implies that the surface-space matrix in brackets has to
be of zero determinant. 

Alternatively, using the $q$-limit of Eq. \eqref{eq:Pireg+sing} and 
\eqref{eq:defGamma0}, we can express the bare total interaction amplitude
$\underline \Gamma^\rho \equiv \underline \Gamma^0 + \underline \Gamma^{1-2}$
as 
\begin{equation}
\underline \nu \underline \Gamma^{\rho} \underline \nu = - \underline \nu - \frac{\det  \underline \Pi^q}{\Pi^q_{11} +\Pi^q_{22}+2\Pi^q_{12}} \left (\begin{array}{cc}
1 & -1 \\
-1 & 1
\end{array} \right ). \label{eq:barevalueGamma}
\end{equation}
From Eq. \eqref{eq:barevalueGamma} the following statement immediately follows:
\begin{equation}
\det \left [\underline \nu + \underline \nu \underline \Gamma^{\rho} \underline \nu\right ] = 0
\end{equation}
This relationship is equivalent to Eq. \eqref{eq:Coulombconditionappendix}.

\subsection{Total density-density response}
\label{sec:Piredappendix}

Here we analyze the one-Coulomb-line-reducible (which we will
also term ``total'') density-density response $\underline \Pi^{{\rm RPA}}$. It
is defined as 
\begin{equation}
\underline \Pi^{{\rm RPA}}\left (K\right ) = \underline\Pi \left (K\right ) +
\underline\Pi \left (K\right )\underline U_0\left (\v q\right )
\underline\Pi^{{\rm RPA}}\left (K\right ).
\label{Pi-RPA}
\end{equation}
Equation (\ref{Pi-RPA}) implies that $\Pi^{{\rm RPA}}$ is obtained a
resummation of the RPA-type series, hence the corresponding superscript. 

For the present {case} we want to obtain $\underline \Pi^{{\rm RPA}}$ in the
diffusive regime. The very idea of dirty FL {is based on} replacing
dynamic section $\tilde \Delta_{s,P}$ according to the following prescription:
\begin{equation}
\frac{\omega_m}{\omega_m +i \v v^F_s \cdot \v q} \quad \rightarrow\quad   \frac{\omega_m}{Z_s \omega_m + D_s \v q^2} ,
\end{equation}
with $Z_s = 1$ at the bare level.
By using definitions \eqref{eq:Pist+dynamic} and \eqref{eq:defGamma0}, the total
density-density response can be written as
\begin{equation}
\underline{\Pi}^{{\rm RPA}}\left (K\right ) = \left [\underline{\Pi}^q -
\underline{\nu\Gamma}^0 \underline \nu\right ]\left [1+ \omega_m \underline
\Delta^{\Gamma}\left [\underline{\Pi}^q - \underline{\nu\Gamma}^0 \underline
\nu\right ]\right ] ,
\end{equation}
where
\begin{equation}
\Delta^{\Gamma} \equiv \Delta^{\Gamma} \left (\omega_m, \v q\right ) = \left [\underline{\nu D} \v q ^2 + \left ( \underline {\nu Z} + \underline{\nu} \underline{\Gamma}^{0+1-2} \underline{\nu}\right ) \omega_m\right ]^{-1}. 
\end{equation}
These equations are used in the main text (Sec. \ref{sec:Densityresponse}) 
to provide a link between the bosonized NL$\sigma$M and the dirty FL
theory.

\subsection{Bare NL$\sigma$M coupling constants}
\label{sec:BareGamma}

According to Eqs. \eqref{eq:effectiveU} and \eqref{eq:barevalueGamma} the
bare values of the interaction amplitudes are fully determined by $\nu_{1}$,
$\nu_{2}$ and  
\begin{equation}
\underline \Pi^q = - \underline \nu \left (1 + \underline F\underline \nu\right )^{-1}, \label{eq:PiofF}
\end{equation}
where
\begin{equation}
\underline F = \left (\begin{array}{cc}
F_{11} & F_{12} \\
F_{12} & F_{22}
\end{array} \right )
\end{equation}
are the FL constants in the density channel(s). It is convenient to
express $\underline \nu \underline \Gamma^{\rho} \underline \nu$ in Eq.
\eqref{eq:barevalueGamma} through $\underline F$ by means of the identity
\eqref{eq:PiofF}: 
\begin{equation}
\frac{\det  \underline \Pi^q}{\Pi^q_{11} +\Pi^q_{22}+2\Pi^q_{12}} = \frac{-1}{1/\nu_1 + 1/\nu_2 + F_{11} + F_{22} -2 F_{12}} . \label{eq:barevalueGamma2}
\end{equation}

In appendix \ref{sec:Elstat} we derived the general expression for FL
constants $\underline{F}^{(d)} \equiv \frac{2\pi}{\epsilon_2}d \underline f$
induced by the finite thickness of the topological insulator {film}. Assuming {that} there
is no {additional} short range interaction one can deduce 
the bare value of interaction constants for the NL$\sigma$M. This is equivalent
to {the} RPA estimate (valid if $\alpha\ll 1$).

In the following we consider two limits. As in the main text, the inverse single surface
screening length is denoted by $\kappa_s = 2\pi\nu_s / \epsilon_2$. The
first limit is the case of equal surfaces $\nu_1 = \nu_2$ in a symmetric setup
$\epsilon_1 = \epsilon_2 = \epsilon_3$. Then the effective FL amplitude is 
\begin{equation}
\underline F^{(d)} = - \frac{2 \pi d}{\epsilon_2} \left (\begin{array}{cc}
0 & 1 \\
1 & 0
\end{array} \right ).
\end{equation}
 
The bare value of the interaction constant is
\begin{equation}
\gamma_{11} = \gamma_{22} = \frac{\left [\underline \nu \underline \Gamma^\rho \underline \nu\right ]_{11}}{\nu_1} = -\frac{1}{2} \left [1 + \frac{\kappa d}{1 + \kappa d}\right ]
\end{equation}
Note that in the limit $\kappa d \rightarrow \infty$ ($ \kappa d \rightarrow 0$) the bare value of $\gamma_{11} = \gamma_{22}$ is {equal to} $-1$ ($-1/2$).

The second limit is the experimentally relevant situation with $\epsilon_1 \ll
\epsilon_2, \: \epsilon_3$. In this limit, we find
\begin{equation}
\underline F^{(d)} = \frac{4\pi d}{\epsilon_2} \left (\begin{array}{cc}
1-\left (\epsilon_2/\epsilon_3\right)^2 & -\left (\epsilon_2/\epsilon_3\right)^2 \\
-\left (\epsilon_2/\epsilon_3\right)^2 & -\left (\epsilon_2/\epsilon_3\right)^2
\end{array} \right ). \label{eq:Fforasymmetric}
\end{equation}
It follows from Eqs. \eqref{eq:barevalueGamma2} and \eqref{eq:Fforasymmetric}
that the bare values for interaction constants are $\epsilon_3$-independent:
\begin{subequations}
\begin{equation}
\gamma_{11} = -1 +  \frac{1}{1 + \frac{\kappa_1}{\kappa_2} + 2\kappa_1 d}
\end{equation}
and
\begin{equation}
\gamma_{22} = -1 +  \frac{1}{1 + \frac{\kappa_2}{\kappa_1} + 2\kappa_2 d}.
\end{equation}
\label{eq:gammabareexp}
\end{subequations}
In view of Eq. \eqref{eq:Finv}] following from the $\mathcal F$-invariance, 
is not surprising that the coupling constants are equal as long as $\nu_1 =
\nu_2$ even in the case of asymmetric dielectric environment. 

\section{Detailed derivation of RG equations}
\label{sec:RGderiv}

In this section we present the detailed derivation of the {one-loop corrections to conductivity}.

\subsubsection{Correlator $B_1$}

In the one loop approximation we can use $Q = \Lambda+ \delta Q$ with $\delta Q = \left (\begin{array}{cc}
0 & q \\
q^T & 0
\end{array} \right )$. {Then we} directly single out the classical contribution in $B_1$, Eq. \eqref{eq:defB1} {and obtain}
\begin{equation}
B_1^{s} = \sigma_s - \frac{\sigma_s}{4n} \sum_{\mu=0,2}\tr \left \langle I_n^\alpha \tilde \tau_\mu \delta Q [ I^\alpha_{-n} \tilde \tau_\mu^T\delta Q -  I^\alpha_n \tilde \tau_\mu\delta Q ]  \right \rangle .\label{eq:calcB1}
\end{equation}
In addition, we write $q = \sum_{\nu=0}^3 q^{(\nu)} \tilde \tau_\nu^T$.
When performing the trace in $\tau$-space it turns out that the two diffuson contributions ($\nu = 0,2$) cancel up. This is a consequence of the opposite sign of $\tau_0$ and $\tau_2$ under transposition. The cooperons ($\nu=1,3$) contribute only to the {last term in Eq.~}\eqref{eq:calcB1}. {Then we find}
\begin{eqnarray}
B_1^s &=& \sigma_s + \frac{\sigma_s}{4n} \sum_{\nu=1,3} \left  \langle \tr I_n^\alpha \left (\begin{array}{cc}
0 & q_\nu \\
q^T_\nu & 0
\end{array} \right )   I_n^\alpha \left (\begin{array}{cc}
0 & q_\nu \\
q^T_\nu & 0
\end{array} \right ) \right  \rangle \notag \\
&=& \sigma_s + {2}\int_{\v p } D_s(\omega_n,\v p).
\end{eqnarray}

\subsubsection{Correlator $B_2$}

The second term $B_2$, Eq. \eqref{eq:defB2}, does not contribute on the
classical level. Expanding to second order in $q$ we obtain the tree level
contribution which also vanishes:
\begin{equation}
\left .B_2^{ss'}\right \vert_{\text{tree level}} = - \frac{\sigma_s\delta_{ss^\prime}}{4} \int\limits_{\v x - \v x'} e^{i \v p \left (\v x- \v x '\right )} \v p^2  D^c_{ss}\left ( \v p , \omega \right ) = 0.
\end{equation}

The quartic order in $q$ provides the one-loop corrections to the {correlator $B_2$}. We will first analyze the effect of diffusons. Exploiting {the relation} $\left \langle q^{(0)}q^{(0)} \right \rangle = \left \langle q^{(2)}q^{(2)} \right \rangle$, we can simplify the {expression for} $B_2$ ($^\bigstar$ and $^\blacklozenge$ denote Wick contractions):
\begin{eqnarray}
B_2^{(ss')} &=& -  \frac{\sigma_s\sigma_{s'}}{8n} \int_{\v x - \v x '}\sum_{\mu = x,y} \notag \\
&& \left [\tr \left (I_n^\alpha q^\blacklozenge_0 \partial _\mu q_0^{T \bigstar}\right )_{s,\v x} \tr \left (I_n^\alpha q^\blacklozenge_0 \partial' _\mu q_0^{T \bigstar}\right )_{s',\v x'} \right . \notag \\
&& +\left .\tr \left (I_n^\alpha q_0^{T \blacklozenge} \partial_\mu q_0^\bigstar\right )_{s,\v x} \tr \left (I_n^\alpha q_0^{T \blacklozenge} \partial'_\mu q_0^\bigstar\right )_{s',\v x'} \right . \notag \\
&&  +2 \left .\tr \left (I_n^\alpha q_0^{T \bigstar} \partial_\mu q_0^\blacklozenge\right )_{s,\v x} \tr \left (I_n^\alpha q^\blacklozenge_0 \partial' _\mu q_0^{T \bigstar}\right )_{s',\v x'} \right ] . \notag \\&& \label{eq:B2calc1}
\end{eqnarray}
{The} Wick contraction produces three types of terms for each of the three terms in \eqref{eq:B2calc1} (see Eq. \eqref{eq:totaldiffusonpropagator}).
First there is the interference term $ D_s D_s$. It contains an additional sum over replicas and hence vanishes in the replica limit. 
Second, there can be a term $\left ( \underline{D\Gamma D^c}\right )_{ss'}\left ( \underline{D\Gamma D^c}\right )_{ss'}$. It vanishes due to its {structure in the Matsubara space}.
The only remaining term is {$\left ( \underline{D\Gamma D^c}\right )_{ss}D_s$ which yields}
\begin{align}
B_2^{(ss')} &=  \frac{32 \pi T \delta_{ss'}}{\sigma_s n}\int_{\v p} \v p^2 \sum_{n_{12}=0}^{N_M} n_{12}\notag \\
&\times \Bigl [ \left ( \underline{D\Gamma D^c}\right )_{ss} \left (\omega_{n_{12}}, {\v p}\right )D_s \left (\omega_{n_{12}+n}, {\v p}\right ) \notag \\
& - \left ( \underline{D\Gamma D^c}\right )_{ss} \left (\omega_{n_{12}+n}, {\v
p}\right )D_s \left (\omega_{n_{12}+2n}, {\v p}\right ) \Bigr ] .
\label{eqAppi3}
\end{align}

At this stage we can send $N_M \rightarrow \infty$. Furthermore, note that,
because disorder is surface uncorrelated, there is no correction to the
transconductance $\sigma_{12}$.
{Since we} are interested in the zero temperature {limit, Eq.~\eqref{eqAppi3} becomes}
\begin{equation}
B_2^{(ss')} =  \frac{16 \delta_{ss'}}{\sigma_s}\int_{\v p} \v p^2 \int\limits_{0}^{\infty} d\omega\, \left ( \underline{D\Gamma D^c}\right )_{ss} \left (\omega, {\v p}\right )D_s \left ( \omega, {\v p}\right ). \label{eq:B2Kuboformula}
\end{equation}
We use the relation
\begin{align}
\left ( \underline{D\Gamma D^c}\right )_{ss} \left (\omega, {\v p}\right )&D_s \left ( \omega, {\v p}\right ) =  \Gamma_{ss} D^2_s\left (\omega, {\v p} \right ) \overline{D}_s\left (\omega, {\v p} \right )  \notag \\
& - \frac{4\omega  \Gamma_{12}^2 D_s\left (\omega, {\v p} \right )\overline{D}_s\left (\omega, {\v p} \right )}{ \sigma_{(-s)} \det \left [\left (\underline D^c \left (\omega, {\v p} \right )\right )^{-1}\right ]} 
\end{align}
in order to split Eq.~\eqref{eq:B2Kuboformula} into the single surface and intersurface contributions. Here {$\sigma_{(-1)} = \sigma_{2}$, $\sigma_{(-2)} = \sigma_{1}$ and }
\begin{equation}
\overline{D}_s\left (\omega, \v p \right ) = \Bigl [\v p ^2 + L^{-2}+ 4 \left (z_s + \Gamma_{ss}\right ) \omega/\sigma_s \Bigr ]^{-1} .
\end{equation}
The single surface induced correction is {given as}
\begin{align}
\left . B_2^{ss'} \right \vert_{\text{single}}&=  \frac{16 \delta_{ss'}}{\sigma_s} \int_{\v p} \v p ^2 \int_{0}^{\infty} d\omega\, \Gamma_{ss} D^2_s\left (\omega, {\v p} \right ) \overline{D}_s\left (\omega,{\v p}\right ) \notag \\
 &= - 4 \delta_{ss'} f\left (\Gamma_{ss}/z_s\right ) \int_{\v p} \v p ^2 D^2_s(0, \v p). \label{eq:B2singlelayer}
\end{align}
{Here we} introduced the function 
\begin{equation}
f\left ( \v x \right ) = 1 - (1+1/x) \ln \left ( 1 +x \right ).
\end{equation}
For the intersurface interaction induced term we separate the poles of $\left [\det \left (\underline D^c \left (\omega, {\v p} \right )\right )^{-1}\right ]$. {It} yields
\begin{align}
\left . B_2^{ss'} \right \vert_{\text{inter}}&= \frac{64 \delta_{ss'} \Gamma_{12}^2}{\det \left (\underline z + \underline \Gamma\right )} \int_{\v p} \v p ^2 D_s(0, \v p) \int\limits_{0}^{\infty} d\omega\, \omega\,  D_s\left (\omega, {\v p} \right )   \notag \\
&  \times  \frac{\overline{D}_s\left (\omega, {\v p} \right )}{d_{+}- d_-} \sum_{\varsigma=\pm} \frac{\varsigma}{d_\varsigma \left (\v p^2 + L^{-2}\right )+ 4 \omega} \notag \\
&=\frac{2 \sigma_s^2\Gamma_{12}^2\delta_{ss'}}{z_s \left ( z_s+ \Gamma_{ss}\right ) \det \left ( \underline z + \underline \Gamma \right ) \left (d_+ - d_-\right )}\Biggl [\sum_{\varsigma=\pm} \varsigma  \notag \\
&\times f_2\left
(\frac{\sigma_s}{z_s},\frac{\sigma_s}{z_s+\Gamma_{ss}},d_\varsigma\right )
\Biggr ] \int_{\v p} \v p^2 D^2_s(0, \v p) ,
\end{align}
where 
\begin{equation}
d_\pm  = \frac{(z_1 \sigma_2 + \sigma_1 z_2)}{2\det \left (\underline z +
\underline \Gamma\right )}\left [ 1  \mp \sqrt {1 -\frac{4 \sigma_1\sigma_2\det
\left (\underline z + \underline \Gamma\right )}{(z_1 \sigma_2 + \sigma_1
z_2)^2}} \right ] 
\end{equation}
and
\begin{equation}
f_2  \left (a,b,c\right ) =  2 \frac{(c-b)a \ln a+(a-c)b\ln b+(b-a)c \ln c}{\left (b-a\right )(c-a)(c-b)}.
\end{equation}

In the {case of the long-range Coulomb interaction the condition} $\det \left (\underline z + \underline \Gamma\right ) = 0$ {holds. Therefore,} $d_-$ diverges and as a consequence $f_2\left (\frac{\sigma_s}{z_s},\frac{\sigma_s}{z_s+\Gamma_{ss}},d_-\right ) \rightarrow 0$. The contribution {due to} $d_+$ is then, in the exemplary case $s = 1$, {given as}
\begin{align}
\left . B_2^{11} \right \vert_{\text{inter}} &= -4 \left (1 + \frac{\Gamma_{11}}{z_1}\right )\left [\frac{\ln\left (1 + \frac{\Gamma_{11}}{z_1}\right )}{\frac{\Gamma_{11}}{z_1}}\right . \notag \\
&\left . - \frac{\ln\left (1 + \frac{\Gamma_{11}}{z_1} + \frac{\sigma_1 \left
(z_2 + \Gamma_{22}\right )}{\sigma_2 z_1}\right )}{\frac{\Gamma_{11}}{z_1}+
\frac{\sigma_1 \left (z_2 + \Gamma_{22}\right )}{\sigma_2 z_1}} \right ]\int_{\v
p} \v p^2 D^2_s(0,\v p) .
\end{align}

Finally, we consider the effect of cooperons in $B_2$. Due to the absence of
interaction amplitudes in the cooper channel all contributions are of the type
$D_s D_s$ and, in analogy with the corresponding diffuson terms, vanish in the
replica limit. 

\section{Stability of the fixed plane of equal surfaces}
\label{sec:stableequalsurfaces}

We discuss here the stability of the fixed plane of identical surfaces with
respect to small perturbations. As anticipated, it hosts the overall attractive
fixed point of the four dimensional RG flow (see also Sec.
\ref{sec:attractiveplane}) and thus is itself attractive. However, the
parameters describing the deviation from equal surfaces ($\delta t = t - 1$ and
$\delta \gamma = \gamma_{11} - \gamma_{22}$) flow towards zero in a
quite nontrivial manner.

From the general RG equations \eqref{eq:RGeqssigmatot} we obtain the {equations}
for small
deviations
\begin{equation}
 \frac{d}{dy}\left( \begin{array}{c}
\delta t \\
\delta \gamma
\end{array} \right ) =  \mathbf{M}\left (\gamma\right )
\left ( \begin{array}{c}
\delta t \\
\delta \gamma
\end{array} \right ) ,
\end{equation}
with the $\gamma$-dependent matrix
\begin{equation}
\mathbf{M} \left (\gamma\right ) = -\frac{2}{\pi \sigma} \left(
\begin{array}{cc}
 \frac{ (3+4 \gamma )G(\gamma)}{  (1+2 \gamma )^2} & \frac{2 G(\gamma)}{  (1+2
\gamma )^2} \\
 - \left(\gamma +\gamma ^2\right) &1+2 \gamma  \\
\end{array}
\right),
\end{equation}
and $G(\gamma) =  -1-2 \gamma +\left (2+2\gamma\right ) \ln (2+2 \gamma
)$.
The eigenvalues of the matrix $\mathbf{M}(\gamma)$ are shown in Fig.
\ref{fig:EWs}. They turn out to be complex in most of the interval $\gamma \in
\left [-1,0\right ]$ (except for a narrow region of very small $\gamma$).
This implies a curious oscillatory scale dependence of the difference of
conductivities $\delta t = 2(\sigma_1-\sigma_2)/\sigma$.
Although the fixed plane of equal surfaces
is repulsive in the regime $\gamma < \gamma_* \approx -0.64$ one should keep in mind that $\gamma$ itself is subjected to
renormalization, flowing towards zero and therefore, the plane of identical
surfaces becomes ultimately attractive.

\begin{figure}
\includegraphics[scale=0.9]{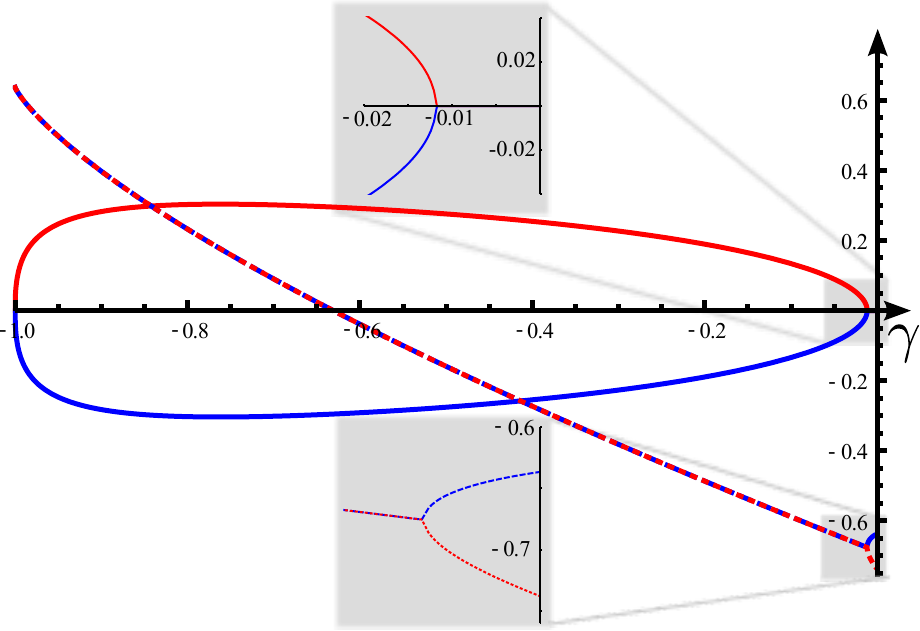}
\caption{Eigenvalues of $\mathbf M\left (\gamma\right )$ (in units of
$1/\sigma$) as a function of $\gamma $. Dashed  lines: real part, solid lines:
imaginary part.}
\label{fig:EWs}
\end{figure}

\section{RG flow for externally screened interaction}
\label{sec:shortrange}

If the single layer screening length $\kappa_s^{-1}$ and the typical length
scale $L_E$ (e.g. the thermal length) exceed the distance to the electrostatic
gates, the external screening of interactions can no longer be neglected.
Effectively, the interactions become short ranged. This implies the breakdown of
$\mathcal F$-invariance. As a consequence, the {relations for} NL$\sigma$M
parameters $\det \left (\underline z + \underline \Gamma\right ) = 0$ and
$(z_1 + \Gamma_{11})/(z_2 + \Gamma_{22}) = 1$ (derived in the case of
long-range interaction in Sec. \ref{sec:Finv} and Appendix
\ref{sec:appendixCoulomb}) {are} no longer true. Note that the invariance under renormalization of $\left (\underline z + \underline \Gamma\right ) $ is not a consequence of $\mathcal F$-invariance and still holds. 

Here we present general RG equations that allow us to describe the crossover
between the cases of  long-range Coulomb interaction and of no interaction:
\begin{subequations}
\begin{align}
\T \B \frac{d\sigma_1}{dy} & = \frac{2}{\pi} \Biggl [\frac{1}{2} -  f\left (\frac{\Gamma_{11}}{z_1}\right )  \notag \\
& - \frac{\sigma_1^2\Gamma_{12}^2 \sum\limits_{\varsigma=\pm} \varsigma f_2\left (\frac{\sigma_1}{z_1},\frac{\sigma_1}{z_1+\Gamma_{11}},d_\varsigma\right )}{2 z_1 \left ( z_1+ \Gamma_{11}\right ) \det \left ( \underline z + \underline \Gamma \right ) \left (d_- - d_+\right )} \Biggr ] \\
\T \B \frac{d\sigma_2}{dy} & = \frac{2}{\pi} \Biggl [\frac{1}{2} -  f\left
(\frac{\Gamma_{22}}{z_2}\right ) , 
\notag \\
& - \frac{\sigma_2^2\Gamma_{12}^2 \sum\limits_{\varsigma=\pm}\varsigma f_2\left
(\frac{\sigma_2}{z_2},\frac{\sigma_2}{z_2+\Gamma_{22}},d_\varsigma\right )}{2
z_2 \left ( z_2+ \Gamma_{22}\right ) \det \left ( \underline z + \underline
\Gamma \right ) \left (d_- - d_+\right )}\Biggr ] , \\
\T \B \frac{dz_1}{dy} &= -\frac{d\Gamma_{11}}{dy} = \frac{\Gamma_{11}}{\pi \sigma_1},  \\
\T \B \frac{dz_2}{dy} &= -\frac{d\Gamma_{22}}{dy} = \frac{\Gamma_{22}}{\pi \sigma_2}.
\end{align}
\label{eq:RGeqsshortrange}
\end{subequations}
In contrast to the Coulomb case (Eq. \eqref{eq:RGeqs}), these RG equations can
not be expressed in terms of the parameter $\gamma_{ss} =
\Gamma_{ss}/z_s$. Further, we emphasize that the RG equations {for}
$\Gamma_{ss}$ and $z_{s}$ are exactly the same as in the Coulomb case. In
particular, $\Gamma_{12}$ is not renormalized, since the general arguments
exposed in Sec. \ref{sec:Densityresponse} hold also in the case of short ranged
interactions. It is worthwhile to repeat that $0 \leq \vert \Gamma_{ss} \vert
\leq z_s$ and typically $\vert \Gamma_{12}\vert \leq \max_{s=1,2} \vert
\Gamma_{ss} \vert $. 

For sufficiently strong interactions, the RG flow implies localizing behaviour
of the conductivities. However, as the RG flow predicts decreasing interaction
amplitudes, the system undergoes a crossover to the free-electron
weak-antilocalization effect. (Note that also $\Gamma_{12}/z_s$ decreases.) 
Accordingly, similar to the case of Coulomb {interaction}, in the case of strong
short range interactions we also predict a non-monotonic conductivity behaviour.
The quantitative difference is the steeper antilocalizing slope in the final
stage of the flow.

\end{document}